%% file: main.tex
\documentclass[acmsmall, screen]{acmart}

\def\mode{revision} %

\usepackage{pdflscape}
\usepackage{makecell}
\usepackage{enumerate}
\usepackage[utf8]{inputenc}
\usepackage{CJKutf8}

\usepackage[flushleft]{threeparttable}
\usepackage{supertabular,booktabs}
\usepackage[figuresright]{rotating}
\usepackage{pifont} %
\usepackage[normalem]{ulem}
\useunder{\uline}{\ul}{}
\usepackage{bbm}
\usepackage{multirow}
\usepackage{booktabs}
\usepackage{subfigure}
\usepackage{graphicx}

\usepackage{color}
\usepackage{colortbl}
\usepackage{url}

\usepackage{tablefootnote}

\colorlet{BLUE}{blue} 
\colorlet{ORANGE}{orange}
\colorlet{BLACK}{black}
\usepackage[normalem]{ulem}

\newcommand{\revise}[1]{{\color{black}{#1}}}
\newcommand{\delete}[1]{}
\def\val{highlight}
\ifx\mode\val
    \newcommand{\newrevise}[1]{{\color{blue}{#1}}}
    \newcommand{\newdelete}[1]{{\color{orange}{\sout{#1}}}}
\else
    \newcommand{\newrevise}[1]{{\color{black}{#1}}}
    \newcommand{\newdelete}[1]{}
\fi

\usepackage{xspace}
\newcommand{\ie}{\textit{i.e.,}\xspace}
\newcommand{\eg}{\textit{e.g.,}\xspace}
\newcommand{\etc}{\textit{etc.}\xspace}
\newcommand{\etal}{\textit{et al.}\xspace}

\usepackage[most]{tcolorbox}
\newcommand{\find}[1]{
\begin{tcolorbox}[leftrule=1mm,toprule=0mm,bottomrule=0mm,left=1pt,right=2pt,top=2pt,bottom=2pt]%
\em #1
\end{tcolorbox}
}

\newcommand{\summary}[1]{
\begin{center}
\begin{tcolorbox}[colback=gray!15, colframe=black, boxsep=-0.15cm, middle=-0.15cm,breakable]
\textbf{\ding{46} Summary}
$\blacktriangleright$
{#1}
$\blacktriangleleft$
\end{tcolorbox}
\end{center}
}

\AtBeginDocument{
  \providecommand\BibTeX{{
    \normalfont B\kern-0.5em{\scshape i\kern-0.25em b}\kern-0.8em\TeX}}}

\setcopyright{acmcopyright}
\copyrightyear{2022}
\acmYear{2023}

\acmJournal{TOSEM}
\acmVolume{0}
\acmNumber{0}
\acmArticle{1}
\acmMonth{0}

\begin{document}

\title{A Survey of Learning-based Automated Program Repair}

\author{Quanjun Zhang}    \email{quanjun.zhang@smail.nju.edu.cn}
\orcid{0000-0002-2495-3805}
\affiliation{
  \institution{State Key Laboratory for Novel Software Technology, Nanjing University}
  \city{Nanjing}
  \state{Jiangsu}
  \country{China}
  \postcode{210093}
}

\author{Chunrong Fang} 
\email{fangchunrong@nju.edu.cn}
\orcid{0000-0002-9930-7111}
\authornote{\textbf{Chunrong Fang and Zhenyu Chen are the corresponding authors.}}
\affiliation{
  \institution{State Key Laboratory for Novel Software Technology, Nanjing University}
  \city{Nanjing}
  \state{Jiangsu}
  \country{China}
  \postcode{210093}
}

\author{Yuxiang Ma} \email{502022320009@smail.nju.edu.cn}
\orcid{0009-0007-6841-8410}
\affiliation{
  \institution{State Key Laboratory for Novel Software Technology, Nanjing University}
  \city{Nanjing}
  \state{Jiangsu}
  \country{China}
  \postcode{210093}
}

\author{Weisong Sun} \email{weisongsun@smail.nju.edu.cn}
\orcid{0000-0001-9236-8264}
\affiliation{
  \institution{State Key Laboratory for Novel Software Technology, Nanjing University}
  \city{Nanjing}
  \state{Jiangsu}
  \country{China}
  \postcode{210093}
}

\author{Zhenyu Chen} 
\authornotemark[1]
\orcid{0000-0002-9592-7022}
\email{zychen@nju.edu.cn}
\affiliation{
  \institution{State Key Laboratory for Novel Software Technology, Nanjing University}
  \city{Nanjing}
  \state{Jiangsu}
  \country{China}
  \postcode{210093}
}

\begin{abstract}
Automated program repair (APR) aims to fix software bugs automatically and plays a crucial role in software development and maintenance.
With the recent advances in deep learning (DL), an increasing number of APR techniques have been proposed to leverage neural networks to learn bug-fixing patterns from massive open-source code repositories.
Such learning-based techniques usually treat APR as a neural machine translation (NMT) task, where buggy code snippets (\ie source language) are translated into fixed code snippets (\ie target language) automatically.
Benefiting from the powerful capability of DL to learn hidden relationships from previous bug-fixing datasets, learning-based APR techniques have achieved remarkable performance.

In this paper, we provide a systematic survey to summarize the current state-of-the-art research in the learning-based APR community.
We illustrate the general workflow of learning-based APR techniques and detail the crucial components, including fault localization, patch generation, patch ranking, patch validation, and patch correctness phases.
We then discuss the widely adopted datasets and evaluation metrics and outline existing empirical studies.
We discuss several critical aspects of learning-based APR techniques, such as repair domains, industrial deployment, and the open science issue.
We highlight several practical guidelines on applying DL techniques for future APR studies, such as exploring explainable patch generation and utilizing code features.
Overall, our paper can help researchers gain a comprehensive understanding about the achievements of the existing learning-based APR techniques and promote the practical application of these techniques.
Our artifacts are publicly available at the repository:
\url{https://github.com/iSEngLab/AwesomeLearningAPR}.

\end{abstract}



\maketitle

\section{Introduction}
\label{s1}

Modern software systems continuously evolve with inevitable bugs due to the deprecation of old features, adding of new functionalities, and refactoring of system architecture~\cite{wang2017qtep}.
These inevitable bugs have been widely recognized as notoriously costly and destructive, such as costing billions of dollars annually across the world~\cite{boulder2019university,weiss2007long}.
The recorded quantity of bugs is increased at a tremendous speed due to the increasing scale and complexity of software systems~\cite{gazzola2019automatic}.
\delete{Manual debugging can be an extremely time-consuming and error-prone task in the software development and maintenance process.}
\revise{It is an extremely time-consuming and error-prone task for developers to fix detected bugs manually in the software development and maintenance process.}
For example, previous reports show that software debugging accounts for over 50\% of the cost in software development~\cite{britton2013reversible}.
Considering the promising future in relieving manual \delete{debugging}\revise{repair} efforts, automated program repair (APR), which aims to automatically fix software bugs without human intervention, has been a very active area in academia and industry.

As a promising research area, APR has been extensively investigated in the literature and has made substantial progress on the number of correctly-fixed bugs~\cite{monperrus2018automatic}.
\delete{A living APR review reports~\cite{monperrus2020living}}\revise{A living APR review~\cite{monperrus2020living} reports} that a growing number of papers get published each year with various exquisitely implemented APR tools being released.
Over the past decade, researchers have proposed a variety of APR techniques to generate patches~\cite{liu2020efficiency}~\cite{benton2020effectiveness}~\cite{wang2020automated}, including \textit{heuristic-based}, \textit{constraint-based} and \textit{pattern-based}.
Among these traditional techniques, \textit{pattern-based} APR employs pre-defined repair patterns to transform buggy code snippets into correct ones and has been widely recognized as state-of-the-art~\cite{xia2022less,liu2019tbar,xia2022practical}.
However, existing pattern-based techniques mainly rely on manually designed repair templates, which require massive effort and professional knowledge to craft in practice.
Besides, these templates are usually designed for specific types of bugs (\eg null pointer exception) and thus are challenging to apply to unseen bugs, limiting the repair effectiveness.

Recently, inspired by the advance of deep learning (DL), a variety of learning-based APR techniques have been proposed to learn the bug-fixing patterns automatically from large corpora of source code~\cite{tufano2019empirical}.
Compared with traditional APR techniques, learning-based techniques can be applied to a wider range of scenarios (\eg multi-languages~\cite{xia2022less} \revise{and multiple multi-hunks~\cite{chen2022neural}}) with \delete{parallel bug-fixing pairs}\revise{pairs of the buggy and corresponding fixed code snippets}.
\revise{For example, CIRCLE~\cite{yuan2022circle} is able to generate patches across multiple programming languages with multilingual training datasets.} 
These learning-based techniques handle the program repair problem as a neural machine translation (NMT) task\revise{~\cite{yuan2022circle,lutellier2020coconut,jiang2021cure,xia2022practical,xia2022less}}, which translates a code sequence from a source language (\ie buggy code snippets) into a target language (\ie correct code snippets).
Existing NMT repair models are typically built on the top of the \textit{encoder-decoder} architecture~\cite{vaswani2017attention}.
The \textit{encoder} extracts the hidden status of buggy code snippets with the necessary context, and the \textit{decoder} takes the encoder's hidden status and generates the correct code snippets~\cite{long2016automatic,jiang2018shaping, li2020dlfix}.
Thanks to the powerful ability of DL to learn hidden and intricate relationships from massive code corpora, learning-based APR techniques have achieved remarkable performance in the last couple of years.

The impressive progress of learning-based APR has shown the substantial benefits of exploiting DL for APR and further revealed its promising future in follow-up research.
However, a mass of existing studies from different organizations (\eg academia and industry) and communities (\eg software engineering and artificial intelligence) make it difficult for interested researchers to understand state-of-the-art and improve upon them.
More importantly, compared with traditional techniques, learning-based techniques heavily rely on the quality of code corpora and model architectures, posing several challenges (\eg code representation and patch ranking) in developing mature NMT repair models.
For example, most learning-based techniques adopt different training datasets, and there exist various strategies \revise{available} to process the code snippets (\eg the code context, abstraction, and tokenization).
Besides, researchers design different code representations (\eg sequence, tree, and graph) to extract code features, which require corresponding encoder-decoder architectures (\eg RNN, LSTM, and transformer) to learn the transformation patterns.
Furthermore, execution-based (\eg plausible and \delete{correctness}\revise{correct} patches) and match-based (\eg accuracy and BLUE) metrics are adopted in different studies.
Such multitudinous design choices hinder developers from conducting follow-up research on the learning-based APR direction.

In this paper, we summarize existing work and provide a retrospection of the learning-based APR field after years of development. 
Community researchers can have a thorough understanding of the advantages and limitations of the existing learning-based APR techniques.
We illustrate the typical workflow of learning-based APR and discuss different detailed techniques that appeared in the papers we collected. 
Based on our analysis, we point out the current challenges and suggest possible future directions for learning-based APR research.
Overall, our work provides a comprehensive review of the current progress of the learning-based APR community, enabling researchers to obtain an overview of this thriving field and make progress toward advanced practices.

\textbf{{Contributions.}}
To sum up, the main contributions of this paper are as follows:
\begin{itemize}
    \item Survey Methodology.
    We conduct a detailed analysis of 112 relevant studies that used DL techniques in terms of publication trends, distribution of publication venues and languages.

    \item \textit{Learning-based APR.}
    We describe the typical framework of leveraging advances in DL techniques to repair software bugs and discuss the key factors, including fault localization, data pre-processing, patch generation, patch ranking, patch validation and patch correctness.
    
    \item \textit{Dataset and Metric.}
    We perform a comprehensive analysis of the critical factors that impact the performance of DL models in APR, including 53 collected datasets and evaluation metrics in two categories.
    
    \item \textit{Empirical studies}.
    We detail existing empirical studies performed to better understand the process of learning-based APR and facilitate future studies.
    
    \item \textit{Some Discussions.}
    We discuss some other crucial areas (\eg security vulnerability and syntax error) where learning-based APR techniques are applied, as well as certain known industrial deployments.
    We demonstrate the trend of employing pre-trained models on APR recently.
    We list the available learning-based tools and reveal the essential open science problem.
    
    \item \textit{{Outlook and challenges.}}
    We pinpoint open research challenges of using DL in APR and provide several practical guidelines on applying DL for future learning-based APR studies.
\end{itemize}  

\textbf{{Comparison with Existing Surveys.}}
Gazzola~\etal~\cite{gazzola2019automatic} present a survey to organize the repair techniques published up to January 2017.
Monperrus~\etal~\cite{monperrus2018automatic} present a bibliography of behavioral and state repair papers.
Unlike existing surveys mainly covering traditional techniques, our work focuses on the learning-based APR, particularly the integration of DL techniques in the repair phase (\eg patch generation and correctness), repair domains (\eg vulnerability and syntax errors), and challenges.
Besides, our survey summarizes the existing studies until Nov 2022.

\textbf{{Paper Organization.}}
The remainder of this paper is organized as follows. 
\delete{Section \ref{sec:methodology} presents the research methodology about how we collect relevant papers from several databases following specific keywords.
Section \ref{sec:bk} introduces some common concepts encountered in the learning-based APR field. 
Section \ref{sec:apr} presents the typical workflow of learning-based APR and discusses the vital components of the workflow in detail.
Section \ref{sec:evaluation} extends the discussion on the collection of datasets and standard evaluation metrics of learning-based APR techniques.
Section \ref{sec:dis} details some discussions, including repair applications, industrial deployments, employment of pre-trained models and the open science problem.
Section \ref{sec:guide} provides practical guidelines.
Section \ref{sec:con} draws the conclusions.}
\revise{Section ~\ref{sec:methodology} presents the research methodology about how we collect relevant papers from several databases following specific keywords.
Section~\ref{sec:bk} introduces some common concepts encountered in the learning-based APR field. 
Section ~\ref{sec:apr} presents the typical workflow of learning-based APR and discusses the vital components of the workflow in detail, as well as some representative approaches across different repair domains.
Section~\ref{sec:pre-trained} focuses on pre-trained model-based APR, which is the recent hot topic in the learning-based APR community.
Section~\ref{sec:evaluation} extends the discussion on the empirical evaluation, including common datasets, standard evaluation metrics, and existing empirical studies of learning-based APR techniques.
Section~\ref{sec:dis} details some discussions, including industrial deployments, traditional APR equipped with learning-based techniques, and the crucial open science problem.
Section~\ref{sec:guide} provides some practical guidelines.
Section~\ref{sec:con} draws the conclusions.}

\revise{
{\bf Availability.} All artifacts of this study are available in the following public repository:
\begin{center}
\url{https://github.com/iSEngLab/AwesomeLearningAPR}
\end{center}
}

\section{Survey Methodology}
\label{sec:methodology}
In this section, we present details of our systematic literature review methodology following \revise{Petersen~\etal~\cite{petersen2015guidelines} and Kitchenham~\etal~\cite{kitchenham2007guidelines}}.

\textit{Search Process.}
For this survey, we select papers by mainly searching the Google Scholar repository, ACM Digital Library, and IEEE Explorer Digital Library at the end of November 2022.
Following existing DL for SE surveys~\cite{wang2022machine,yang2022survey}, we divide the search keywords used for searching papers into two groups: (1) \delete{a}\revise{an} APR-related group containing some commonly used keywords related to program repair;
and (2) a DL-related group containing some keywords related to deep learning or machine learning.
Considering a significant amount of relevant papers from both SE and AI communities, following Zhang~\etal~\cite{zhang2011identifying}, we first try to collect some papers from the community-driven website\footnote{\url{http://program-repair.org/bibliography.html}} and the living review of APR by Monperrus~\cite{monperrus2020living}, and then conclude some frequent words in the titles of these papers. 
The search strategy can capture
the most relevant studies while achieving better efficiency than a purely manual search.
Finally, we identify a search string including several DL-related terms frequently appearing in APR papers that make use of DL techniques, listed as follows.

\begin{center}
    \setlength{\fboxrule}{0.5pt}
    \fbox{
    \parbox{0.95\textwidth}{%
    \textit{
    (``program repair'' OR ``software repair'' OR ``automatic repair'' OR ``code repair'' OR ``bug repair'' OR ``bug fix'' OR ``code fix'' OR ``automatic fix'' OR ``patch generation'' OR ``fix generation'' OR ``code transformation'' OR ``code edit'' OR ``fix error'')
    AND (``neural'' OR ``machine'' OR ``deep'' OR ``learning'' OR ``transformer/transformers'' OR ``model/models'' OR ``transfer'' OR ``supervised'')
    }
    }%
    }
\end{center}

\begin{figure}[t]
    \centering
    \graphicspath{{graphs/}}
    \includegraphics[width=0.8\linewidth]{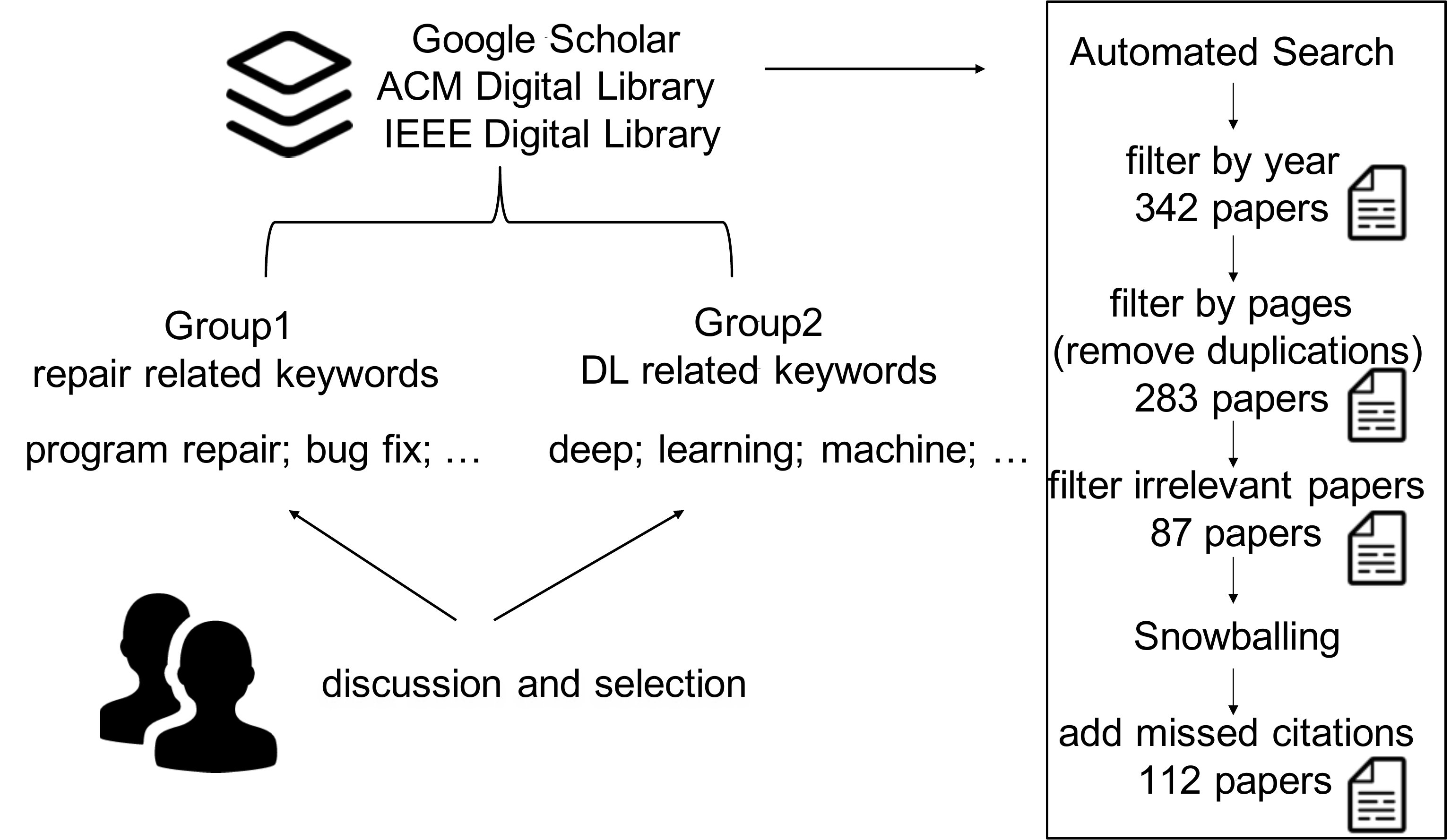}
    \caption{General workflow of the paper collection}
    \label{fig:paper collection procedure}
\end{figure}

\textit{Study selection.}
Once the potentially relevant studies based on our search strategy are collected, we perform a filtering and deduplication phase to exclude papers not aligned with the study goals.
We first attempt to filter out the papers before 2016, considering that Long~\etal~\cite{long2016automatic} propose the first learning-based APR study in 2016.
We then filter out any paper less than 7 pages and \delete{the }duplicated papers, resulting in 283 papers in total.
We then scrutinize the remaining papers manually to decide whether \delete{it is}\revise{they are} relevant to the learning-based APR field. 
We obtained 87 papers at last.
\delete{
To be as much comprehensive as possible, we include the relevant papers that we miss with our searches but were cited in the papers we selected.}
\revise{
To ensure that the collected papers are as comprehensive as possible, we further perform the common practice snowballing to manually include other relevant papers that are missed in our search process~\cite{watson2022systematic}.
In particular, we look at every reference within the collected papers and determine if any of those references are relevant to our study.}
\revise{For example, the title of SampleFix~\cite{hajipour2021samplefix} does not contain any keywords we mention above in the two groups, but it is an APR approach targeting syntax errors, so we include it in our survey.}
We manually analyzed all these cited papers by scanning the papers and finally collected 112 papers in our survey.
The general workflow of how we collected papers is shown in Figure \ref{fig:paper collection procedure}.

\begin{figure}[htbp]
\centering

\begin{minipage}[t]{0.48\textwidth}
\centering
    \includegraphics[width=6cm]{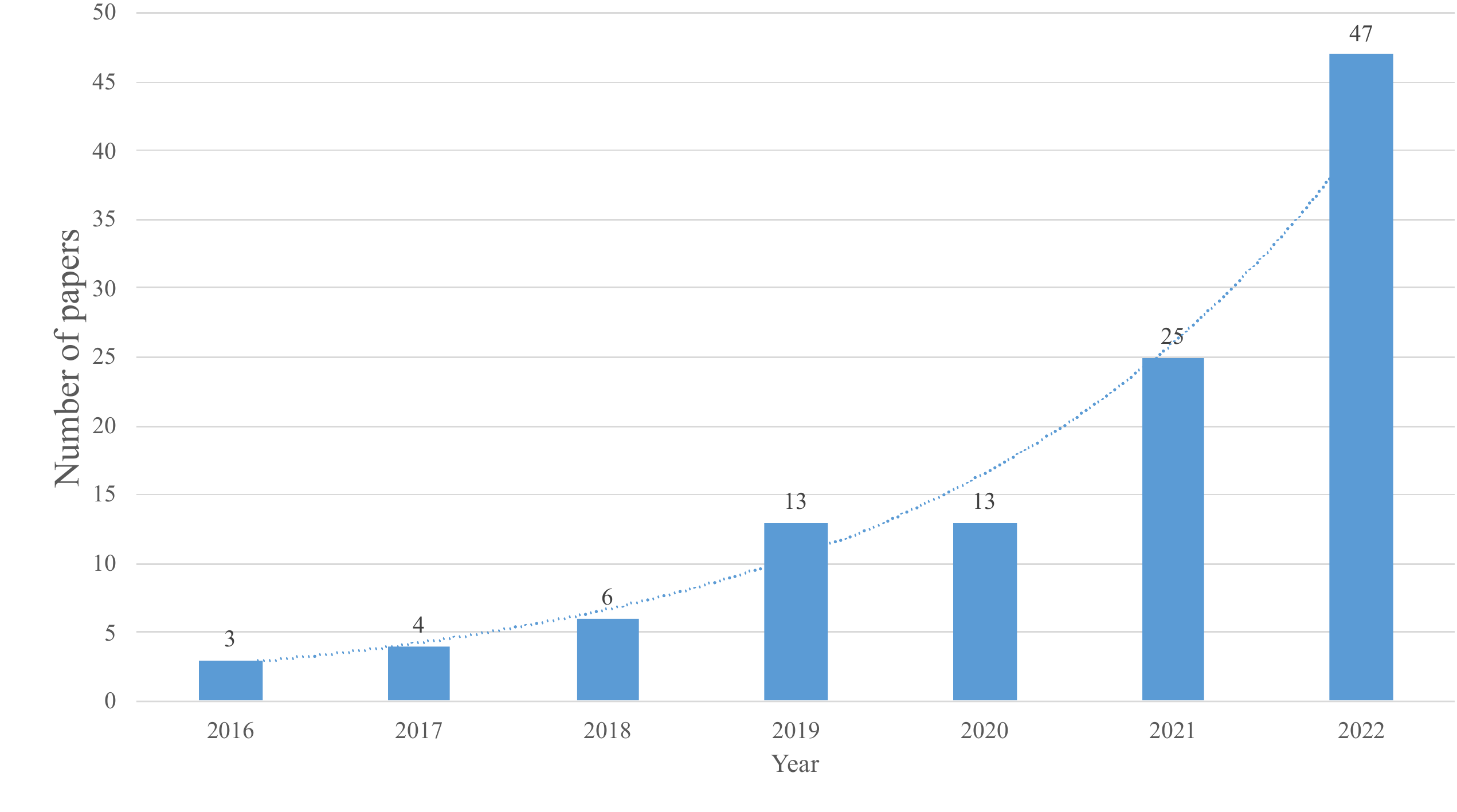}
    \caption{\revise{Collected learning-based APR papers from 2016 to 2022}}
    \label{fig:learning-based apr papers}
\end{minipage}
\begin{minipage}[t]{0.48\textwidth}
\centering
    \includegraphics[width=6cm]{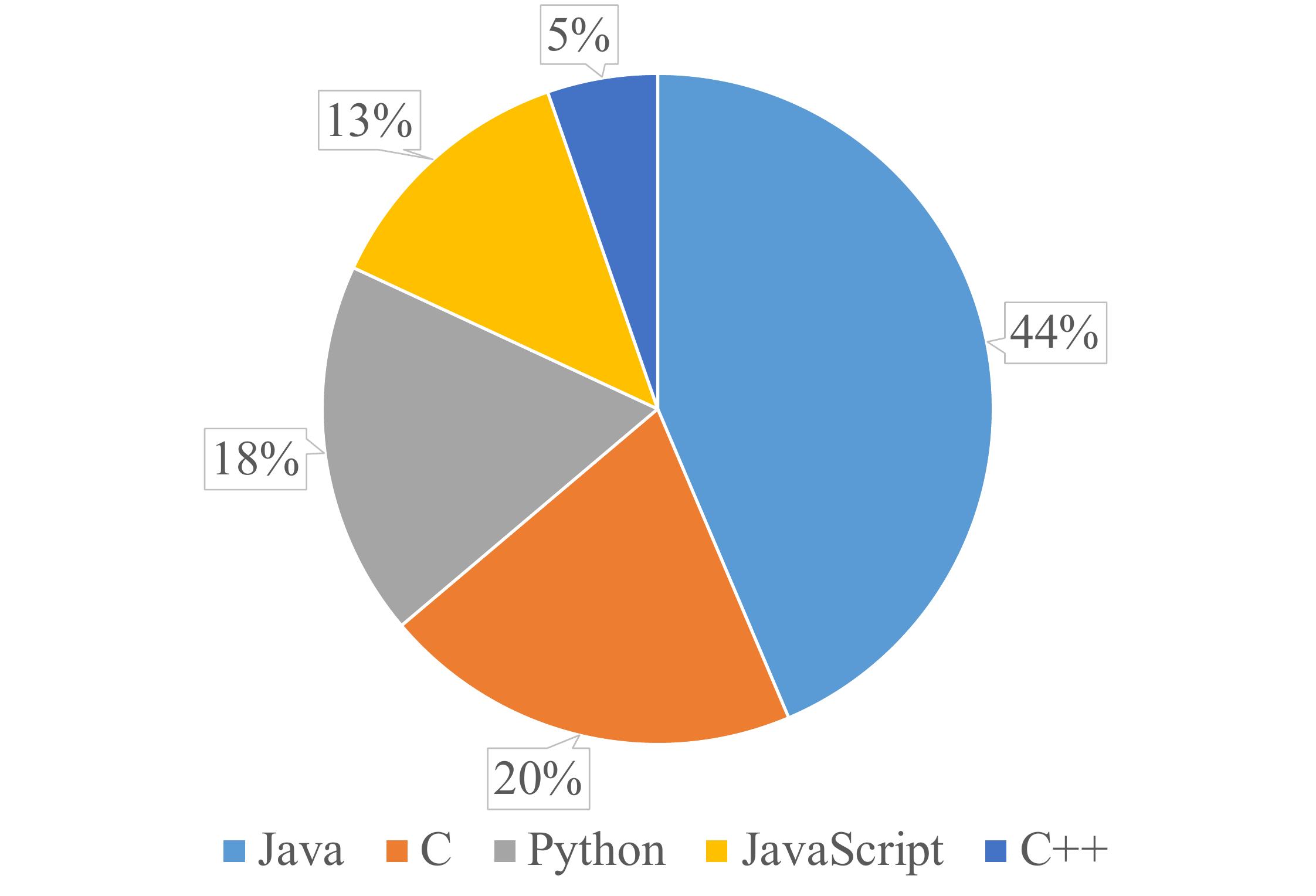}
    \caption{\revise{Paper distribution on programming languages}}
    \label{fig:paper_distribution}
\end{minipage}

\end{figure}

\textit{Trend Observation.}
Figure \ref{fig:learning-based apr papers} shows the collected papers from 2016 to 2022. 
It is found that the number of learning-based APR papers has increased rapidly since 2020, indicating that more researchers are considering DL as a promising solution to fixing software bugs.
\revise{One reason behind this phenomenon is that traditional APR techniques have reached a plateau~\cite{yang2022transplantfix,lutellier2020coconut} and researchers hope to find a brand-new way to address the problem.}
Another non-negligible reason is that DL has proved its potential in various tasks, including natural language translation, which is similar to bug fixing to some extent.
Figure \ref{fig:paper_distribution} presents an overview of the programming languages targeted by learning-based APR techniques in our survey.
We can find Java occupies a large proportion, which is understandable as Java is widely adopted in modern software systems nowadays and the most targeted language in existing mature datasets (\revise{\eg Defects4J~\cite{just2014defects4j}}).
We also find that the collected papers cover a wide range of programming languages (\ie Java, JavaScript, Python, C, and C++).
For example, there exist several papers~\cite{yuan2022circle,lutellier2020coconut} involving multiple programming language repair.
The probable reason may be that learning-based APR techniques usually regard APR as an NMT problem, independent of programming languages.

\section{Background and Concepts}
\label{sec:bk}
In this section, we will introduce some background information and common concepts in the learning-based APR field.

\begin{figure}[t]
\centering
\graphicspath{{graphs/}}
    \includegraphics[width=0.7\linewidth]{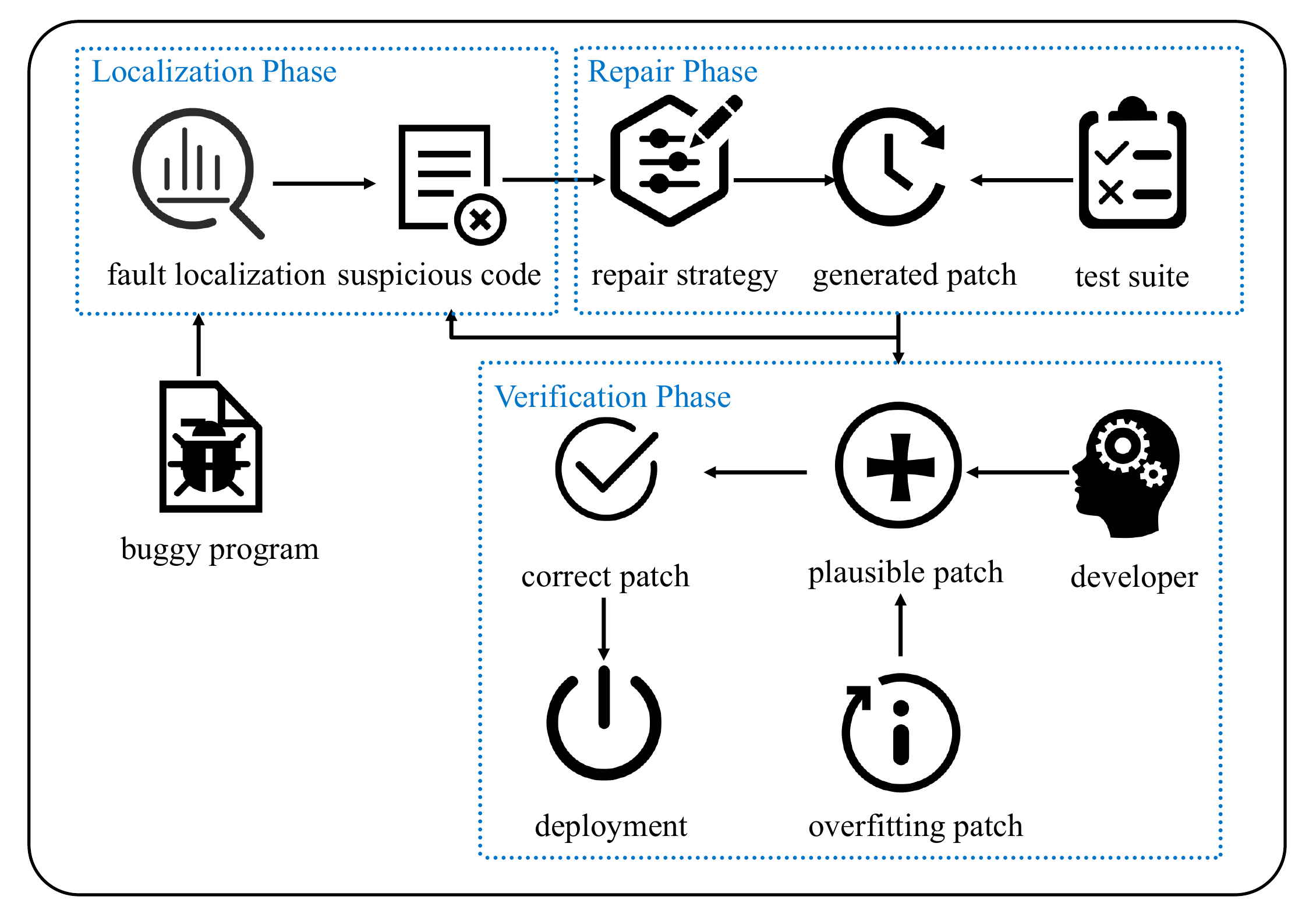}
    \caption{Overview of APR}
    \label{fig:apr}
\end{figure}

\subsection{Automated Program Repair}
The primary objective of APR techniques is to identify and fix software bugs without human intervention.
In the software development and maintenance process, after a designed functionality is implemented, developers usually write some test suites (\eg Junit test cases) to check the functionality.
If there exist test suites that make the functionality fail, developers adopt the failing test suites to analyze the symptoms and the root cause of the bug, and attempt to fix the bug by making some changes to suspicious code elements.
More generally, \revise{according to Nilizadeh\etal~\cite{nilizadeh2022realistic}}, we can give the following definition.

\find{
\begin{definition}
\ding{45} \textbf{APR:}
Given a buggy program $P$, \delete{the corresponding specification $S$ that makes $P$ fail}\revise{the corresponding specification $S$ that $P$ does not satisfy}, the transformation operators $O$ and the allowed maximum edit distance $\epsilon$,
APR can be formalized as a function $APR(P,S,O, \epsilon)$.
$PT$ is the set of its all possible program variants by enumerating all operators $O$ on $P$.
The problem of APR is to find a program variant $P'$ ($P'\in PT$) that satisfies $S$ and the changes satisfies $\epsilon$ ($distance(P,P') \leq \epsilon$).
\end{definition}
}

The specification $S$ denotes a relation between inputs and outputs and most APR techniques usually adopt a test suite as a specification.
In other words, \textit{APR aims to find a minimal change to $P$ that passes all available test suites}.
The maximum edit distance $\epsilon$ limits the range of changes based on the \textit{competent programmer hypothesis}~\cite{offutt1994empirical}, which assumes that experienced programmers are capable of writing almost correct programs and most bugs can be fixed by small changes.
\delete{If $\epsilon$ is set to $\infty$, $APR(P,S,O, \epsilon)$ becomes a program synthesizing problem that aims to synthesize a program to satisfy $S$.}
\revise{For example, if $\epsilon$ is set to $0$, $APR(P,S,O, 0)$ becomes a program validation problem that aims to identify if $P$ satisfys $S$.
On the contrary, if $\epsilon$ is set to $\infty$, $APR(P,S,O, \epsilon)$ becomes a program synthesizing problem that aims to synthesize a program to satisfy $S$.
}

The typical workflow of APR techniques is illustrated in Figure \ref{fig:apr}, which is usually composed of three parts: 
(1) off-the-shelf fault localization techniques are applied to outline the buggy code snippets~\cite{abreu2007accuracy}~\cite{ayewah2008using}; 
(2) these snippets are modified based on a set of transformation rules or patterns to generate new various program variants (\ie \textit{candidate patches}); 
(3) the original test suite is adopted as the oracle to verify all candidate patches.
Specifically, a candidate patch passing the original test suite is called a \textit{plausible} patch.
A plausible patch, which is also semantically equivalent to the developer patch, denotes a \textit{correct} patch.

However, such specifications (\ie test suites) are inherently incomplete as programs have infinite domains.
It is fundamentally challenging to ensure the correctness of the plausible patches (\ie overfitting issue) due to the weak test suites in practice.
Existing studies have demonstrated that manually identifying the overfitting patches is time-consuming and may harm the debugging performance of developers~\cite{tao2014automatically, smith2015cure}.
The overfitting issue is a critical challenge in both traditional and learning-based APR techniques.
We will discuss the issue in Section \ref{sec:pat_correctness}.

\subsubsection{Patch Generation Techniques}
In the literature, numerous traditional APR techniques have been proposed to generate patches from different aspects, which can be categorized into three classes.
We list them as follows.

\begin{itemize}
 
\item
\emph{Heuristic-based repair techniques}.
These techniques usually apply heuristic strategies (\eg genetic algorithm) to build search space from previous patches and generate valid patches by exploring the search space~\cite{le2012genprog, martinez2016astor,yuan2018arja}.
For example, SimFix~\cite{jiang2018shaping} builds an abstract search space from existing patches and a concrete search space from similar code snippets in the buggy project.
SimFix then utilizes the intersection of the above two search spaces to search the final patch by basic heuristics (\eg syntactic distance).

\item
\emph{Constraint-based repair techniques}.
\delete{These techniques usually treat APR as a constraint-solving task and rely on SMT solvers to return a feasible solution.}
\revise{These techniques usually focus on a single conditional expression and employ advanced constraint-solving or synthesis techniques to synthesize candidate patches~\cite{martinez2018ultra,durieux2016dynamoth, mechtaev2016angelix}.} 
For example, Nopol~\cite{Xuan2016Nopol} relies on an SMT solver to solve the condition synthesis problem after identifying potential locations of patches by angelic \delete{fix}\revise{fault} localization and collecting test execution traces of the program.
\revise{Besides, Cardumen~\cite{martinez2018ultra} synthesizes candidate patches at the level of expressions with its mined templates from the program under repair to replace the buggy expression.}

\item
\emph{Pattern-based repair techniques}.
These techniques usually design certain repair templates by manually analyzing specific software bugs and generating patches by applying such templates to buggy code snippets~\cite{koyuncu2020fixminer,Liu2019Avatar,liu2019tbar}.
For example, TBar~\cite{liu2019tbar} revisits the effectiveness of pattern-based APR techniques by systematically summarizing a variety of repair patterns from the literature. 
\end{itemize}

\revise{In addition to the above traditional APR techniques, researchers attempt to fix software bugs enriched by DL techniques due to the large-scale open-source source code repositories~\cite{tufano2019empirical,zhu2021syntax}.}
Such learning-based techniques have demonstrated promising results and are getting growing attention recently, which is the focus of our work (introduced in Section \ref{sec:NMT}).

\subsection{Neural Machine Translation}
\label{sec:NMT}

Sequence-to-sequence (Seq2Seq) is an advanced DL framework widely used in some NLP tasks (\eg machine translation~\cite{johnson2017google} and text summarization~\cite{nallapati2016abstractive}).
A Seq2Seq model usually consists of two components (\ie an encoder and a decoder) to learn mappings between two sequences.
Inspired by the success of Seq2Seq models in text generation tasks, program repair can be formulated as an NMT task.
The learning-based APR problem is formally defined as follows:

\find{
\begin{definition}\textbf{\ding{45} Learning-based APR:} 
Given a buggy code snippet $X_i=\left[x_1, \ldots, x_n\right]$ with $n$ code tokens and a fixed code snippet $Y_i=\left[y_1, \ldots, y_m\right]$ with $m$ code tokens, the problem of program repair is formalized to maximize the conditional probability \revise{(\ie the likelihood of $Y$ being the correct fix)}: $P\left(Y \mid X \right) = \prod_{i=1}^{m} P\left(y_i \mid y_1, \ldots, y_{i-1}; x_1, \ldots, x_n\right)$.
\end{definition}
}

In other words, the objective of an NMT repair model is to learn the mapping between a buggy code snippet $X$ and a fixed code snippet $Y$.
Then the parameters of the model are updated by using the training dataset, so as to optimize the mapping (\ie maximizing \revise{the conditional probability} $P$).
In the literature, recurrent neural network architecture (RNN) is widely used in existing learning-based APR techniques~\cite{chen2019sequencer,gupta2017deepfix,tufano2019empirical, tufano2019learning}.
Besides, researchers use long short-term memory (LSTM) architecture to capture the long-distance dependencies among code sequences~\cite{Chakraborty2020codit, meng2022improving}.
Recently, as a variant of the Seq2Seq model, Transformer~\cite{vaswani2017attention} has been considered the state-of-the-art NMT repair architecture due to the self-attention mechanism~\cite{fu2022vulrepair,chi2022seqtrans,chen2022neural}.

\section{Learning-based APR}
\label{sec:apr}
In this section, we will discuss the workflow of learning-based APR tools and introduce some popular learning-based APR techniques with several examples.

\begin{figure}[t]
    \centering
    \graphicspath{{graphs/}}
    \includegraphics[width=0.95\linewidth]{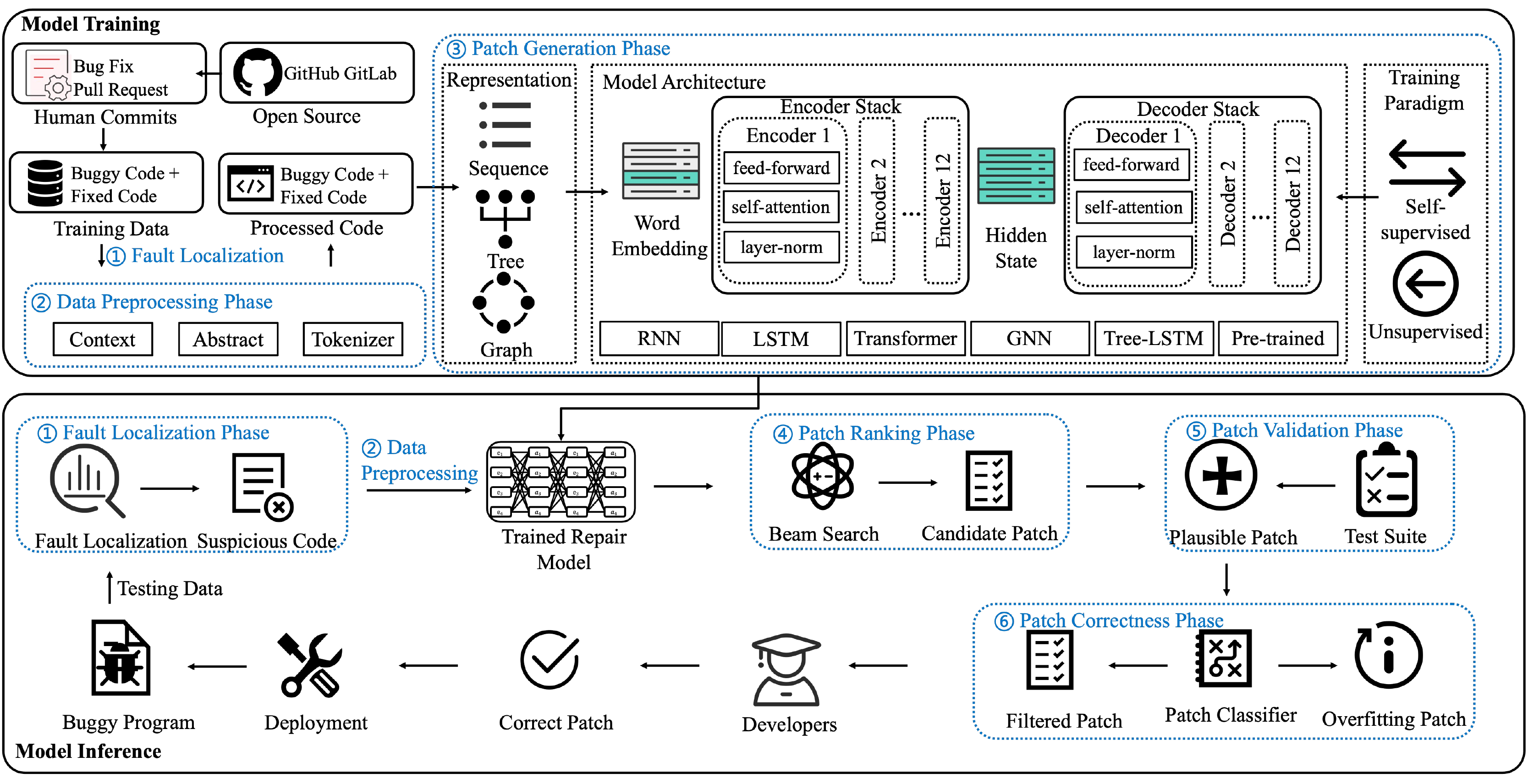}
    \caption{\revise{Detailed workflow of Learning-based APR}}
    \label{fig:detailed DL apr workflow}
\end{figure}

\subsection{Overall Workflow}

Figure \ref{fig:detailed DL apr workflow} illustrates the typical framework of existing learning-based APR techniques.
The framework can be generally divided into \delete{seven phrases}\revise{six phases}: \textit{fault localization}, \textit{data pre-processing}, \textit{input encoding}, \textit{output decoding}, \textit{patch ranking}, \textit{patch validation}, and \textit{patch correctness assessment}.
We now discuss the \delete{phrases}\revise{phases} in detail as follows.

\begin{itemize}

\item[\ding{172}] \textbf{In the fault localization phase}, 
a given buggy program is taken as the input and a list of suspicious code elements (\eg statements or methods) is returned~\cite{wong2016survey}, \revise{detailed in Section~\ref{sec:fl}}.

\item[\ding{173}] \textbf{In the data pre-processing phase},
a given software buggy code snippet (\eg buggy statement) is taken as the input and the processed code tokens are returned.
According to existing learning-based APR studies~\cite{chen2022neural,chi2022seqtrans}, there generally exist three potential ways to pre-process the buggy code: code context, abstraction, and tokenization.
First, \textit{code context} information refers to other correlated non-buggy lines within the buggy program~\cite{namavar2022controlled}.
Previous work has demonstrated that NMT-based repair models reveal diverse code changes to fix bugs under different contexts~\cite{chen2019sequencer}.
Second, \textit{code abstraction} renames some special words (\eg string and number literals) to a pool of predefined tokens, which has been proven to be an effective method in reducing the vocabulary size\revise{~\cite{tufano2019empirical}}.
Third, \textit{code tokenization} splits source code into words or subwords, which are then converted to ids through a look-up table~\cite{fu2022vulrepair}. \revise{These pre-processing methods are detailed in Section~\ref{sec:preprocessing}}.

\item[\ding{174}] \textbf{In the patch generation phase}, 
the processed code tokens are first fed into a word embedding stack to generate representation vectors, which can capture the semantic meaning of code tokens and their position within a buggy code.
Then an encoder stack is implemented to derive the encoder's hidden state, which is further passed into a decoder stack.
Similar to the encoder stack, a decoder stack is implemented to take the hidden states provided by the encoder stack and previously generated tokens as inputs, and returns the probability distribution of the vocabulary.
\revise{There exist two training paradigms to learn bug-fixing patterns automatically, \ie unsupervised learning~\cite{tufano2019empirical,ciniselli2021empiricaltse} and self-supervised learning~\cite{ye2022selfapr,yasunaga2021break}, \revise{detailed in Section~\ref{sec:patch_generation}}.}

\item[\ding{175}] \textbf{In the patch ranking phase},
after the NMT-based repair model is well-trained, a rank strategy (\eg beam search) is leveraged to prioritize the candidate patches as prediction results based on the probability distribution of the vocabulary~\cite{smith2015cure}.
\revise{
Particularly, beam search~\cite{chen2019sequencer,yuan2022circle,allamanis2021self} is a common practice to select several most high-scoring candidate patches by iteratively ranking top-$k$ probable tokens based on their estimated likelihood scores, detailed in Section~\ref{sec:patch_ranking}.
}

\item[\ding{176}] \textbf{In the patch validation phase},
the generated candidate patches are then verified by the available program specification, such as functional test suites or static analysis tools~\cite{benton2022towards}, \revise{detailed in Section~\ref{sec:patch_validation}}.

\item[\ding{177}] \textbf{In the patch correctness assessment phase},
the plausible patches (\ie passing the existing specification) are assessed to predict their correctness (\ie whether the plausible are overfitting)~\cite{wang2020automated}, which are finally manually checked by developers for deployment in the software pipeline, \revise{detailed in Section~\ref{sec:pat_correctness}}.

\end{itemize}

\subsection{Fault Localization}
\label{sec:fl}

Fault localization aims to diagnose buggy program elements (\eg statements and methods) without human intervention and has been extensively studied to facilitate the program repair process~\cite{wong2016survey}.
As a crucial start in the learning-based APR pipeline, fault localization provides the repair model with information about where a software bug is and directly influences the performance of the repair model.
For example, the repair accuracy under normal fault localization is usually lower than the circumstance under perfect fault localization.

In the literature, fault localization techniques often leverage various static analysis or dynamic execution information to compute suspiciousness scores (\ie probability of being faulty) for each program element.
Program elements are then ranked in descending order of their suspiciousness scores, based on which APR techniques can further be applied.
Researchers have proposed a variety of fault localization techniques, such as spectrum-based~\cite{zhang2019empirical, pearson2017evaluating}, mutation-based~\cite{papadakis2015metallaxis,li2017transforming}, slicing-based~\cite{mao2014slice, bayati2020smbfl} and learning-based~\cite{lou2021boosting,li2021fault} techniques.
Among them, spectrum-based fault localization (SBFL) has been extensively utilized as a general mechanism to localize the statements that are likely to be faulty in the APR literature.

\subsubsection{Localization Techniques}
Similar to traditional APR techniques, some learning-based APR techniques rely on existing SBFL fault localization approaches to localize the revealed bug.
For example, DLFix~\cite{li2020dlfix} adopts Ochiai algorithm to identify a buggy line and extracts all AST nodes (including intermediate ones) related to that buggy line as a replaced subtree for patch generation.
Recoder~\cite{zhu2021syntax} also assumes the faulty location of a bug is unknown to APR tools and uses Ochiai algorithm with GZoltar~\cite{riboira2010gzoltar}, which is widely used in existing APR tools, such as RewardRepair~\cite{ye2022neural} and AlphaRepair~\cite{xia2022less}.
Such SBFL techniques exploit runtime information to recognize the program elements that are likely to be faulty when the buggy program is executed by the available test suite.
The crucial insight is that (1) the program elements executed by more failing test suites and fewer passing test suites are likely to be faulty;
and (2) the program elements executed by more passing test suites and fewer failing suites are likely to be correct.
In particular, SBFL produces a list of program elements ranked according to their likelihood of being faulty based on the analysis of the program entities covered by passing and failing tests (\eg Ochiai and Tarantula~\cite{liu2019you}).

However, Liu~\etal~\cite{liu2019you} have demonstrated that the fault localization techniques may introduce a significant bias in the evaluation of APR techniques.
The vast majority of learning-based APR techniques consider repairing software bugs under perfect-based fault localization techniques.
Perfect-based fault localization techniques assume that the genuine localization of the bug is known.
Thus, perfect-based fault localization can provide a fair assessment of APR techniques and the assessment is independent of the localization techniques.
For example, CoCoNut~\cite{lutellier2020coconut} manually checks the bug-fixing pairs in Defects4J benchmark and extracts the changed statements as inputs to the repair model.
Subsequently, recent learning-based APR techniques adopt the same or similar processing method to conduct perfect localization, such as CIRCLE~\cite{yuan2022circle}, CURE~\cite{jiang2021cure}, SelfAPR~\cite{ye2022selfapr} and AlphaRepair~\cite{xia2022less}.

Besides, there exist some techniques attempting to perform fault localization on their own.
For example, DeepFix~\cite{gupta2017deepfix} proposes an end-to-end approach in which the network reports a ranked list of potentially erroneous lines with a beam search mechanism.
Similarly, Prophet~\cite{long2016automatic} designs a fault localization algorithm to return a ranked list of program candidate lines to modify by analyzing dynamic execution traces of the test suite.
Szalontai~\etal~\cite{szalontai2021detecting} first localize the nonidiomatic code snippets by LSTM networks and predict the nonidiomatic pattern by a feed-forward neural network, which is fixed by a high-quality alternative.
Recently, \delete{Meng et al .\cite{meng2022improving}}\revise{Meng~\etal~\cite{meng2022improving}} build a novel fault localization technique based on deep semantic features and transferred knowledge, which is further fed to a fix template prioritization model and a template-based APR technique \revise{TBar~\cite{liu2019tbar}}.

\subsubsection{Localization Granularity}
APR techniques consider program elements of different granularities, thus determining the scope of the fault localization.
In other words, APR and fault localization usually work at the same granularity level.
For example, if APR techniques focus on repairing buggy statements (or methods), the fault localization also works at the level of program statements (or methods).
In the literature, a majority of fault localization techniques adopted in learning-based APR techniques usually record the line of a buggy code snippet~\cite{yuan2022circle,zhu2021syntax,lutellier2020coconut,jiang2021cure,li2020dlfix,li2022dear}.
There also exists little work considering other granularity.
For example, Tufano~\etal~\cite{tufano2019empirical} adopt the NMT-based repair model to learn the translation from buggy to fixed code at the method level.

\summary{
\revise{
As a preceding step in the learning-based APR workflow, fault localization has a significant impact on the performance of the patch generation, which cannot generate a correct patch with a wrong suspicious code element.
Most learning-based APR techniques follow the common practice in the traditional APR field to generate patches by integrating spectrum-based fault localization techniques.
There are also some repair techniques that are starting to use perfect localization (\ie the ground-truth buggy code element), to avoid the noise introduced by the off-the-shelf fault localization techniques.
Besides, thanks to the code comprehension capabilities of DL models, some learning-based APR techniques can generate patches with coarse-grained fault localization, \eg only the buggy method is provided.}
}

\subsection{Data Pre-processing}
\label{sec:preprocessing}

Data pre-processing phase aims to analyze and parse the identified buggy code snippets, which are then passed into neural networks for training and inference.
In the data pre-processing phase, a given software buggy code snippet (\eg a buggy function) is taken as the input and the processed code tokens are returned.
According to existing learning-based repair studies~\cite{chen2022neural,chi2022seqtrans}, 
the data pre-processing phase generally consists of three parts: \delete{\textit{code abstraction}, \textit{code context} and \textit{code tokenization}}\revise{\textit{code context}, \textit{code abstraction} and \textit{code tokenization}}.

\subsubsection{Code Context}

Code context generally refers to other correct statements around the buggy lines.
In the manual repair scenario, the context of the buggy code plays a significant role in understanding faulty behaviors and reasoning about potential repairs. 
Developers usually identify the buggy lines, and then analyze how they interact with the rest of the method’s execution, and observe the context (\eg variables and other methods) in order to come up with the possible repair and pick several tokens from the context to generate the fixed line~\cite{ko2006exploratory}.
In learning-based APR, the NMT model mimics this process by extracting the code context and the buggy line into a certain code representation to preserve the necessary context that allows the model to predict the possible fixes.

Existing learning-based APR techniques typically consider the surrounding source code relevant to the buggy statement as context.
These techniques typically employ context in various ways, such as extracting code near the buggy statement within the buggy method, class, and even file.
On the one hand, a broad context contains plenty of essential fix ingredients, while such a large vocabulary size introduces noise that negatively affects the repair performance of the NMT model due to the tricky long-term dependency problem in NMT models\revise{~\cite{chen2019sequencer}}.
\revise{
In particular, long-term dependency refers to the situation that the meaning of a token depends on another token that is far apart from it in a code snippet~\cite{vaswani2017attention}.
As a result, NMT repair models often struggle to capture long-term dependencies when dealing with tokens that appear over long code snippets~\cite{yuan2022circle}.}
On the other hand, a narrow context contains too little information to capture the proper semantics of the buggy statement and leads to incorrect patches generation due to a lack of necessary vocabulary.
There seems to be a trade-off relationship between vocabulary size and context size.
\delete{In the literature, our survey concludes the code context into four granularities: context-free, line-level context, method-level context, and class-level context.}
\revise{Our survey concludes the code context of existing learning-based APR studies into four granularities: context-free, line-level context, method-level context, and class-level context.}

\begin{itemize}
    \item \delete{\textit{Context-free} means that NMT models only consider buggy statements without any additional context information.}
    \revise{\textbf{Context-free.}
    This granularity refers to the scenario where NMT repair modes only take buggy statements without any additional code snippets as inputs~\cite{hata2018learning,ding2020patching,mashhadi2021applying}.}
    For example, Mashhadi~\etal~\cite{mashhadi2021applying} consider single statement bugs from the ManySStuBs4J dataset and extract the buggy statement as a source side and the fixed statement as a target side from bug-fixing commits.
    Ding~\etal~\cite{ding2020patching} provide NMT models with a single program line that contains a buggy statement.
    However, previous work demonstrates that fixing nearly 90\% of bugs requires new vocabulary relative to the buggy code.
    \revise{Therefore,} NMT repair models suffer from capturing enough information from the buggy statements alone.
    
    \item \delete{\textit{Statement-level context} means that the buggy code and several statements around it are fed to \delete{MNT}\revise{NMT} repair models.}
    \revise{\textbf{Statement-level context}.
    This granularity refers to the scenario where NMT repair models take the buggy statements and several statements that the buggy code and some surrounding correct statements as inputs~\cite{berabi2021tfix,chi2022seqtrans}.}
    For example, TFix~\cite{berabi2021tfix} extracts the two neighboring statements of the buggy code as the code context.
    Chi~\etal~\cite{chi2022seqtrans} extract statement-level code changes by the ``\delete{git dif}\revise{git diff}'' command and employ data-flow dependencies to capture more critical information around the context.

    \item \delete{\textit{Method-level context} means that the method to which the buggy line belongs is fully fed into the model.}
    \revise{\textbf{Method-level context}.
    This granularity refers to the scenario where NMT modes take the whole method to which the buggy statements belong as inputs~\cite{tufano2019empirical,lutellier2020coconut,yuan2022circle}.}
    It is the most commonly used type of context in literature as it often contains enough information for repairing the bug, such as the type of variables and the function of this method.
    For example, Tufano \delete{et al .}\revise{\etal}~\cite{tufano2019learning} focus on the method-level context since
    (1) the functionality to be fixed is usually implemented in program methods;
    (2) the methods provide neural networks with meaningful abundant context information, such as literals and variables.
    Similarly, CoCoNuT~\cite{Chakraborty2020codit} extracts the entire method of the buggy code as context, which is encoded as a separate input. 

    \item \delete{\textit{Class-level context} means that the class to which this buggy code belongs is fed into the NMT model.}
    \revise{\textbf{Class-level context}.
    This granularity refers to the scenario where NMT repair models take the class to which the buggy statements belong to as inputs.}
    It is a relatively broad context, while it can provide the model with rich information.
    For example, SequenceR~\cite{chen2019sequencer} considers the class-level context and conducts abstract buggy context from the buggy class, which captures the most important context around the buggy source code and reduces the complexity of the input sequence to 1,000 tokens.
    Hoppity~\cite{dinella2020hoppity} takes the whole buggy file as the context with a length limit of 500 \delete{tokens }nodes in the AST.

\end{itemize}

\begin{figure}[t]
\centering
    \subfigure[raw buggy code]
    {
        \includegraphics[width=0.47\linewidth]{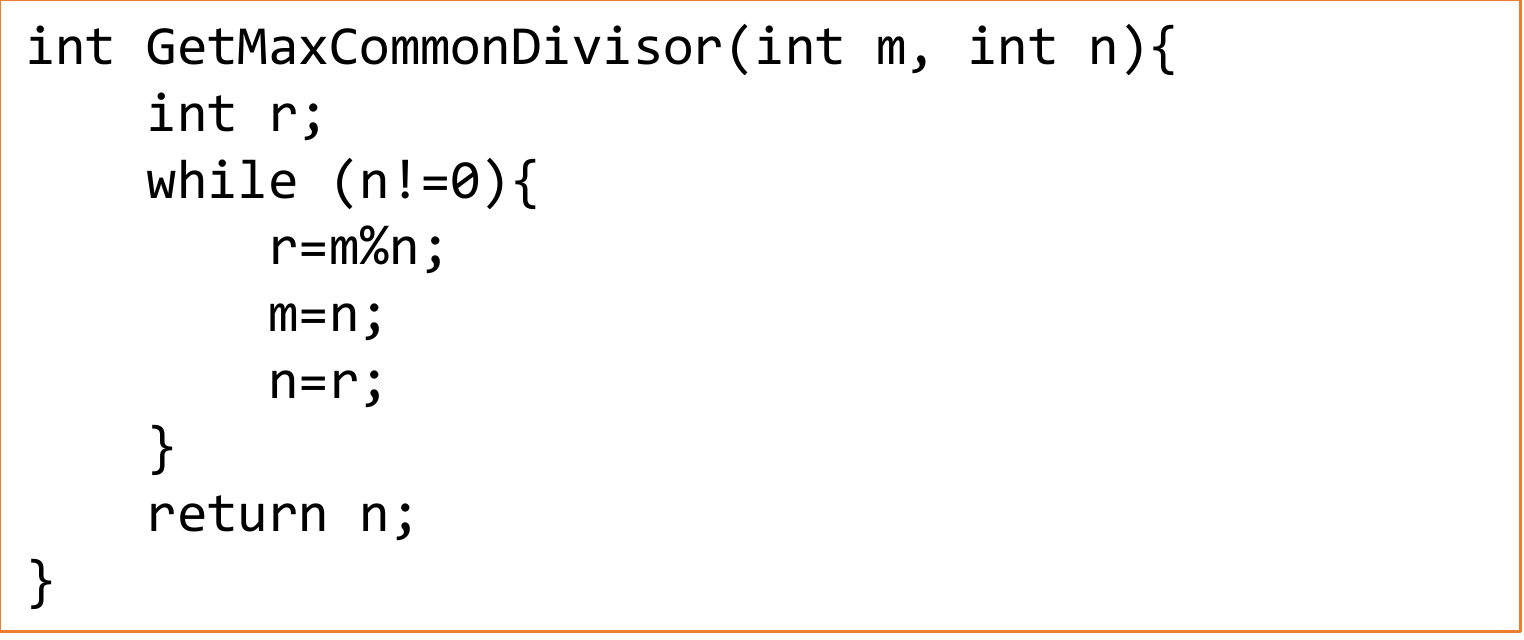}
        \label{fig:raw_buggy}
    }
    \subfigure[raw fixed code]
    {
        \includegraphics[width=0.47\linewidth]{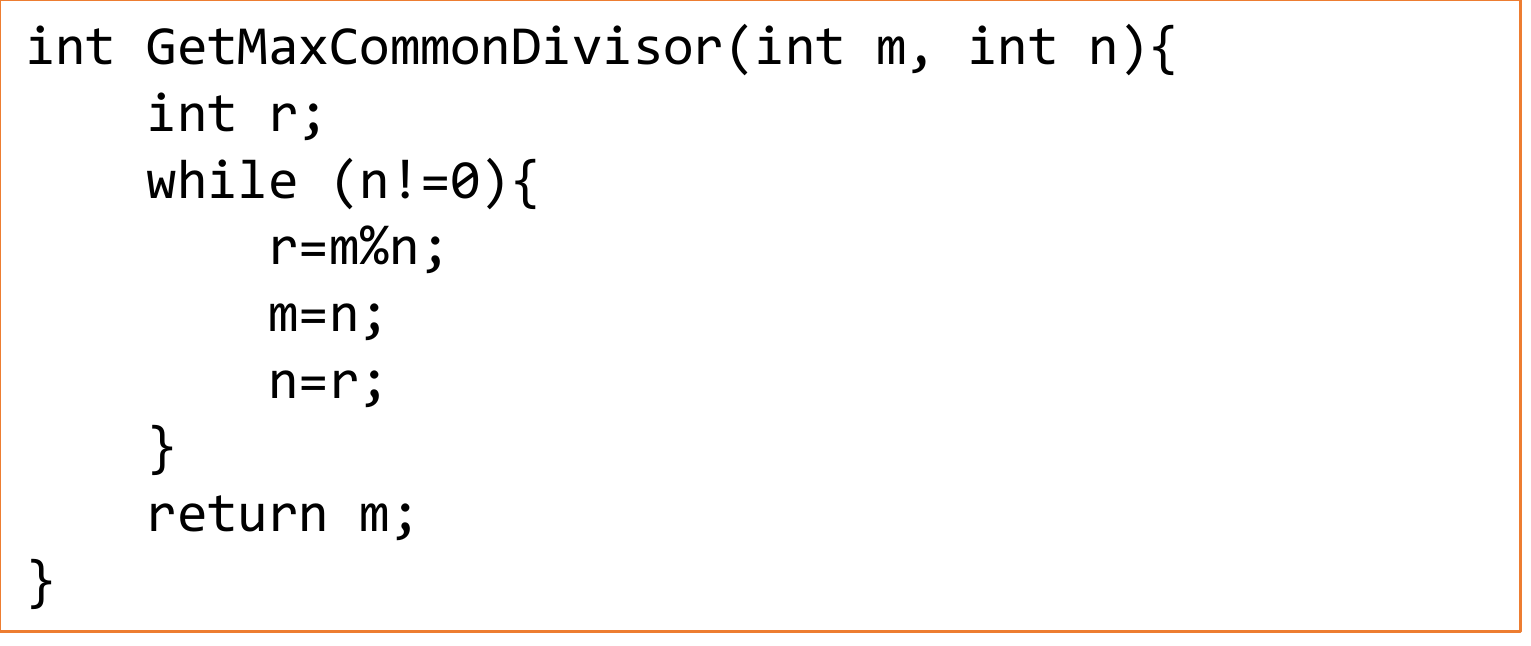}
        \label{fig:raw_fix}
    }
    \subfigure[abstracted buggy code]
    {
        \includegraphics[width=0.47\linewidth]{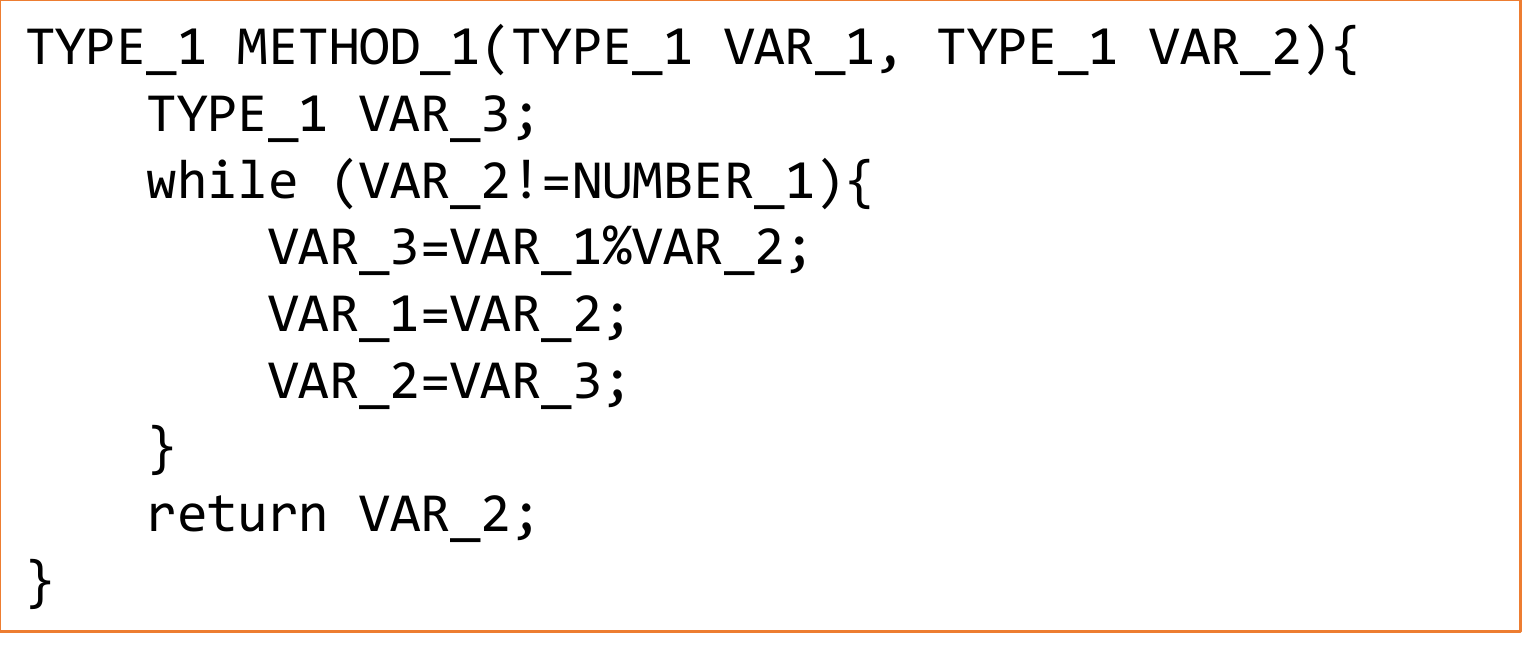}
        \label{fig:abs_raw}
    }
    \subfigure[abstracted fixed code]
    {
        \includegraphics[width=0.47\linewidth]{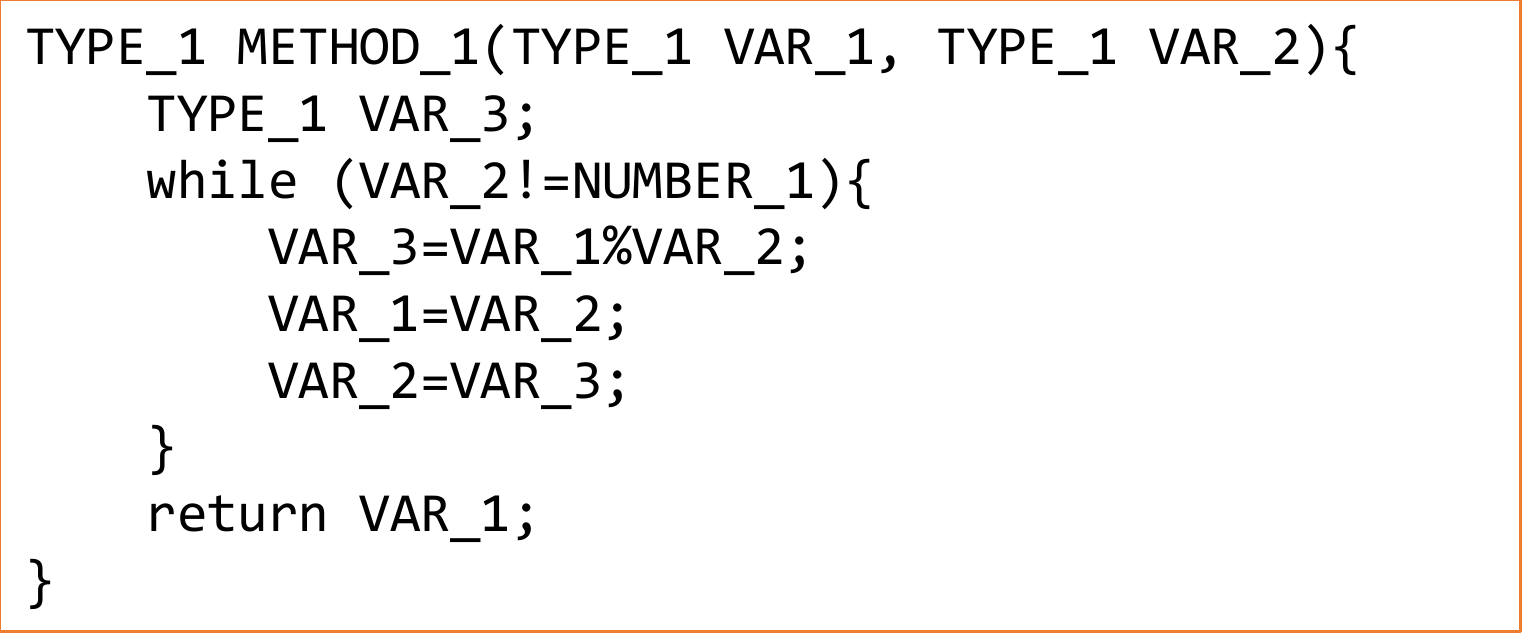}
        \label{fig:abs_fix}
    }
    \caption{A simple example of code abstraction}
    \label{fig:abstraction_example}
\end{figure}

\subsubsection{Code Abstraction}

Code abstraction aims to limit the number of words the NMT models need to process by renaming raw words (\eg function names and string literals) to a set of predefined tokens.
Previous work demonstrates that it is challenging for NMT models to learn bug-fixing transformation patterns due to the huge vocabulary of source code~\cite{tufano2019empirical}.
In particular, NMT models usually employ a beam-search decoding strategy to output repair candidates by a probability distribution over all words.
The search space can be extremely large with many possible words in the source code, resulting in inefficient patch generation.

In our survey, a considerable number of learning-based papers we collect employ the abstracted source code to tackle this problem. 
Such abstraction operation means the original source code is not directly fed into the NMT model.
Benefiting from the abstracted code, we can
(1) reduce the size of vocabulary significantly and the frequency of specific tokens;
(2) filter out irrelevant information and improve the efficiency of the NMT model.
Generally, the natural elements (\eg identifiers and literal) in the source code are renamed, while the core semantic information (\eg idioms) should be preserved.
For example, Tufano~\etal~\cite{tufano2019empirical} propose the first code abstraction approach in the learning-based APR field by
(1 ) adopting a lexer to tokenize the raw source code as a stream of tokens based on Another Tool for Language Recognition (ANTLR)~\cite{parr2011ll};
(2) passing the stream of tokens into a parser to identify the role of each identifier and literals (\eg whether it represents a variable, method, or type name);
(3) replacing each identifier and literal with a unique ID to generate the abstracted source code.
Besides, they extract the idioms (\ie tokens that appear many times) and keep their original textual tokens in the abstraction process because such idioms contain beneficial semantic information.
The typical code abstraction example is presented in Figure \ref{fig:abstraction_example}.
Similarly, CoCoNut~\cite{lutellier2020coconut} and CURE~\cite{jiang2021cure} only abstract string and number literals except for the frequent numbers (\eg $0$ and $1$).
DLFix~\cite{li2020dlfix} adopts a novel abstraction strategy to alpha-rename the variables, so as to learn the fix between methods with similar scenarios while having different variable names.
DLFix also keeps the type of the variable to avoid accidental clashing names and maintains a mapping table to recover the actual names.
Recoder~\cite{zhu2021syntax} abstracts infrequent identifiers with placeholders to make the neural network learns to generate placeholders for these identifiers.

\delete{
Although a variety of techniques adopt the code abstraction strategy (such as Tufano~\etal~\cite{tufano2019empirical}) to limit the vocabulary size and make the transformer concentrate on learning common patterns from different code changes, we still find some repair techniques prefer raw source code~\cite{yuan2022circle,zhu2021syntax}.
For example, developers may name one function as \textit{SetHeightValue} to indicate that this function can set the value of height as they want. 
If this name is abstracted directly as \textit{func\_1}, critical semantic information would be missed.
Instead of renaming rare identifiers through a custom abstraction process, SequenceR~\cite{chen2019sequencer} utilizes the copy mechanism to generate candidate patches with a large set of tokens.
Chen~\etal~\cite{chen2022neural} adopt the raw source code as they think abstracted code may hide valuable information about the variable that can be learned by word embedding.
A strategy similar to Chen~\etal~\cite{chen2022neural} is also implemented in other learning-based APR techniques, such as in CODIT~\cite{Chakraborty2020codit}, CIRCLE~\cite{yuan2022circle} and TFix~\cite{berabi2021tfix}.
}

\revise{
Although a variety of learning-based APR techniques adopt the code abstraction strategy (such as Tufano~\etal~\cite{tufano2019empirical}) to limit the vocabulary size and make the transformer concentrate on learning common patterns from different code changes, we still find some repair techniques prefer raw source code~\cite{yuan2022circle,zhu2021syntax}  because it contains semantic information.
For example, developers may name one function as \textit{SetHeightValue} to indicate that this function can set the value of height as they want. 
If this name is abstracted directly as \textit{func\_1}, the critical semantic information would be missed, resulting in suboptimal repair training.
Thus, instead of renaming rare identifiers through a custom abstraction process, SequenceR~\cite{chen2019sequencer} utilizes the copy mechanism to generate candidate patches with a large set of tokens.
During programming, developers are not restricted by a set vocabulary (\eg English) when defining names for variables or methods, resulting in an extremely large vocabulary with many rare tokens.
The copy mechanism seeks to copy some rare input tokens to the output and is effective in reducing the required vocabulary size~\cite{chen2019sequencer}.
Besides, Chen~\etal~\cite{chen2022neural} adopt the raw source code as they think abstracted code may hide valuable information about the variable that can be learned by word embedding.
A strategy that is similar to Chen~\etal~\cite{chen2022neural} is also implemented in other learning-based APR techniques, such as in CODIT~\cite{Chakraborty2020codit}, CIRCLE~\cite{yuan2022circle} and TFix~\cite{berabi2021tfix}.
}

\subsubsection{Code Tokenization}

Code tokenization aims to split source code into a stream of tokens, which are then converted to ids through a look-up table\footnote{\url{https://huggingface.co/Salesforce/codet5-base/blob/main/vocab.json}}. 
These id numbers are in turn used by the repair models for further processing and training.
A simple tokenization approach can be conducted by dividing the source code into individual characters.
The core concept of this char-level tokenization is that although the source code has many different words, it has a limited number of characters. 
This approach is straightforward and leads to an exceeding small vocabulary.
However, it leads to a relatively long tokenized sequence with the splitting of each world into all characters.
More importantly, it is pretty difficult for repair models to meaningful input representations as characters alone do not have semantic meaning.
Generally, there exist two main granularities of code tokenizers used in \revise{existing} learning-based APR techniques: \delete{word-level tokenizers and~\cite{fu2022vulrepair} and subword-level tokenization~\cite{chi2022seqtrans}}\revise{word-level tokenization~\cite{fu2022vulrepair} and subword-level tokenization~\cite{chi2022seqtrans}}.

\delete{
The word-level tokenization means that a sentence is divided according to its words (\eg space-separated), which is widely used in NLP tasks.
However, different from natural language, words (\eg variable names) in programming languages can be created arbitrarily, leading to a more irregular vocabulary.
This kind of granularity often causes the out-of-vocabulary (OOV) problem for infrequent tokens, and the model could be more efficient for an excessively large vocabulary size.
\revise{VRepair~\cite{chen2022neural}} employs a word-level tokenization to tokenize C source code and the copy mechanism to deal with the out-of-vocabulary
problem.
For example, CoCoNut~\cite{lutellier2020coconut} designs a code-aware space-separated tokenization algorithm that is specific to programming languages by
(1) separating operators from variables as they might not be space-separated;
(2) considering underscores, camel letters, and numbers as separators as many words are composed of multiple words without separation (\eg \textit{SetHeightValue});
(3) introducing a new token \textit{<CAMEL>} to mark where the camel case split occurs to regenerate source code from the list of tokens generated correctly.
}

\revise{
The word-level tokenization means that a sentence is divided according to its words (\eg space-separated), which is widely used in NLP tasks~\cite{raffel2019t5}.
However, different from natural language (\eg English dictionary), words (\eg variable and method names) in programming languages can be created by developers arbitrarily.
As a result, there may exist some rare words not available in the vocabulary (\ie the out-of-vocabulary problem), resulting in unknown tokens in patch generation.
To address this issue, 
\revise{VRepair~\cite{chen2022neural}} employs a word-level tokenization to tokenize C source code and the copy mechanism to deal with the out-of-vocabulary
problem.
Similarly, CoCoNut~\cite{lutellier2020coconut} designs a code-aware space-separated tokenization algorithm that is specific to programming languages by
(1) separating operators from variables as they might not be space-separated;
(2) considering underscores, camel letters, and numbers as separators as many words are composed of multiple words without separation (\eg \textit{SetHeightValue});
(3) introducing a new token \textit{<CAMEL>} to mark where the camel case split occurs to regenerate source code from the list of tokens generated correctly.
}

The subword-level tokenization splits rare tokens into multiple subwords instead of directly adding full tokens into the vocabulary.
Besides, the frequent words should not be split into smaller subwords.
This kind of granularity can reduce the vocabulary size significantly and is widely used in the learning-based APR field.
Technically, there exist several subword-level tokenization techniques, such as byte-pair encoding (BPE), byte-level byte-pair encoding (BBPE)~\cite{sennrich2016neural} and SentencePiece~\cite{kudo2018sentencepiece}, listed as follows. 

\begin{enumerate}
    \item  BPE tokenizer generally needs to be trained upon a given dataset by 
    (1) leveraging a pre-tokenizer to splits the dataset into words by space-separated tokenization; 
    (2) creating a set of unique words and counting the frequency of each word in the dataset; 
    (3) building a base vocabulary with all symbols that occur in the set of unique words and learning merge rules to form a new symbol from two symbols of the base vocabulary; 
    (4) repeating the above process until the vocabulary is reduced to a reasonable size, which is a pre-defined hyperparameter, before training the tokenizer.
    For example, VulRepair~\cite{fu2022vulrepair} employs a BPE algorithm to train a subword tokenizer on eight different programming languages (\ie Ruby, JavaScript, Go, Python, Java, PHP, C, C\#)~\cite{wang2021codet5} and is suitable for tokenizing source code.
    In the learning-based APR literature, a majority of repair studies adopt BPE as the tokenization technique, such as CURE~\cite{jiang2021cure}, CoCoNut~\cite{lutellier2020coconut}, SeqTrans~\cite{chi2022seqtrans}.
    The results have demonstrated the effectiveness of BPE in reducing vocabulary size and mitigating the OOV problem by extracting the most frequent subwords and merging the most frequent byte pair iteratively.

    \item BBPE refines BPE by employing bytes as the base vocabulary,  ensuring that every base character is included with a proper vocabulary size.
    For example, AlphaRepair~\cite{xia2022less} builds a BBPE-based tokenizer to reduce the vocabulary size by breaking uncommon long words into meaningful subwords.
    
    \item SentencePiece contains the space in the base vocabulary and utilizes the existing BPE algorithm (\eg BPE) to create the desired vocabulary by regarding the source code as a raw input stream.
    In the literature, before entering source code into the neural network, several learning-based APR techniques use SentencePiece to divide words into a sequence of subtokens, such as \revise{SelfAPR~\cite{ye2022selfapr}}, RewardRepair~\cite{ye2022neural} and CIRCLE~\cite{yuan2022circle}.
\end{enumerate}

\summary{\revise{
Data preprocessing is responsible for processing the code snippets into a suitable format and feeding it to the NMT repair models for training.
Different learning-based APR techniques employ diverse data pre-processing methods, learning to complex experimental settings in the literature.
For example, \textit{code abstraction} involves raw code or abstracted code; \textit{code context} involves context-free, statement-level, method-level and class-level context; \textit{code tokenization} involves BPE, BBPE and SentencePiece tokenizers.
On the one hand, these different configurations may introduce bias to the evaluation of existing learning-based APR techniques. 
On the other hand, the optimal combination of these configurations requires further exploration, and it is also important to consider their interactions with other factors, such as the model architectures and the types of software bugs being fixed.
}}

\subsection{Patch Generation}
\label{sec:patch_generation}

In the learning-based APR context, to apply NMT repair models to high-level programming languages, the code snippets need to be converted to embedding vectors.
Then an NMT repair model is built on top of the encoder-decoder architecture~\cite{vaswani2017attention} to learn the repair patterns automatically.
Finally, the mapping from buggy code to fixed code is optimized by updating the parameters of the designed model.
Thus, it is crucial to determine (1) how to represent the source code (with which format) as input for word embedding, referred to as \textit{code representation};
and (2) how to design the specific architecture (with which neural network) as encoder-decoder for repair transformation learning, referred to as \textit{model architecture}.

In the literature, various strategies have been proposed to represent the source code as the input for NMT repair models, which can be categorized into three classes: \textit{sequence-based}, \textit{tree-based} and \textit{graph-based representation}.

\subsubsection{Sequence-based Generation}
These techniques divide the textual source code as a sequence of tokens and treat APR as a token-to-token translation task based on a \delete{sequence-to-sequence}\revise{Seq2Sep} model.

\textit{Code Representation}.
Considering the buggy lines and the context, there generally exist four different ways to sequence the textual code tokens.
\begin{enumerate}
    \item \textit{Raw representation}.
    
    \newrevise{Similar to NMT, which translates a sentence from one source language (\eg English) to another target language (\eg Chinese)}, most sequence-based techniques directly feed the model with the buggy code snippet~\cite{tufano2019empirical}.
    For example, Tufano~\etal~\cite{tufano2019empirical} extract the buggy method and train an NMT model for method-to-method translation.
    The size of this code snippet depends on the choice of the buggy code and code context. 
    However, the raw representation is unaware of the difference between the buggy code and the code context, as these two parts are sent into the encoder together.
    As a result, the transformation rules may be applied in some correct lines, limiting the repair performance.

    \item \textit{Context representation.}

    The context representation splits the buggy code and the code context, then feeds them into two encoders separately.
    Under this circumstance, the model is aware of the difference between buggy code and the corresponding context.
    For example, Lutellier~\etal~\cite{lutellier2020coconut,jiang2021cure} attempt to encode these two parts separately and then merge the encoding vectors.
    However, it is challenging to merge the two separated encoding vectors and eliminate the semantic gaps between the two encoders.
    
    \item \textit{Prompt representation.}
    
    The prompt representation refers to a text-in-text-out input format and can effectively concatenate different input components with some prefixed prompt~\cite{raffel2019t5}.
    The prefixed prompt is a piece of tokens inserted in the input, so that the original task can be formulated as a language modeling task.
    For example, Yuan~\etal~\cite{yuan2022circle} employs manually designed prompt template to convert buggy code and corresponding context into a unified fill-in-the-blank format.
    In particular, they employ ``Buggy line:'' and ``Context:'' to denote the buggy code and code context, and then employ ``The fixed code is:'' to guide the NMT model to generate candidate patches according to the previous input.
    This mechanism has been proven effective in bridging the gap between pre-trained tasks and the downstream task, facilitating fine-tuning pre-trained models for APR.

    \item \textit{Mask representation.}
    
    The mask representation replaces the buggy code with mask tokens and queries NMT models to fill the masks with the correct code lines.
    This mechanism views the APR problem as a cloze task and usually adopts the pre-trained model as the query model in the learning-based APR. 
    For example, Xia~\etal~\cite{xia2022less} transform the original buggy code into a comment and generate multiple mask lines with templates.
    The input is represented by comment buggy code, context before buggy code, mask lines and context after buggy code.
    In particular, the buggy code is masked randomly from one token to the whole line, and researchers expect to generate every possible patch for different situations within a limited candidate patch size.
    Compared with the above three representation strategies, the mask representation can adopt pre-trained models to predict randomly masked tokens to perform cloze-style APR without any additional training on the bug-fixing dataset.

\end{enumerate}

\textit{Model Architecture.}
\delete{Sequenced-based}\revise{Sequence-based} techniques usually treat the source code as a sequence of tokens and adopt existing sequence-to-sequence architectures in the NLP field instead of designing new network architectures.
For example, CoCoNut~\cite{lutellier2020coconut} adopts two fully convolutional (FConv) encoders to represent the buggy lines and the context separately.
One common encoder architecture is long short-term memory (LSTM), and it resolves the long-term dependency problem of the RNN module by introducing the gate mechanism and ensures that short-term memory is not neglected.
For example, SequenceR~\cite{chen2019sequencer} is based on an LSTM encoder-decoder architecture with copy mechanism.
As a powerful kind of DL architecture, \revise{the} transformer can model global dependencies between input and output effectively thanks to the attention mechanism and has been adopted in existing APR studies, such as Bug-Transformer~\cite{yao2022bug}, SeqTrans~\cite{chi2022seqtrans} and VRepair~\cite{chen2022neural}.

Recently, the usage of pre-trained models has gradually attracted the attention of researchers in the learning-based APR community.
Such models are first pre-trained by self-supervised training on a large-scale unlabeled corpus (\revise{\eg CodeSearchNet~\cite{husain2019codesearchnet}}), and then transferred to benefit multiple downstream tasks by fine-tuning on a limited labeled corpus.
For example, Mashhadi~\etal~\cite{mashhadi2021applying} employ CodeBERT, a bimodal pre-trained language model for both natural and programming languages, to fix Java single-line bugs by fine-tuning on the \revise{ManySStuBs4J small and large datasets~\cite{le2015manybugs}}.
CURE~\cite{jiang2021cure} applies a pre-trained GPT \delete{module} \revise{model} to further revise an NMT-based APR architecture (\ie CoCoNut).
CIRCLE~\cite{yuan2022circle} proposes a T5-based program repair framework equipped with continual learning ability across multiple languages.
We will discuss the application of pre-trained models in Section \ref{sec:pre-trained}.

\input{sec/sec_SOTA_seq}

\subsubsection{Tree-based Generation}
Sequence-based APR techniques usually adopt \delete{sequence-to-sequence} \revise{Seq2Seq} models for patch generation.
However, these techniques ignore code structure information because they are designed for NLP, which is significantly different from programming language with strict syntactic and grammatical rules.
The generated patches of these techniques may suffer from syntax errors that cause compilers to fail.
As a result, researchers recently propose various tree-based generation techniques by considering the syntactic structure of source code.
These techniques treat the APR problem as a tree transformation learning task.

\textit{Code Representation}.
A common solution is to parse the source code into an AST and adopt a tree-aware model to perform patch generation, \ie \textit{structure-aware representation}.
For example, given a bug-fixing method pair $M_b$ and $M_f$ representing the buggy and fixed method, DLFix~\cite{li2020dlfix} first extracts a buggy AST for $M_b$ (\ie $T_b$), a fixed AST for $M_f$ (\ie $T_f$), a buggy sub-AST (i.e, $T_b^s$) and a fixed sub-AST (\ie $T_f^s$) between $T_b$ and $T_f$.
DLFix then adopts an existing summarization model to \delete{encoder}\revise{encode} $T_b^s$ as a single node $S_b^s$.
Finally, the buggy method $M_b$ can be represented as a context tree by replacing $T_b^s$ in $T_b$ with $S_b^s$ and a sub-changed tree $T_b^s$.
The fixed method $M_f$ is represented in a similar way.

As tree-based representation contains the structure information, which cannot be directly deployed to \delete{sequenced-based}\revise{sequence-based} neural models.
Thus, an additional code representation strategy is utilized to parse the tree representation as a sequential traverse sequence, \ie \textit{sequential-traverse representation}.
For example, Tang~\etal~\cite{tang2021grammar} parse the source code into AST representation, which is further translated into a sequence of rules.
The sequence of rules can be processed by the vanilla transformer~\cite{vaswani2017attention} while capturing the grammar and syntax information.
\delete{CODIT~\cite{Chakraborty2020codit} first identifies the edited AST nodes (\ie the inserting, deleting, and updating) and selects the minimal subtree of each AST.
CODIT then collects the edit context by including the nodes that connect the root of the method to the root of the changed tree. CODIT expands the considered context until the context exceeds a maximum tree size. 
Given each bug-fixing method pair, CODIT extracts a buggy AST and fixed AST, and then converts the ASTs to their tree representation.}
\revise{Similarly, CODIT~\cite{Chakraborty2020codit} first represents code snippets as AST by (1) identifying the edited AST nodes (\ie the inserting, deleting, and updating); (2) selecting the minimal subtree of each AST and (3) collecting the edit context by including the nodes that connect the root of the method to the root of the changed tree.
CODIT then employs a tree-based model to learn the structural changes in the form of a sequence of grammar, which is finally used to predict the fixed code sequence with a standard Seq2Seq model.}

\textit{Model Architecture.}
Most NMT-based APR models treat patch generation as a machine translation from buggy code to a fixed one.   
However, such models could not capture the information of code structures and suffer from handling the context of the code.
Tree-based encoders consider the structure features of source code, such as AST.
\delete{For example, DLFix~\cite{li2020dlfix} parses the source code to AST and adopts a tree-based LSTM to represent the changed and context sub-trees.}
\revise{For example, DLFix~\cite{li2020dlfix} represents source code as ASTs and employs tree-based RNN models to encode the context tree and sub-changed trees.}
Besides, Devlin~\etal~\cite{devlin2017semantic} encode the AST with a sequential bidirectional LSTM by enumerating a depth-first traversal of the nodes.

\input{sec/sec_SOTA_tree}

\subsubsection{Graph-based Generation}
\delete{These techniques transform source code into graph representations with contextual information and frame the APR problem in terms of learning a sequence of graph transformations.}
\revise{These techniques transform source code into graph representations with contextual information and frame the APR problem in terms of learning a sequence of graph transformations based on graph-based models.
Instead of directly manipulating the source code, such graph-based APR techniques aim to learn a sequence of transformations on the graph representation that would correspond to a corrected version of the original code.}

\textit{Code Representation}.
To capture the neighbor relations between AST nodes, Recoder~\cite{zhu2021syntax} treats AST as a directional graph where the nodes denote AST nodes and the edges denote the relationship between each node and its children and left sibling.
Besides, Xu~\etal~\cite{xu2022m3v} consider the context structure by data and control dependencies captured by a data dependence graph (\ie DDG) and a control dependence graph (\ie CDG).

\textit{Model Architecture.}
Existing graph-based APR techniques usually design graph neural networks and their variants to capture graph representation and perform patch generation.
For example, Hoppity~\cite{dinella2020hoppity} adopts a gated graph neural network (GGNN) to treat the AST as a graph, where a candidate patch is generated by a sequence of predictions, including the position of graph nodes and corresponding graph edits.
Besides, Xu~\etal~\cite{xu2022m3v} design a graph neural network (GNN) for obtaining a graph representation by first converting DDG and CDG into two graph representations and then fusing them.

\input{sec/sec_SOTA_graph}

\summary{\revise{
As the most crucial phase in the repair workflow, a majority of existing learning-based APR techniques focus on patch generation.
These patch generation techniques typically can be divided into two parts: code representation and the corresponding model architecture.
The key research question lies in how to appropriately represent code snippets and determine the model architecture that can effectively learn the transformation relationship between buggy code and correct code.
Inspired by NLP, incipient repair techniques usually represent the source code as a sequence of tokens, and transform an APR problem into an NMT task on top of a sequence-to-sequence model.
The follow-up techniques represent the source code as a tree or graph representation and adopt tree-aware models (\eg tree-LSTM) or graph-aware models (\eg GGNN) to perform patch generation.
The literature does not demonstrate which code representation or model architecture exhibits the best performance.
An in-depth controlled experiment can be conducted to investigate the performance between different code representations and the corresponding model architectures.
}}

\subsection{Patch Ranking}
\label{sec:patch_ranking}

The patch generation is a search process for the maximum in the combinatorial space.
Given the max output length \emph{l} and the size of vocabulary \emph{V}, the total number of candidate patches that the decoder can generate reaches $V^l$, all of which it is impossible to validate in practice.
Developers may spend a considerable amount of effort to assess the correctness of the generated candidate patches manually.
In such a scenario, only inspecting fewer repair candidates (\eg Top-1 and Top-5) that have a high probability of being correct is more practical and reduces the valuable manual effort.
As a result, a patch ranking strategy is crucial to ensure the inference efficiency of the model and relieve the burden of patch validation.

Beam search is an effective heuristic search algorithm to rank the outputs in previous NMT applications~\cite{wang2021codet5} and is the most common patch ranking strategy in learning-based APR studies, such as CIRCLE~\cite{yuan2022circle}, SelfAPR~\cite{ye2022selfapr}, RewardRepair~\cite{ye2022neural} and Recoder~\cite{zhu2021syntax}.
\delete{In particular, for every iteration, the beam search algorithm selects the $k$ most probable tokens for the patch (corresponding to beam size $k$) and ranks them according to the likelihood estimation score of the next $d$ prediction steps (corresponding to search depth $d$). At last, the top $k$ most likely patches are maintained for validation in the next procedure.}
\revise{In particular, at each iteration, the beam search algorithm selects the $k$ most probable tokens (corresponding to beam size $k$) and ranks them according to the likelihood estimation score of the next $d$ prediction steps (corresponding to search depth $d$).
The iteration repeats until a stopping condition is met, such as reaching a certain sequence length or all sequences ending with an end-of-sequence token.
\newrevise{Finally, the top $k$ high-scoring candidate} patches are generated and ranked for further validation in the next procedure of the overall learning-based APR workflow.}
Beam search provides a great trade-off between repair accuracy versus inference cost via its flexible choice of beam size.

However, the vanilla beam search considers only the log probability to generate the next token while ignoring the code-related information, such as variables.
Thus, it may generate high-score patches with unknown variables, leading to uncompilable candidate patches.
In addition to directly applying the existing beam search strategy, researchers design some novel strategies to filter out low-probability patches. 
For example,
CURE~\cite{jiang2021cure} designs a code-aware beam search strategy to generate more compilable and correct patches based on valid-identifier check and length control components. 
The code-aware strategy first performs static analysis to identify all valid tokens used for sequence generation and then prompts beam search to generate sequences of a similar length to the buggy line.
DLFix~\cite{li2020dlfix} first derives the possible candidate patches by program analysis filtering and ranks the list of possible patches by a CNN-based binary classification model.
The classifier adopts a Word2Vec model as the encoder stack at the char level, followed by a CNN stack as the learning stack (containing a Convolutional layer, pooling, and fully connected layers), and a softmax function as the classification stack.
Then DLFix ranks the given list of patches based on their possibilities of being a correct patch.
Further, DEAR~\cite{li2022dear} applies a set of filters to verify the program semantics and ranks the candidate patches in the same manner as DLFix does.

\revise{In addition to the widely-used beam search and their variants, there are also some self-designed patch ranking methods as a component in the patch generation. 
As early as 2016, Long~\etal~\cite{long2016automatic} propose Prophet, which trains a ranking model to assign a high probability to correct patches based on designed features (detailed in Section~\ref{sec_dl4Tapr}).}
\delete{Besides}\revise{Recently}, AlphaRepair~\cite{xia2022less} designs a patch ranking strategy based on a masked language model.
In particular, given a candidate patch, AlphaRepair calculates its priority score by
(1) extracting all generated tokens;
(2) masking out only one of the tokens; 
(3) querying CodeBERT to obtain the conditional probability of that token;
(4) repeating the same process for all other previous mask tokens; 
and (5) computing the joint score which is an average of the individual token probabilities.

\summary{\revise{
Patch ranking seeks to prioritize candidate patches with a higher probability of being correct in the search space.
As a greedy strategy, beam search is widely adopted in existing learning-based APR techniques to keep $k$ optimal tokens at every iteration according to the likelihood estimation score.
Besides, some advanced patch ranking strategies (\eg a code-aware beam search strategy to consider valid identifiers) are proposed to identify high-probability while low-quality patches, such as uncompilable candidate patches.
Overall, a majority of existing learning-based APR techniques follow the vanilla beam search strategy and the literature fails to see systematic research to delve into the impact of patch ranking strategies on repair performance. 
As a guideline for future work, after summarizing existing patch ranking works, we recommend that a reasonable patch ranking technique needs to consider three aspects: 
\ding{172}\textit{effectiveness}, \ie it should have a sufficiently large search space to encompass the correct patches;
\ding{173}\textit{efficiency}, \ie it should have a fast retrieval speed to find the correct patch in a reasonable amount of time;
and, \ding{174}\textit{priority}, \ie it should prioritize the patch that is more likely to be correct higher based on additional code information, such as code syntactic and semantic features.
}}

\subsection{Patch Validation}
\label{sec:patch_validation}
Patch validation takes a ranked list of candidate patches generated by NMT models as the input and returns the plausible patches for deployment, which is a crucial phase in the learning-based APR pipeline.
However, developers may spend a considerable amount of effort to inspect the candidate patches manually. 
Thus, researchers usually recompile the buggy program with the generated patch to check if it can pass the available test suite. 
In such a scenario, hundreds or even thousands of candidate patches can be filtered automatically (\eg 1000 candidate patches per bug in CIRCLE~\cite{yuan2022circle}), which may benefit its adoption in practice.

Similar to traditional APR techniques, most learning-based techniques adopt a test-based validation strategy (\ie executing available test suites against each candidate patch) to assess patch correctness~\cite{zhu2021syntax,yuan2022circle,li2020dlfix,jiang2021cure,lutellier2020coconut,li2022dear}.
For example, CIRCLE~\cite{yuan2022circle} filters out the candidate patches that do not compile or do not pass available test suites.
There generally exist two criteria for the validation process:
(1) the passing test suites that make the buggy program pass should still pass on the patched program;
and (2) the fault-triggering test suites that fail on the buggy program should pass on the patched program.
All candidate patches are checked until a plausible patch (\ie a patch passing all test suites) is found.
Finally, CIRCLE stops the validation process and reports the plausible patch for manual investigation.
\revise{Similar test-based validation strategies are also employed by other learning-based APR appraoches, such as Recoder~\cite{zhu2021syntax}, CoCoNut~\cite{lutellier2020coconut} and CURE~\cite{jiang2021cure}.}

However, it can be extremely time-consuming to compile a large number of candidate patches and repeat all test executions to identify plausible patches.
For example. CURE~\cite{jiang2021cure} generates 10,000 candidate patches per bug and validates the top 5,000 ones considering the overhead time.
Similarly, AlphaRepair~\cite{xia2022less} returns at most 5,000 candidate patches for each bug.
To reduce the validation cost, some learning-based APR techniques return an acceptable amount of candidate patches.
For example, RewardRepair configures the beam size as 200 and outputs the 200 best patches per bug.
Similarly, SelfAPR adopts a beam search size of 50 and Recoder generates 100 valid candidate patches for validation.
Besides, similar to traditional APR techniques~\cite{jiang2018shaping,liu2019tbar}, there exist several learning-based ones limiting maximum time for validation.
For example, DEAR~\cite{li2022dear} and DLFix~\cite{li2020dlfix} set a 5-hour running-time limit for patch generation and validation.

\revise{In addition to the above strategies in patch validation, the learning-based APR community benefits from some optimizations to speed up the dynamic execution.
For example, AlphaRepair~\cite{xia2022less} adopts the UniAPR~\cite{chen2021fast} tool to validate the candidate patches on-the-fly.
For example, Inspired by the PraPR work~\cite{ghanbari2019practical}, Chen~\etal~\cite{chen2021fast} present UniAPR as the first unified on-the-fly patch validation framework to speed up APR techniques for JVM-based languages at both the source and byte-code levels. 
\revise{They leverage the JVM HotSwap mechanism and Java Agent technology to implement this framework. 
Besides, they apply the JVM resetting technique based on the ASM byte-code manipulation framework.}
Since previous work shows that on-the-fly patch validation can be imprecise, they reset the JVM state right after each patch execution to address such an issue.} 

\revise{
Orthogonal to UniAPR, Bento~\etal~\cite{benton2022towards} introduce SeAPR, the first self-boosted patch validation tool. Based on the idea that patches similar to earlier high-quality/low-quality patches should be promoted/degraded, they leverage the patch execution information on its similarity with the executed patches to update each patch’s priority score. The evaluation shows that SeAPR can substantially speed up the studied APR techniques and its performance is stable under different formulae for computing patch priority.
Besides, the literature has seen the emergence of several patch validation studies.
For example, as early as 2012, Qi~\etal~\cite{qi2012making} propose WAutoRepair, a repair system that combines Genprog with a recompilation technique called weak recompilation to reduce time cost and make program repair more efficient.
WAutoRepair views a program as a set of components and for each candidate patch, only one component is modified.
After that, the changed component is compiled to a shared library to reduce the time cost.
In 2013, inspired by regression test prioritization, Qi~\etal~\cite{qi2013efficient} propose TrpAutoRepairto prioritize test case execution based on the faults information in the repair process.
Although these works have achieved commendable results, most of them have all been applied to traditional APR techniques, \eg GenProg~\cite{le2012genprog}. 
However, considering that the patch validation phase is designed to compile and execute the candidate patch, which is independent of the specific patch generation techniques, such patch validation techniques have the potential to be extended to learning-based repair techniques in the future.
}

\delete{
Since previous APR techniques often neglect the impact of test selection for each patch, Lou~\etal~\cite{lou2021does} conduct an extensive study to investigate the impact of Regression Test Selection (RTS) on APR. They explore three representative RTS techniques for 12 \delete{state-of-the-art} APR systems at different levels (\ie class/method/statement levels) with over 2M patches. Results show that all studied RTS techniques can substantially improve APR efficiency and should be considered in future APR work. Besides, method- and statement-level RTS substantially outperform class-level RTS, and are more recommended for APR.}

\summary{\revise{
Dynamic execution is a common practice to automatically validate the code's compilability and functional correctness of programs in the SE community.
However, it is time-consuming to compile and execute such programs, especially in the field of APR where thousands of candidate patches and a mass of functional test cases are involved.
Most learning-based APR techniques rely on a test-based validation strategy to identify plausible patches, which is a standard step in both traditional and learning-based APR communities.
Besides, recently there have been some advanced techniques proposed specifically to validate these candidate patches more quickly, such as JVM HotSwap.
Currently, there is no distinct differentiation in patch validation research between the traditional and learning-based APR communities.
The possible reason lies in that, the patch validation phase aims to identify high-quality patches that pass the available test cases.
This objective can be achieved by directly executing the test cases, without concerning whether the patches come from traditional or learning-based APR techniques.
We encourage more research on patch validation that is specific to learning-based APR techniques, discussed in Section~\ref{sec:guide}.
}}

\subsection{Patch Correctness}
\label{sec:pat_correctness}

Patch correctness is an additional phase for developers to further filter out overfitting patches after patch validation, so as to improve the quality of returned patches.
As discussed in Section \ref{sec:patch_validation}, a majority of existing learning-based APR techniques usually leverage the developer-written test suites as the program specification to assess the correctness of the generated patches.
However, the test suite is an incomplete specification as it only describes a part of the program's behavioral space.
As a result, it is fundamentally difficult to achieve high precision for returned patches due to the incomplete program specification~\cite{le2019reliability}.
The plausible patch passing the available test suites may not generalize to other potential test suites, leading to a long-standing challenge of APR (\ie \textit{the overfitting issue})~\cite{le2019reliability,zhang2023boosting}.
\delete{Previous studies~\cite{qi2015analysis,long2016automatic} have demonstrated that a majority of the overfitting patches are equivalent to a single modification that deletes the buggy functionality and does not actually fix the detected bugs.}

\revise{For example, Qi~\etal~~\cite{qi2015analysis} have demonstrated that a majority of the overfitting patches generated by previous APR approaches (\eg GenProg~\cite{le2012genprog}) for 105 C language bugs are equivalent to a single modification that deletes the buggy functionality and does not actually fix the detected bugs.}
Under the circumstances, it takes enormous time and effort to manually filter out the overfitting patches, even resulting in a negative debugging performance~\cite{tao2014automatically, zhang2022program1}.
Different from some traditional APR techniques that guide the repair process to generate patches with a high probability of being correct, DL techniques lead to an end-to-end repair mechanism and the patches are generated in a black-box manner.
The overfitting issue in learning-based APR is more \revise{significant and severe~\cite{wang2022attention}}.

In the literature, researchers have proposed a mass of automated patch correctness assessment (APCA) techniques to identify whether a plausible patch is indeed correct or overfitting~\cite{tian2020evaluating}.
\revise{
Xiong~\etal~\cite{xiong2018identifying} propose PATCH-SIM to identify correct patches based on the similarity of test case execution traces on the buggy and patched programs.
PATCH-SIM has been acknowledged as a foundational work in this field~\cite{tian2020evaluating}, providing crucial guidance for the conception and development of follow-up works~\cite{ye2021automated,yang2020daikon}.
}
There are usually two types of traditional APCA techniques based on the employed patch features: \textit{static and dynamic}~\cite{wang2020automated}.
The former focuses on the transformation patterns or the static syntactic similarity (\revise{\eg Anti-pattern~\cite{tan2016anti}}), while the latter relies on the dynamic execution outcomes by additional test suites from automated test generation tools (\revise{\eg PATCH-SIM~\cite{xiong2018identifying}}).
Recently, inspired by large-scale patch benchmarks being released, some learning-based APCA techniques have been proposed to predict patch correctness with the assistance of DL models~\cite{ye2021automated,tian2020evaluating, tian2022predicting}.
In general, such learning-based APCA techniques extract the code features \delete{by code embedding}\revise{(\eg static representation or dynamic execution traces)} and build a classifier model to directly perform patch \revise{correctness} prediction.
We view patch correctness as an essential component of the learning-based APR pipeline and focus on such APCA techniques that employ DL models.

\input{tab/tab_APCA}

Table~\ref{tab:apca} presents a summary of existing learning-based techniques to predict patch correctness automatically in the literature.
The first and second columns list the APCA technique and the time of publication.
The third column lists the targeted programming languages.
The fourth column lists the features adopted by the APCA technique.
The remaining columns list the employed datasets and the public repositories.
\delete{Now, we list \revise{and summarize} the existing learning-based techniques to predict patch correctness \delete{automated}\revise{automatically} as follows.}
\revise{Now, we discuss and summarize these individual approaches as follows.}

\input{APR/APCA/2020-IBF-Utilizing}

\input{APR/APCA/2020-ASE-Evaluating}

\input{APR/APCA/2021-TSE-ODS}

\input{APR/APCA/2021-TOSEM-Cache}

\input{APR/APCA/2022-TOSEM-BATS}

\input{APR/APCA/2022-ASE-QUATRAIN}

\input{APR/APCA/2022-IST-Crex}

\input{APR/APCA/2022-ISSTA-Shibboleth}

\summary{\revise{
The overfitting issue has become a key focus in the field of APR, which has led to the emergence and rapid development of recent APCA techniques.
DL techniques have been gradually used to predict the correctness of patches by learning features from historical corpora.
Compared to traditional dynamic and static APCA, learning-based APCA has shown impressive performance in prediction accuracy and recall.
We provide a summary of the existing learning-based APCA techniques in Table~\ref{tab:apca}.
In the literature, most existing APCA techniques employ a two-component pipeline, \ie the feature extractor and the classifier.
The former extracts the features from the source code of patches, \eg hand-crafted features, static representation features and dynamic execution features, while the latter trains a classifier to perform binary prediction, \eg Random Forest and Decision tree.
Despite increasing research efforts being put into this phase and encouraging progress being made, the problem of patch overfitting still hinders the application and deployment of repair techniques in practice.
Therefore, the APR community needs more advanced APCA techniques to improve the correctness of returned patches, \eg patch-aware feature extraction and more powerful pre-trained models.
}}

\subsection{\revise{State-of-the-Arts}}
\label{sec:sota}

\revise{
In the learning-based APR field, semantic error (\ie test-triggering error) has attracted considerable attention from researchers, which is the most general application of repair techniques discussed in the previous sections.
A living review~\cite{monperrus2020living} summarizes and categorizes existing APR techniques into different repair scenarios during software development, including static errors, concurrency errors, \etc
In this section, we attempt to summarize existing representative learning-based APR techniques across different scenarios where learning-based APR is most applied.

Table~\ref{tab:apr} presents a summary of existing representative learning-based APR techniques. The first and second columns list the investigated repair techniques and the time when these techniques are presented.
The third and fourth columns list the targeted bug types and programming languages.
The fifth column lists the adopted fault localization technique.
The sixth, seventh, and eighth columns list the detailed data pre-processing methods, \ie code context, code abstraction and code tokenization.
The ninth and tenth columns list the detailed code representation and employed models.
The last column lists the employed patch ranking strategy.
In the following, we discuss these learning-based APR techniques according to the repair scenarios.
}

\input{tab/tab_APR}

\revise{\input{sec/sec_sematic}}

\revise{\input{sec/sec_syntax}}

\revise{\input{sec/sec_vul}}

\revise{\input{sec/sec_programming}}

\newrevise{In addition to the above-mentioned studies, the community has also seen an increasing number of pre-trained model-based APR studies.
Considering the growing popularity and potential impact of pre-trained models on the learning-based APR community, we deserve a separate section to thoroughly discuss such stat-of-the-arts for a more comprehensive insight, detailed in Section~\ref{sec:pre-trained}.
}

\summary{\revise{Overall, learning-based APR is generally applicable to different types of bugs thanks to the end-to-end black-box NMT training, including semantic bugs, syntax bugs, and security vulnerabilities.
On top of encoder-decoder architecture, the APR problem can be abstracted as an MMT task, which translates a buggy code snippet into a correct one.
Thus, researchers can train NMT models to learn code transformations automatically from a mass of bug-fixing code pairs without considering the specific bug types.
For example, CodeT5 is fine-tuned to repair semantic bugs~\cite{xia2022practical}, syntax bugs~\cite{berabi2021tfix}, and security vulnerabilities~\cite{fu2022vulrepair}, respectively.
Despite its high scalability, the learning-based APR is currently mostly focused on the aforementioned typical domains. 
Future work can be conducted to explore the performance of the learning-based APR in more repair scenarios, \eg API misuse, detailed in Section~\ref{sec:guide}.
}}

\input{sec/sec_domain_old}

\section{\revise{Pre-trained Model-based Repair}}
\label{sec:pre-trained}

\delete{In this section, we will discuss the existing studies of pre-trained models on the ARP task.}

Pre-trained models have significantly improved performance across a wide range of natural language processing (NLP) and code-related tasks, such as machine translation, defect detection and code classification~\cite{lu2021codexglue,guo2020graphcodebert}.
Typically, the models are pre-trained to derive generic vector representation by self-supervised training on a large-scale unlabeled corpus and then are transferred to benefit multiple downstream tasks by fine-tuning on a limited labeled corpus~\cite{kenton2019bert}.
The application of existing pre-trained models to program repair is usually divided into two categories: \textit{universal and specific pre-trained model-based APR techniques}.
The former aims to propose universal pre-trained models for multiple code-related tasks (including program repair), while the latter only focuses on program repair by designing a novel APR technique based on pre-trained models.

\subsection{Universal Pre-trained Model-based APR Techniques}

Existing pre-trained models generally adopt the encoder-decoder transformer architecture, which can be classified into three types: encoder-only, decoder-only, and encoder-decoder models.
Encoder-only models (\eg CodeBERT~\cite{feng2020codebert}) usually pre-train a bidirectional transformer where tokens can attend to each other.
Encoder-only models are good at understanding tasks (\eg code search), but their bidirectionality nature requires an additional decoder for generation tasks.
Decoder-only models (\eg CodeGPT~\cite{brown2020gpt}) are pre-trained using unidirectional language modeling that only allows tokens to attend to the previous tokens and themselves to predict the next token.
Decoder-only models are good at auto-regressive tasks like code completion, but the unidirectional framework is sub-optimal for understanding tasks.
Encoder-decoder models (\eg CodeT5~\cite{raffel2019t5}) often make use of denoising pre-training objectives that corrupt the source input and require the decoder to recover them.
Compared to encoder-only and decoder-only models that favor understanding and auto-regressive tasks, encoder-decoder models can support generation tasks like code summarization.

Inspired by the success of pre-trained models in NLP, many recent attempts have been adopted to boost numerous code-related tasks (such as program repair) with pre-trained models (\eg CodeBERT)~\cite{feng2020codebert, guo2020graphcodebert}.
In the context of APR, an encoder stack takes a sequence of code tokens as input to map a buggy code $X_i=\left[x_1, \ldots, x_n\right]$ into a fixed-length intermediate hidden state, while the decoder stack takes the hidden state vector as an input to generate the output sequence of tokens $Y_i=\left[y_1, \ldots, y_n\right]$.
Researchers treat the APR problem as a generation task, and consider encoder-decoder or encoder-only (with an additional decoder) \delete{pre-traiend}\revise{pre-trained} models, which are usually evaluated by the BFP dataset from Tufano~\etal~\cite{tufano2019empirical}.

We summarize existing pre-trained models involving the program repair task as follows.

\delete{Feng~\etal~\cite{feng2020codebert} present a bimodal pre-trained model (\ie \textit{CodeBERT}) for natural language and programming language with a transformer-based architecture.
CodeBERT utilizes two pre-training objectives (\ie masked language modeling and replaced token detection) to support both code search and code documentation generation tasks.
To support program repair task, 
Lu~\etal~\cite{lu2021codexglue} leverage CodeBERT as the encoder, which is connected with a randomly initialized decoder.}

\delete{
Wang~\etal~\cite{wang2021codet5} present a pre-trained encoder-decoder model (\ie \textit{CodeT5}) that considers the code token type information based on T5 architecture.  
CodeT5 employs a unified framework to support code understanding (\eg clone detection) and generation tasks (\eg program repair) and allows for multi-task learning.
The most crucial feature of CodeT5 is that the code semantics of identifiers are taken into consideration. 
Assigned by developers, identifiers often convey rich code semantics and thus a novel identifier-aware objective is added to the training of CodeT5.}

\delete{
Guo~\etal~\cite{guo2020graphcodebert} present the first structure-aware pre-trained model (\ie \textit{GraphCodeBERT}) that learns code representation from source code and data flow.
Unlike existing models focusing on syntactic-level information (\eg AST), GraphCodeBERT takes semantic-level information of code (\eg data flow) for pre-training with a transformer-based architecture.
The results on BFP datasets~\cite{tufano2019empirical} demonstrate the advantage of leveraging code structure information to repair software bugs.}

\delete{Zhang~\etal~\cite{zhang2022coditt5} propose CoditT5, a pre-trained language model for software-related edit tasks. CoditT5 is pre-trained on both program languages and natural language comments. Zhang~\etal~fine-tune it for three down-streaming tasks: comment updating, bug fixing, and automatic code review. 
For bug-fixing, they fine-tune it with Java datasets BFP small and BFP medium. The evaluation shows that CoditT5 outperforms other \delete{state-pf-art tools} \revise{APR tools like CodeT5 and PLBART} on three down-streaming tasks.}

\delete{Mastropaolo~\etal~\cite{mastropaolo2022using} propose \revise{a} pre-trained text-to-text transfer transformer (T5) to address four code-related tasks, namely automatic bug fixing, injection of code mutants, generation of assert statements in test methods, and code summarization. They apply BFP small and BFP medium datasets to train and evaluate the bug-fixing task, and then compare other \delete{state-of-the-art} learning-based APR tools on the same benchmark. Moreover, they have done single-task fine-tuning and multi-task fine-tuning to fully evaluate the function of the pre-trained T5 model. Although multi-task fine-tuning does not improve the result of code-related tasks, single-task fine-tuning does prove that this model outperforms other tools \revise{(\eg Tufano~\etal~\cite{tufano2019empirical})} on the same benchmarks.}

\delete{
Niu~\etal~\cite{niu2022spt} propose a Seq2Seq pre-trained model (\textit{\ie SPT}) by three code-specific tasks (code-AST prediction, masked sequence to sequence and method name generation) and fine-tune on the generation tasks (\ie code summarization, code completion, program repair and code translation) and classification task (\ie code search).
\revise{Results show that SPT outperforms CodeBERT~\cite{feng2020codebert}, GraphCodeBERT~\cite{guo2020graphcodebert} and Tufano~\etal~\cite{tufano2019empirical} on the bug-fixing datasets \newrevise{BFP-small} and BFP-medium.}
}

\revise{The most commonly used model type is the T5-like encoder-decoder architecture, which can naturally support the program repair task as a code generation problem.
For example, in 2021, Wang~\etal~\cite{wang2021codet5} present a pre-trained encoder-decoder model (\ie \textit{CodeT5}) that considers the code token type information based on T5 architecture.  
CodeT5 employs a unified framework to support code understanding (\eg clone detection) and generation tasks (\eg program repair) and allows for multi-task learning.
The most crucial feature of CodeT5 is that the code semantics of identifiers are taken into consideration. 
Assigned by developers, identifiers often convey rich code semantics and thus a novel identifier-aware objective is added to the training of CodeT5.} 

\revise{In 2022, Mastropaolo~\etal~\cite{mastropaolo2022using} propose \revise{a} pre-trained text-to-text transfer transformer (T5) to address four code-related tasks, namely automatic bug fixing, injection of code mutants, generation of assert statements in test methods, and code summarization. They apply BFP small and BFP medium datasets to train and evaluate the bug-fixing task, and then compare other \delete{state-of-the-art} learning-based APR tools on the same benchmark. Moreover, they have done single-task fine-tuning and multi-task fine-tuning to fully evaluate the function of the pre-trained T5 model. Although multi-task fine-tuning does not improve the result of code-related tasks, single-task fine-tuning does prove that this model outperforms other tools \revise{(\eg Tufano~\etal~\cite{tufano2019empirical})} on the same benchmarks.
Besides, Niu~\etal~\cite{niu2022spt} propose a Seq2Seq pre-trained model (\textit{\ie SPT}) by three code-specific tasks (code-AST prediction, masked sequence to sequence and method name generation) and fine-tune on the generation tasks (\ie code summarization, code completion, program repair and code translation) and classification task (\ie code search).
\revise{Results show that SPT outperforms CodeBERT~\cite{feng2020codebert}, GraphCodeBERT~\cite{guo2020graphcodebert} and Tufano~\etal~\cite{tufano2019empirical} on the bug-fixing datasets \newrevise{BFP-small} and BFP-medium.}
Unlike previous general-purpose pre-trained models considering various tasks, Zhang~\etal~\cite{zhang2022coditt5} propose CoditT5, a pre-trained language model only for code-related edit tasks. CoditT5 is pre-trained on both program languages and natural language comments. Zhang~\etal~fine-tune it for three down-streaming tasks: comment updating, bug fixing, and automatic code review. 
For bug-fixing, they fine-tune it with Java datasets BFP small and BFP medium. The evaluation shows that CoditT5 outperforms other \delete{state-pf-art tools} \revise{APR tools like CodeT5 and PLBART} on three down-streaming tasks.}

\revise{
There also exist some pre-trained models with BERT-like encoder-only architecture, which usually with an additional decoder ti support program repair.
For example, in 2020, Feng~\etal~\cite{feng2020codebert} present a bimodal pre-trained model (\ie \textit{CodeBERT}) for natural language and programming language with a transformer-based architecture.
CodeBERT utilizes two pre-training objectives (\ie masked language modeling and replaced token detection) to support both code search and code documentation generation tasks.
To support program repair task, 
Lu~\etal~\cite{lu2021codexglue} leverage CodeBERT as the encoder, which is connected with a randomly initialized decoder.
Besides, Guo~\etal~\cite{guo2020graphcodebert} present the first structure-aware pre-trained model (\ie \textit{GraphCodeBERT}) that learns code representation from source code and data flow.
Unlike existing models focusing on syntactic-level information (\eg AST), GraphCodeBERT takes semantic-level information of code (\eg data flow) for pre-training with a transformer-based architecture.
The results on BFP datasets~\cite{tufano2019empirical} demonstrate the advantage of leveraging code structure information to repair software bugs.}

\subsection{Specific Pre-trained Model-based APR Techniques}

In addition to those above-mentioned typical pre-trained models that involve program repair, researchers have adopted pre-trained models to design novel APR techniques.
Table~\ref{tab:tool} presents existing pre-trained model-based APR techniques.
We summarize existing APR studies that employ pre-trained models as follows.

\input{tab/tab_LLM}

\input{APR/syntax/2021-PMLR-TFix}

\input{APR/syntax/2022-TSE-SYNSHINE}

\input{APR/semantic/sequence/2021-ASE-MODIT}

\input{APR/semantic/sequence/2022-ISSTA-CIRCLE}

\input{APR/semantic/sequence/2022-FSE-AlphaRepair}

\input{APR/vul/2022-FSE-VulRepair}

In parallel with these newly proposed approaches equipped with pre-trained models, the community has also seen some studies that empirically explore the actual performance of pre-trained models in different repair scenarios.
We will discuss these empirical studies in Section~\ref{sec:empirical}.

\summary{\revise{
Overall, pre-trained models have significantly influenced a large amount of code-related fields in the SE community, especially program repair.
At the current stage, existing techniques using pre-trained models for program repair are usually divided into three categories.
First, when some pre-trained models are built, they are fine-tuned and evaluated by some downstream tasks, including program repair.
The evaluation experiments are usually conducted by these authors of the pre-trained models and reported in their original papers using the BPF dataset from Tufano~\etal~\cite{tufano2019learning}.
The typical pre-trained models involve CodeT5~\cite{wang2021codet5}, T5Learning~\cite{mastropaolo2021t5learning} and SPT~\cite{niu2022spt}.
Second, researchers have proposed some novel pre-trained model-based APR techniques.
The first typical one is the fine-tuning scenario, \eg CIRCLE~\cite{yuan2022circle} is proposed to fine-tune the pre-trained T5 model with continual learning.
The second typical one is the zero-shot scenario, \eg AlphaRepair~\cite{xia2022less} is proposed to use CodeBERT to generate correct code under a cloze-style way. 
Third, there exists an increasing number of empirical studies to evaluate the ability of pre-trained models in program repair.
These empirical studies encompass different pre-trained models~\cite{xia2022practical}, bug types~\cite{huang2022repairing} and programming languages~\cite{kim2022empirical}.
In the future, pre-trained models can further deeply influence various steps of the program repair workflow, such as patch correctness assessment, detailed in Section~\ref{sec:guide}.
}}

\section{Empirical Evaluation}
\label{sec:evaluation}

In this section, we introduce existing widely adopted datasets in the learning-based APR field and discuss common evaluation metrics for evaluating repair performance.

\subsection{Dataset}
\label{sec:dataset}

\input{tab/tab_dataset}

Different from previous APR techniques conducted in a traditional pipeline (\eg generating patches by heuristic strategies), the process of learning-based APR techniques is two-fold
(1) a training process with supervised learning on large labeled datasets (\eg CoCoNut~\cite{lutellier2020coconut}); 
and (2) an evaluation process on \delete{limited} \revise{a small set of} labeled datasets (\eg Defects4J~\cite{just2014defects4j}).
Benefiting from a large amount of research effort in the learning-based APR community, there are several existing benchmarks to evaluate NMT techniques for automatically repairing bugs.
Now we discuss the widely adopted datasets in the literature.

Defects4J~\cite{just2014defects4j} is the most widely adopted benchmark in learning-based APR studies, which contains 395 known and reproducible real-world bugs from six open-source Java projects.
To facilitate reproducible studies, each bug contains a buggy version and a fixed version, as well as a corresponding test suite that triggers that bug.
Defects v2.0 provides 420 additional real-world bugs from 17 Java projects, which is adopted by some recent studies~\cite{zhu2021syntax,xia2022less}.
QuixBugs~\cite{lin2017quixbugs} is a multi-lingual parallel bug-fixing dataset in Python and Java used in~\cite{xia2022less, yuan2022circle}.
QuixBugs contains 40 small classic algorithms with one bug on a single line, along with the test suite.
Bugs.jar~\cite{saha2018bugs} contains 1,158 real bugs from 8 large open-source Java projects, each of which has a fault-revealing test suite.
ManyBugs~\cite{le2015manybugs} contains 185 real-world bugs from 9 open-source C projects and each bug has a corresponding developer patch and test suite.
IntroClass~\cite{le2015manybugs}  consists of 998 bugs in six small student-written programming assignments for C language.
Due to a well-defined test suite, these datasets are effective in evaluating the correctness of generated patches by dynamic program behavior.

However, NMT-based APR techniques employ neural network techniques to learn the bug-fixing patterns from the training dataset.
\revise{The training of a reliable NMT repair model is hindered by the scarcity of high-quality test datasets, which require extensive manual effort to produce.}
To make experiment results more persuasive, lots of large-scale datasets have been \delete{conducted} \revise{curated} recently.
Such datasets contain bug-fixing code pairs for the model to learn how to transform a buggy code into the expected fixed code. 
In particular, researchers usually mine open-source projects from code platforms (\eg GitHub) and extract the commits by fixing-related keywords.
Then unqualified commits are filtered out by pre-defined rules (\eg non-code changes).
For example, Tufano~\etal~\cite{tufano2019empirical} extract the bug-fixing commits between March 2011 and October 2017 on GitHub and release two BFP datasets for small (\ie 0$\sim$50 tokens) and medium (\ie 50$\sim$100 tokens) methods, consisting of 58k (58,350) and 65k (65,455) bug-fixing samples, respectively.
Recoder~\cite{zhu2021syntax} releases a dataset of 103,585 bug-fixing pairs by crawling Java projects on GitHub between March 2011 and March 2018.
Further, CoCoNut~\cite{lutellier2020coconut} provides five datasets across four languages (\ie Java, Python, C and JavaScript) by extracting commits from GitHub projects, resulting in more than twenty million bug-fixing pairs.

Table \ref{tab:dataset} presents the description of all involved datasets in our survey.
The first two columns list the dataset name and the third column lists the programming languages the dataset covers.
The fourth column lists the number of bugs the dataset contains.
The fifth column indicates whether the dataset has corresponding test suites.
The sixth and seventh columns indicate whether the dataset is used in the training and evaluation process.
The last column lists some learning-based studies employing the dataset.

Among the collected datasets in our survey, we find that training datasets usually only contain bug-fixing pairs for NMT model training, while evaluation datasets may additionally contain some test suites to validate the correctness of generated patches.
For example, existing studies~\cite{lutellier2020coconut,yuan2022circle} generally adopt some datasets like Defects4J as the evaluation datasets while adopting other datasets like CoCoNut as the training datasets.
Besides, we find some studies~\cite{tufano2019empirical,tufano2019learning} adopt the same dataset for training and evaluation without executing test suites.
For example, Tufano~\etal~\cite{tufano2019empirical} split BFP dataset into training and evaluation parts and evaluate the repair performance by match-based metric.

Table \ref{tab:dataset} also presents the programming languages of all datasets.
It can be found that the collected datasets mainly involve five languages (\ie Java, JavaScript, Python, C and C++).
Among them, similar to traditional APR, Java is the most targeted language in the learning-based APR techniques.
Besides, researchers conduct lots of datasets in other languages (\eg Python), indicating that learning-based APR techniques begin to consider more languages in practice.
For Java, researchers prefer the traditionally dominated Defects4J dataset and the recently-released BFP dataset.
For other programming languages, researchers have different choices for datasets due in part to the lack of publicly-accepted datasets.
We also find that some recent datasets involve multi-languages, such as CoCoNut~\cite{lutellier2020coconut} and \revise{QuixBugs~\cite{lin2017quixbugs,ye2019comprehensive}}, while the traditional APR techniques mainly focus on Java language~\cite{durieux2019empirical}.
The possible reasons lie in that
(1) traditional techniques are widely conducted on the same benchmark Defects4J while some additional datasets have been released along with the application of DL;
(2) traditional techniques may rely on language-specific features to generate patches, which is challenging to apply to other languages (\eg PraPR adopting JVM bytecode~\cite{ghanbari2019practical}), while learning-based techniques treat APR as an NMT task similar to NLP, which is independent of specific programming languages.

\summary{\revise{Within the expansive arena of learning-based APR, datasets play a pivotal role in shaping the trajectory of research advancements.
Different from traditional APR techniques, which often leverage heuristic strategies for patch generation, learning-based APR techniques are distinctly split into a two-phase methodology: a supervised training on large-scale labeled datasets and a subsequent evaluation on smaller, selected datasets.
While the traditional APR realm has seen an inclination towards Java-centric datasets like Defects4J, the infusion of DL into the sector has broadened horizons.
One typical trend is the construction of large-scale training datasets, \eg the BPF dataset~\cite{tufano2019learning}.
The other typical trend is the application of multiple programming languages, \eg the CoCoNut dataset~\cite{lutellier2020coconut}.
However, we observe that while ample datasets exist for training—mainly comprising bug-fixing pairs, evaluation datasets often carry the added component of test suites to ascertain patch correctness. 
In summation, as learning-based APR continues to evolve, it is imperative for the community to prioritize the curation of comprehensive, high-quality datasets that cater to both training and evaluation. 
}}

\subsection{Metric}
\label{sec:metric}

\revise{
Evaluation metrics play a crucial role in the development and growth of learning-based APR techniques as they serve as the standard to quantitatively define how good an NMT repair model is.
In this section, we discuss the common evaluation metrics in the learning-based APR community.
}

\subsubsection{Execution-based Metrics}
In general, learning-based APR techniques predict some candidate patches with high probability as the outputs.
The generated patches are evaluated by executing available test suites to determine whether to report them to the developers for deployment.
We list the standard metrics as follows.

\begin{description}
    \item[(1) Compilable Patch.] Such a candidate patch makes the patched buggy program compile successfully.
    \item[(2) Plausible Patch.] Such a compilable patch fixes the buggy functionality without harming existing functionality (\ie passing all available test suites).
    \item[(3) Correct Patch.] Such a plausible patch is semantically or syntactically equivalent to the developer patch (\ie generalizing the potential test suite).
    
\end{description}

\subsubsection{Match-based Metrics}
\delete{However}\revise{Although widely used in the learning-based APR literature}, it is time-consuming to evaluate generated patches on dynamic execution for all available test suites.
Besides, test suites may not always be available in large-scale evaluation datasets.
More recently, an increasing number of studies evaluate the performance by code token matching between the generated patch and the ground truth (\ie developer-written patches), listed as follows.

\begin{description}

    \item[(1) Accuracy.]
    \text{Accuracy} measures the percentage of candidate patches in which the sequence predicted by the model equals the ground truth. 
    As learning-based APR techniques usually employ a beam-search strategy, the beam-search strategy reports the $k$ sequences (\ie sequence of terms representing the fixed code) with the highest probability.
    Researchers consider these $k$ final sequences as candidate patches for a given buggy code snippet.
    Then \text{Accuracy@K} value is defined as follows.

    \begin{equation}
	Accuracy@K = \frac
	{\sum_{i=1}^{n} \mathbbm{1} \{match(\sum_{j=1}^{k} c_i^j)\}}{n}
    \end{equation}
    where $\mathbbm{1}$ denotes whether $C_i$ contains a predicted repair sequence equal to the ground truth repair sequence.
    The sequence accuracy is 1 if any predicted sequence among the $k$ outputs matches the ground truth sequence, and it is 0 otherwise.

    \item[(2) BLEU.] \delete{BLUE}\revise{BLEU} (Bilingual Evaluation Understudy)~\cite{papineni2002bleu} score measures how similar the predicted candidate patch and the ground truth \delete{is}\revise{are}.
    Given a size $n$, BLEU splits the candidate patch and ground truth into n-grams and determines how many n-grams of the candidate patch appear in the reference patch. 
    The BLEU score ranges between 0 (the sequences are completely different) and 1 (the sequences are identical).

\end{description}

Compared with execution-based metrics, accuracy and BLUE evaluate the candidate patch by matching the tokens of the candidate patch and ground truth without dynamic execution.
These two metrics can be employed to evaluate the performance of a mass of candidate patches in a limited time and thus have been commonly adopted in the learning-based APR community~\cite{tufano2019empirical,yuan2022circle,tufano2019learning}.
However, accuracy and BLUE are initially designed in NLP tasks and may be improper to evaluate the program repair task due to the differences between natural language and \revise{programming language. For example,} accuracy refers to the perfect prediction, which ignores that different code snippets may have the same semantic logic.
Besides, BLEU is originally designed for natural language sentences by token-level matching, neglecting important syntactic and semantic features of codes.
To address the above concerns, recently researchers adopt a variant of BLEU (\ie CodeBLEU ~\cite{ren2020codebleu}) to evaluate the performance of learning-based APR techniques~\cite{lu2021codexglue}.
Compared with BLEU, CodeBLEU further considers the weighted n-gram match, the syntactic AST match, and the semantic data-flow match.
In particular, the n-gram match assigns different weights for different n-grams, the syntactic match considers the AST information in the evaluation score by matching the sub-trees, and the semantic match employs a data-flow structure to measure semantic similarity.

\summary{\revise{
Overall, within the realm of learning-based APR, evaluation metrics are of paramount importance in guiding the evolution of repair models.
On the one hand, similar to traditional APR, learning-based APR the APR domain has been inclined towards execution-based metrics, such as plausible patches, which are derived from the field of SE.
On the other hand, unlike traditional APR, are increasingly biased towards match-based metrics, such as BLEU, which are derived from the field of NLP.
The possible reason behind this trend is the lack of test cases in the evaluation datasets, such as the BFP dataset~\cite{tufano2019learning}.
Despite their convenience, these NLP-inspired metrics are not without their pitfalls. 
For example, Accuracy focuses narrowly on perfect predictions, and traditional BLEU might overlook the intricate semantics of source code. 
To sum up, as learning-based APR continues its upward trajectory, the spotlight is increasingly on the development and adoption of nuanced, code-centric evaluation metrics (such as CodeBLUE) that mirror the complexities of the programming domain.
}}

\subsection{Empirical Study}
\label{sec:empirical}

Despite an emerging research area, a variety of learning-based APR techniques have been proposed and continuously achieved promising results in terms of the number of fixed bugs in the literature~\cite{lutellier2020coconut,yuan2022circle}.
In addition to developing new repair techniques that address technical challenges, the learning-based APR research field is benefiting from several empirical studies. 
These empirical studies systematically explore the impact of different components (\eg code representation), providing insights into future learning-based APR work.
\revise{We summarize existing empirical studies in Table~\ref{tab:empirical} and discuss them in detail as follows.}

\input{tab/tab_empirical}

\delete{
Tufano~\etal~\cite{tufano2019empirical} conduct a systematic empirical study to investigate the capability of utilizing NMT models to fix software bugs from open-source bug-fixing commits.
They first mine the bug-fixing commits by message patterns from projects in GitHub repositories and filter out the low-quality commits by specific rules.
They then extract correct and buggy code pairs at the method level by GumTree and fixed code and design a code abstraction strategy to reduce vocabulary size.
Finally, they construct two datasets (\ie small and medium BFPs) and train NMT models to translate the buggy method into the correct method. 
The results demonstrate that NMT models are able to fix a considerable number of buggy methods in the wild, proving the applicability of NMT for APR.}

\revise{
Traditional APR techniques are usually restrained by a relatively limited set of manually crafted repair patterns~\cite{liu2019tbar}.
Inspired by the potential of advanced DL techniques, which have shown impressive performance in tackling several SE tasks~\cite{yang2022survey}, in 2019, Tufano~\etal~\cite{tufano2019empirical} conduct the first systematic empirical study to investigate the capability of utilizing NMT models to fix software bugs from open-source bug-fixing commits.
First, they mine the bug-fixing commits by message patterns from projects in GitHub repositories and filter out the low-quality commits by specific rules.
Second, they identify the list of edit actions performed between the buggy and fixed files using the GumTree~\cite{falleri2014fine} and extract bug-fixing method pairs with at least one edit action.
Third, they design a code abstraction strategy to reduce vocabulary size by only keeping frequent identifiers and literals.
Finally, they construct two datasets (\ie BFP-small and BFP-medium) and train NMT models to translate the buggy method into the corresponding correct method.
The experimental results show that NMT models are able to fix a considerable number of buggy methods in 9\%–50\% of the cases.
More importantly, this study highlights the future of NMT for APR, providing a solid empirical foundation for follow-up studies in the learning-based APR community.
}

\revise{In 2020, Ding~\etal~\cite{ding2020patching} empirically investigate to what extent program repair is like machine translation. 
They reveal that there exist essential differences between \delete{seq2seq}\revise{Seq2Seq} models and translation models in terms of task design and architectural design.
The translation model is inappropriate for program repair due to the lack of vocabulary and immediate context.
Besides, the translation model usually keeps up most tokens from the bug code while replacing only a small number, which is not ideal for program repair.
Finally, they implement an edit-based model by adapting the \delete{seq2seq}\revise{Seq2Seq} models used for translation to generate edits rather than raw tokens, which leads to promising improvement.}

\revise{In 2021, with the rise of pre-trained models in the SE domain, Mashhadi~\etal~\cite{mashhadi2021applying} conduct a  preliminary to apply CodeBERT to Java simple bugs. They fine-tune and evaluate it on ManySStuBs4J datasets and find it is capable of generating patches \revise{in a short time}. Their approach gets rid of the limitation of token length and vocabulary problems, thus this model is more efficient and effective. 
This model can generate patches for different types of bugs and outperform simple Seq2Seq models in terms of the accuracy of generated patches.}
\revise{Similarly, Kolak~\etal~\cite{kolak2022patch} propose to apply large pre-trained language models to generate patches for one-line bugs in Java and Python programs. 
They consider pre-trained models with a wide range of sizes (\eg GPT-2 with 160M, 0.4B, and 2.7B parameters and CodeX 12B parameters) for evaluation and comparison. 
After evaluating these models on the QuixBugs benchmark, they discover that larger language models tend to generate more predictable patches and thus are more promising in guiding patch selection in APR work.}

In 2022, focusing on code representation, Namavar~\etal~\cite{namavar2022controlled} conduct a systematic study to understand the effect of different code representation ways on learning-based APR performance.
In particular, they implement REPTORY as a tool for controlled experiments to assess the accuracy of different code representations (\eg AST variants) and the functionality of four different embeddings (\eg Word2Vec). 
They conduct 21 experiments with different models to evaluate their automatic patchability and perceived usefulness as well as accuracy. 
The results reveal that mixed code representation with Golve embedding outperforms other settings.
Moreover, they find that bug type affects the accuracy of different code representations.

\revise{At the same time,} Xia~\etal~\cite{xia2022practical} present the first extensive evaluation of large programming language models (PLMs) for program repair. 
They select nine state-of-art pre-trained PLMs with different types (\ie infilling and generative models) and parameter sizes (\ie ranging from 125M to 20B).
They design three different repair settings for PLMs (\ie complete function generation, correct code infilling, and single line generation).
They then conduct experiments on 5 datasets across 3 different languages to compare different PLMs in the number of bugs fixed, \revise{generation speed and compilation rate.}
They also compare the performance of PLMs against \delete{state-of-the-art} \revise{existing} APR techniques \revise{(\eg Recoder~\cite{zhu2021syntax} and CURE~\cite{jiang2021cure})} and results demonstrate the promising future of directly adopting PLMs for APR.

\input{APR/APCA/2022-arxiv-Attention}

\revise{
Different from previous empirical studies~\cite{xia2022practical,tufano2019empirical} focusing on semantic bugs triggered by test cases, Kim~\etal~\cite{kim2022empirical} conduct an empirical study to investigate the performance of existing learning-based APR techniques in fixing defects detected by a static analysis tool.
They employ the pre-trained TFix~\cite{berabi2021tfix} model as the representative APR technique to fix defects from industrial Samsung Kotlin projects.
The experimental results demonstrate the original TFix model can fix 94 out of 1,961 defects.
They also find that a fine-tuned TFix model using the defect-fixing dataset can fix 289 more defects than the original TFix model.
Besides, the TFix model with additional transfers performed using the bug-fixing dataset fixes 211 more defects than the model transferred using only defect-fixing knowledge.
More importantly, as the first work to apply TFix to an industrial software project, this empirical study demonstrates the potential of transfer learning when applying existing learning-based APR techniques to industrial software.}

\input{APR/vul/2022-DSN-W-Repairing}

\summary{\revise{
As the APR research community embraces an influx of learning-based APR approaches, there is a parallel rise in empirical studies aimed at scrutinizing the progression and subtleties of these techniques.
These empirical studies explore the actual performance of existing approaches from different aspects, such as the impact of code representation, the ability of pre-trained models, and the potential in repairing vulnerabilities.
However, considering that there exist different repair phases, and each process can introduce various specific techniques, the community urgently needs more and deeper empirical studies to illuminate the landscape of learning-based APR.
For example, future work can empirically explore whether mature dynamic program execution techniques from other domains (\eg mutation testing and fuzzing) can be used to accelerate the patch validation, detailed in Section~\ref{sec:guide}.
}}

\section{\revise{Application and }Discussion}
\label{sec:dis}
\delete{In this section, we will discuss several prevalent applications of learning-based repair and list some papers for reference.}
\revise{In this section, we will discuss and summarize several crucial aspects of the learning-based APR community.}

\subsection{Industrial Deployment}

As a promising field, APR has been extensively studied in academia and even has drawn growing attention from industry~\cite{bader2019getafix}.
For example, Marginean~\etal~\cite{marginean2019sapfix} present \textsc{SapFix}, the first end-to-end deployment of industrial APR in Meta.
\textsc{SapFix} is implemented in a continuous integration environment and deployed into six production systems with tens of millions of code lines.
Similar industrial practices can also be found in other companies, such as Fujitsu~\cite{saha2017elixir}, Bloomberg~\cite{kirbas2021introduction} and Alibaba~\cite{zhang2020precfix}.
In addition to the above-mentioned traditional deployment, the industry recently explored the feasibility of deploying learning-based APR tools.
For example, GitHub launches a product Copilot\footnote{\url{https://github.com/features/copilot}}, which can provide code suggestions (\eg fixing bugs) for more than a dozen programming languages.
Copilot is deployed in multiple IDEs, such as VS Code, Visual Studio, Neovim, and JetBrains.
Besides, Microsoft recently released a new tool Jigsaw\footnote{\url{https://www.microsoft.com/en-us/research/blog/jigsaw-fixes-bugs-in-machine-written-software/}} to fix bugs in machine-written software.

Now, we summarize the existing learning-based APR techniques and industrial deployment from enterprises.

\revise{As early as in 2019,} Bader~\etal~\cite{bader2019getafix} present Getafix, the first industrially-deployed automated bug-fixing tool for Java programs. To be fast enough to suggest fixes in time, this model produces a ranked list of fix candidates based entirely on past fixes and on the context in which a fix is applied. 
Besides, it leverages the hierarchical clustering technique for discovering repetitive fix patterns. 
Moreover, They apply a statistical ranking technique to enable the model to predict human-like fixes among the top few suggestions. 
An evaluation with a large dataset containing six types of common bugs and their experience of deploying Getafix within Facebook shows that the approach accurately predicts human-like fixes for various bugs, reducing the time developers have to spend on fixing recurring kinds of bugs.

\revise{In 2020,} Hellendoorn~\etal~\cite{hellendoorn2019global} from Google conduct experiments for two different \delete{models}\revise{model} architectures that leverage both local and global information. 
They propose sandwich models that apply different message-passing techniques and GREAT models that add extra information to a transformer. 
Both architectures achieve high results and outperform \delete{state-of-art tools} \revise{ both RNN and transformer architectures}, proving that a hybrid model with global information and incorporating structural bias helps improve accuracy.

\revise{In 2021,} Baudry~\etal~\cite{baudry2021software} present R-HERO, a novel software repair robot to automatically repair bugs on the single platform GitHub/Travis CI. R-HERO contains six main blocks: a) Continuous integration, b) Fault localization, c) Patch generation, d) Compilation \& Test execution, e) Overfitting prevention, and f) Pull-request creation. It receives and analyzes the events from a continuous integration (CI) system. R-HERO leverages continual learning to acquire bug-fixing strategies from the platform mentioned above. It shows that developers and bots can cooperate fruitfully to produce high-quality, reliable software systems.

\revise{Different from previous works with supervised learning,} Allamanis et al~\cite{allamanis2021self} from Microsoft propose BUGLAB to detect and repair software bugs automatically by self-supervised learning.
Similar to BIFI~\cite{yasunaga2021break}, BUGLAB employs a detector model to repair bugs and a selector model to generate buggy code snippets as the training data \revise{for} the detector.
The authors create a dataset PYPIBUGS of 2374 real-world bugs from the PyPI packages.
The results show that BUGLAB can fix a number of software bugs and detect some previously unknown bugs in open-source software.

\revise{In parallel to BUGLAB,} Tang~\etal~\cite{tang2021grammar} from Microsoft introduce a grammar-guided end-to-end approach to generate patches, which treats APR as the transformation of grammar rules.
They apply structure-aware modules and \revise{design} three different types of strategies for grammar-based inference algorithms. 
They also leverage two encoders and enhance the model with a new tree-based self-attention.
The experimental results on BFP datasets~\cite{tufano2019empirical} demonstrate that the proposed technique outperforms \delete{other state-of-art APR techniques} \revise{previous RNN-based techniques (\eg Tufano~\etal~\cite{tufano2019empirical})}.

\revise{Considering the raise of pre-trained models,} Drain~\etal~\cite{drain2021generating} from Microsoft introduce DeepDebug, a span-masking pre-trained encoder decoder transformer as a tool to fix Java methods. The model is pre-trained from BART which is pre-trained in English. 
They conduct three pre-training experiments to verify the feasibility of the model and test it on the Java benchmarks from Tufano~\etal~\cite{tufano2019empirical}. 
\delete{DeepDebug  many state-of-art  APR tools}
\revise{Results show that DeepDebug outperforms existing APR tools (\eg CodeBERT~\cite{feng2020codebert} and Tufano~\etal~\cite{tufano2019empirical})}, and adding syntax embeddings along with the standard positional embeddings helps improve the model.

\revise{In 2022, similar to DeepDebug, }Hu~\etal~\cite{hu2022fix} from AWS AI propose NSEdit to generate patches for Java programs. Given only the buggy code, NSEdit uses the pre-trained CodeBERT as the encoder and CodeGPT as the decoder to address the \delete{sequence-to-sequence}\revise{Seq2Seq} NMT problem. 
Moreover, it uses a pointer network to select content-based edit locations. 
They apply beam search and design a novel technique to fine-tune the reranker to re-rank the top-k patches for the buggy code. 
The results on BFP benchmarks~\cite{tufano2019empirical} indicate that NSEdit outperforms \delete{state-of-art APR tools} \revise{CodeBERT~\cite{feng2020codebert}} and \revise{the ablation study} demonstrates the effectiveness of each component of the model.

\revise{Meanwhile,} Wang~\etal~\cite{wang2022leveraging} from Ping An Technology propose CPR, short for causal program repair, as a tool to utilize data augmentation strategy for input perturbations. This model can generate patches for Java, Python, JavaScript, and C based on causally related input-output tokens. Besides, it can offer explanations by transforming code into explainable graphs on various Seq2Seq models in APR. They conduct experiments on four programming languages and prove that APR models can be utilized as causal inference tools.

\summary{\revise{
The APR domain has witnessed an unprecedented surge in industrial adoption.
With giants like Meta, Fujitsu, Bloomberg, and Alibaba exploring and harnessing its potential, learning-based APR has undoubtedly established its foothold in real-world applications.
Emphasis has notably shifted to learning-based APR tools, as exhibited by GitHub's Copilot and Microsoft's Jigsaw, which underscore the blend of machine learning with traditional programming paradigms.
Noteworthy contributions emerge from global tech titans including Microsoft, Google, and AWS AI. 
From tools like Getafix, R-HERO, and BUGLAB, which emphasize speed, collaboration, and self-supervised learning respectively, to models like DeepDebug and NSEdit that push the envelope of program repair using state-of-the-art machine learning architectures, industry-affiliated research has been at the forefront.
As the APR community moves forward, the collaboration between academia and industry in the APR domain is poised to shape the next generation of repair tools and methodologies.
The trend demonstrates the desire to harness advanced DL techniques to address recurrent software bugs, thereby alleviating the developers' workload in the industry.
}}

\subsection{DL for Traditional APR}
\label{sec_dl4Tapr}

\delete{These approaches attempt to boost traditional APR techniques by utilizing deep learning or machine learning.}
\revise{
In addition to the increasing number of end-to-end learning-based APR techniques, there has been growing interest in leveraging these learning technologies to improve and refine the capabilities of traditional APR techniques.
These studies usually treat machine learning as a component to address the inherent limitation in the original APR workflow.
Table~\ref{tab:dl4apr} presents existing studies that attempt to boost traditional APR techniques by utilizing deep learning or machine learning.
The first and second columns list the summarized studies and the years.
The third column denotes the traditional APR techniques targeted by these summarized studies.
The remaining two columns list the targeted languages and a brief description.}

\input{tab/tab_dl4tapr}

\revise{As early as 2016,} Long~\etal~\cite{long2016automatic} propose Prophet, a patch-generation system for repairing \delete{defects}\revise{bugs}.
It uses dynamic analysis on the given test suite to get the program points for the patch to modify.
Then, the SPR~\cite{long2015staged} is used to generate search space.
With a trained probabilistic model, Prophet ranks the candidate patches, which are validated by executing the test suites.
They collect eight projects from \delete{Github}\revise{GitHub} and get 777 patches to train their model and test it on a benchmark~\cite{Fan2016benchmark}.
The result shows that Prophet can generate patches correctly with the learned knowledge compared with previous patch generation systems.
\revise{From the perspective of community development, while Prophet may not be an end-to-end NMT-based patch generation approach like CoCoNut~\cite{lutellier2020coconut}, its pioneering integration of machine learning into the repair process offers invaluable insights for subsequent research endeavors.}

\revise{In 2017, }Xiong~\etal~\cite{xiong2017precise} introduce ACS, which aims to generate precise conditions at faulty statements.
During the condition synthesis process, ACS selects what variables should be used in the conditional expression and decides what predicate should be performed on the variables.
The predicates are mined from existing projects, and sorted based on their frequencies in contexts similar to the target condition.
The results on Defects4J show that ACS is the first APR approach that achieves a precision higher than 70\% (the precision of previous approaches is below 40\%).
ACS employs a learning component to infer which predicates should be used with the current variable.
Although the learning component (just counting the frequencies in a corpus of source code) is very simple, it is still learning.
Thus, we regard ACS as one of the earliest learning-based APR techniques.

\revise{
\revise{At the same time,} Long~\etal~\cite{long2017automatic} present a new system, Genesis, that processes human patches to automatically infer code transforms for automatic patch generation.
They first extract transforms from the training set to obtain a pair containing a program before a change and a program after a change.
For each transformation, they create a template that defines the AST changes.
They then collect templates to create AST template forests which contain template variables to match any appropriate AST subtrees.
Given a set of training pairs, Genesis will select from the inference search space to obtain potential transforms.
They design an algorithm to reach a trade-off between search space coverage and tractability.
Finally, from these transforms they obtain a set of candidate patches.
They then evaluate Genesis on a dataset collected from GitHub Java programs covering null pointer (NP), out-of-bounds (OOB), and class cast (CC) bugs.
Results show that Genesis outperforms another patch generation technique PAR~\cite{kim2013automatic} that leverages manually defined templates.
}

\input{APR/2019-SANER-DeepRepair}

In 2022, Chen~\etal~\cite{chen2022program} propose a \delete{novel} search-based technique called LIANA, which is based on a designed learning-to-rank prioritization mode. 
It is based on the idea of repeatedly updating a statistical model online based on the intermediate validation results of an ongoing program repair process. The model is first trained offline and updated repeatedly after the generating progress starts. The most up-to-date model is used to generate fixes and prioritize those that are more likely to include the correct ingredients.

\revise{To improve the template-based APR,} Wang~\etal~\cite{meng2022improving} propose TRANSFER, a fault localization and program repair approach with deep semantic features and transferred knowledge which is obtained by a combination of spectrum-based and mutation-based localization techniques. 
They build a fault localization and program repair dataset respectively and employ existing fix templates designed by TBar. 
They also design 11 binary classifications to identify whether one of the 11 bug types they define exists in a statement and a multi-classification to determine which fix template this statement should apply. 
The binary classification, consisting of one embedding layer, one RNN layer, one max pooling layer, and one dense layer, is fed with spectrum-based, mutation-based, and semantic features and outputs the probability of containing specific bugs. 
Although this approach is only tested on Java, it is proven to outperform many state-of-art approaches.

\revise{Similarly, to improve the search-based APR,} Li~\etal~\cite{li2022improving} design a novel framework called ARJANMT to leverage both redundancy assumption and Seq2Seq learning of correct patches to generate fixes for Java methods using NSGA-II algorithm. This framework combines both ARJA and SequenceR into a unified framework. After evaluating ARJANMT on two Java benchmarks, results show that it benefits from search-based and NMT-based techniques and outperforms \delete{other state-of-art} \revise{existing APR} techniques \revise{(\eg CoCoNut~\cite{lutellier2020coconut}, DLFix~\cite{li2020dlfix} and CURE~\cite{jiang2021cure})}.

\revise{To address multiple bugs,} Valueian~\etal~\cite{valueian2022siturepair} propose SituRepair for repairing multiple bugs in C programs based on pre-defined repair patterns. 
It applies a machine learning model to predict the buggy type and localization of the buggy code and then repairs them with situational modifications accordingly. 
SituRepair is evaluated on a C benchmark Code4Bench and it successfully repairs 3,848 multiple-fault programs, \revise{outperforming Genprog~\cite{le2012genprog}}.

\summary{\revise{
Although a mass of research effort has been devoted to end-to-end patch generation, the literature has also seen some orthogonal works utilizing DL to enhance traditional APR techniques.
Different from most learning-based APR techniques that design an NMT-based patch generation model from scratch, these techniques can leverage mature traditional APR techniques and employ DL to improve specific components, such as the selection of repair templates~\cite{meng2022improving}.
Future work can be conducted to address certain limitations of traditional APR techniques, such as the donor code retrieval issue~\cite{liu2019tbar} using pre-trained models.
}}

\subsection{Open \delete{science}\revise{Science}}
\input{tab/tab_tool}

Recent years have witnessed an increasing \delete{use} \revise{application} of DL in traditional SE problems and tasks. 
In particular, software bug is a growing quality concern for modern software, and accordingly, APR has become an actively studied topic in the SE community. 
According to our survey, various learning-based APR techniques have been introduced in the last five years (discussed in Section \ref{sec:methodology}). 
DL brings a new repair paradigm (\ie training and repairing) for the APR problem with promising results.
However, due to the nature of DL, learning-based APR techniques face some concerns in reproducibility, which is quite different from \delete{transitional}\revise{traditional} APR techniques.
For example, it may require a large number of machine resources for researchers to reproduce the NMT model's work.
The cost is unaffordable for most researchers from academia.
Besides, there exists randomness in the neural network training process, which hinders the reproduction results.

Such challenges posed by DL motivate us to further understand the potential issues with open science in the learning-based APR area, so as to advance existing techniques by taking advantage of the general merits of open science. 
Open science advocates that researchers make their artifacts (\eg raw data, dataset, scripts, related models, or any results produced in their work) available to all levels of researchers~\cite{masuzzo2017you}, so knowledge can be shared without boundaries~\cite{nong2022open}.
While a mass of DL techniques are proposed to fix software bugs automatically, more support is needed to investigate the critical open science problem.
In particular, we investigate to what extent the collected papers make their artifacts publicly available and in what way they provide the relevant information.

Table \ref{tab:tool} shows the tool availability results of the investigated papers.
For each paper we collect, we check whether an accessible link for its tool or data is provided in the main text or footnotes of the paper.
We only present the studies that provide the link of publicly available data or tools due to limited space, listed in \textit{the first column}.
We then investigate the following five dimensions in characterizing the availability of each paper:
    
\begin{itemize}
    \item \textbf{Hosting Site.} This information indicates which hosting site the available artifact is uploaded to for public access (\eg GitHub or Google), if the artifact link is presented in the paper.
    The detailed information is listed in \textit{the third column}.

    \item \textbf{Link Accessibility.}
    This information indicates whether the provided link is accessible, such that we can download the artifacts.
    The detailed information is listed in \textit{the fourth column}.

    \item \textbf{Source Code Available \revise{(SA)}.}
    This information indicates whether the source code (\eg training and evaluation scripts) is available in the artifacts.
    The detailed information is listed in \textit{the fifth column}.
    
    \item \textbf{Dataset Available \revise{(DA)}.}
    This information indicates whether the dataset (\eg raw data and training data) is available in the artifacts.
    The detailed information is listed in \textit{the sixth column}.
    
    \item \textbf{Trained Model Available \revise{(TA)}.}
    This information indicates whether the trained model (\eg raw data and training data) is available in the artifacts.
    The detailed information is listed in \textit{the seventh column}.
    
\end{itemize}

We also list the programming languages targeted by the tools in \textit{the second column} and list the accessible \delete{url}\revise{URL} links in \textit{the last column}.
After carefully checking the collected papers, we find that only a few of the papers have made their source code available to the public.
For convenient public access, a majority of papers upload their works to GitHub.
The possible reason is that GitHub is the most popular platform to host open-source code publicly.
Meanwhile, we find that several papers fail to provide the source code, dataset, or already trained model~\cite{tang2021grammar,yuan2022circle}.
The possible reasons may be (1) 
the artifacts need to be refactored or reorganized for public availableness;
(2) the artifacts are used for further studies;
and (3) the artifacts are lost due to some accidents.
We also find while the artifacts are available, some studies cannot be reproduced because
(1) the missing of default hyperparameters\footnote{\url{https://github.com/lin-tan/CoCoNut-Artifact/issues/11}};
(2) the complexity of environment settings for training\footnote{\url{https://github.com/pkuzqh/Recoder/issues/11}};
and (3) the \revise{insufficiency} of documentation to reproduce the experiments\footnote{\url{https://github.com/ICSE-2019-AUTOFIX/ICSE-2019-AUTOFIX/issues/5}}.

\summary{\revise{
Overall, compared with traditional APR, the need for high-quality artifacts in learning-based APR is even more vital for replication and future research.
On the one hand, the learning-based APR usually involves abundant training time and expensive equipment (\eg GPUs) to train a repair model, and thus it is much harder to reproduce existing works.
On the other hand, some learning-based APR models require complex environment settings (\eg the best hyperparameters and the random seed) and some authors may fail to provide high-quality code.
In contrast, traditional APR results are typically more straightforward and deterministic to reproduce when provided with open-source code and data.
Therefore, we hope that researchers in the learning-based APR community can provide high-quality open-source code and detailed instructions to construct a unified repair framework for convenient reproduction.
}}

\subsection{The Latest Advancements}
\newrevise{
While the scope of this survey encompasses literature up to Nov 2022, it is worth noting that there have been significant advancements in the field of learning-based APR during 2023.
For example, notable papers presented at prominent conferences such as ICSE and ASE have introduced innovative learning-based APR approaches, especially many of which leverage pre-trained models to achieve remarkable results.
Although a comprehensive review and analysis of these recent works are beyond the scope of this paper, we acknowledge their contributions to the field and recognize them as pivotal developments that will shape future research in APR.
In the following, we summarize some recent studies for a timely understanding of the latest advancements.

In line with Section~\ref{sec:preprocessing}, code context provides necessary information for repair models to generate correct patches and plays a vital role in the learning-based APR workflow.
However, existing approaches mainly extract code in close proximity to the buggy
statement within the enclosing file, class, or method, without any analysis to find actual relations with the bug.
Sintaha~\etal~\cite{sintaha2023katana} propose
a learning-based APR approach Katana, which employs a program slicing-based approach to analyze code context in program repair.
Particularly, Katana designs a dual slicing strategy to analyze statements that have a control or data dependency on the buggy statement.

In line with Section~\ref{sec:patch_generation},  Zhu~\etal~\cite{zhu2023tare} further propose Tare built upon their previous graph-based APR approach Recoder, a type-aware model for program repair to learn the typing rules.
Compared with Recoder, Tare replaces the grammar in Recoder with a T-Grammar that integrates the type information into a standard grammar, and replaces the neural components of Recoder encoding ASTs with neural components encoding T-Graphs, which is a heterogeneous graph with attributes.
Besides, Jiang~\etal~\cite{jiang2023knod} propose KNOD, a learning-based APR approach based on a three-stage tree decoder and a domain-rule distillation.
The first tree decoder directly generates ASTs of patched code according to the inherent tree structure and 
The second domain-rule distillation leverages syntactic and semantic rules and teacher-student distributions to explicitly inject the domain knowledge into the decoding procedure during both the training and inference phases.

In line with Section~\ref{sec:patch_validation}, Xiao~\etal~\cite{xiaoexpressapr,xiao2023accelerating} systematically investigate whether existing mutation testing acceleration techniques are suitable for general-purpose patch validation.
They then introduce ExpressAPR, a patch validation framework by designing two adaption strategies, \ie execution scheduling and interception-based instrumentation.
The experimental results on four previous APR approaches (including the learning-based one Rcoder) demonstrate that ExpressAPR is able to reduce patch validation time significantly. 

In line with Section~\ref{sec:sota}, some domain approaches are proposed to address the repair problem for various bug types.
For example, So~\etal~\cite{so2023smartfix} propose SmartFix, a learning-based technique for repairing vulnerable smart contracts.
SmartFix employs statistical models to intelligently guide the repair procedure, so as to prioritize candidate patches that are helpful in finding desired safe contracts.
At the same time, Fan~\etal~\cite{fan2023automated} systematically investigate whether existing APR techniques (\eg Recoder~\cite{zhu2021syntax}) can fix the incorrect solutions produced by pre-trained models in LeetCode contests. 
Besides, First~\etal~\cite{first2023baldur} propose Baldur, an automated whole-proof generation and repair approach on top of a large pre-trained model.
Baldur first generates whole formal proofs by a proof generation model trained on natural language text and code and fine-tuned on proofs.
Baldur then combines this proof generation model with a fine-tuned repair model to repair incorrectly generated proofs, further increasing proving power.

In line with Section~\ref{sec:pre-trained}, there exist some recent approaches proposed to explore how to transfer domain bug-fixing knowledge into the pre-trained model-based patch generation process.
The first example is RAP-Gen~\cite{wang2023rap}, a retrieval-augmented program repair approach on top of a pre-trained CodeT5 model.
RAP-Gen retrieves a relevant bug-fixing pair from an external codebase to augment the buggy input for the CodeT5 patch generator. 
The second example is FitRepair~\cite{xia2023revisiting}, a CodeT5-based APR approach that incorporates domain-specific knowledge with the insights of the plastic surgery hypothesis.
FitRepair designs two domain-specific fine-tuning strategies and one prompting strategy to leverage the hypothesis from the buggy projects.
The third example is Repilot~\cite{wei2023copiloting}, which helps pre-trained models generate more valid patches through a completion engine.
Repilot employs the interaction between a pre-trained model and a completion engine to generate candidate patches by first pruning away infeasible tokens suggested by the pre-trained model and then completing the token based on the suggestions provided by the completion engine.

In line with Section~\ref{sec:empirical}, some empirical studies further explore the actual performance of learning-based APR from different aspects.
For example, Jiang~\etal~\cite{jiang2023impact} empirically evaluate the fixing capabilities of pre-trained models with and without fine-tuning for
the APR task, involving ten pre-trained models and four benchmarks.
Zhang~\etal~\cite{zhang2023pre} conduct an extensive empirical study to investigate how pre-trained models are applied to vulnerability repair in the workflow (\ie data pre-processing, model training and repair inference) and further propose an enhanced approach with bug-fixing transfer learning, involving more than 100 variants of fine-tuned models.
Similarly, Wu~\etal~\cite{wu2023effective} conduct an extensive study to evaluate the fixing capabilities of five pre-trained models and four learning-based APR approaches on real-world Java vulnerabilities.

In line with Section~\ref{sec_dl4Tapr}, there exist some approaches proposed to combine traditional APR and recent learning-based APR.
For example, Zhang~\etal~\cite{zhang2023gamma} propose GAMMA,
a template-based program repair approach on top of the advance of fix patterns from traditional template-based APR  and mask prediction from pre-trained models.
Similarly, Meng~\etal~\cite{meng2023template} propose TENURE, a novel template-and-learning-based program repair approach by combining the template-based and NMT-based methods.
Importantly, both GAMMA and TENURE preliminarily demonstrate the prospect of combining the advances of traditional APR and DL models.
At the same time, Parasaram~\etal~\cite{parasaram2023rete} propose RETE, which aims to navigate the search space of patches by learning project-independent information about the program namespace.
RETE first employs repair patterns to generate candidate patches and prioritize patches by learning rich semantic information about the project namespace. 

\summary{\newrevise{Overall, these latest research findings further demonstrate the timeliness and comprehensiveness of our survey.
Importantly, the most apparent trend is the increasing use of pre-trained models, including enhanced pre-trained model-based approaches, empirical studies on diverse bug types, and the combination with traditional APR.
Besides, there exist some studies focusing on optimizing other components of the repair process, such as code context and patch validation acceleration.}}

}
\section{Implication and \delete{Discussion}\revise{Guidelines}}
\label{sec:guide}

\input{sec/sec_guidelines}
\input{sec/sec_guidelines_old}

\section{Conclusion}
\label{sec:con}

APR techniques address the long-standing \delete{the }challenge of fixing software bugs automatically, and \delete{alleviates}\revise{alleviate} manual debugging effort significantly, which promotes software testing, validation, and debugging practices.
In the last couple of years, learning-based APR techniques have achieved promising results, demonstrating the substantial potential of using DL techniques for APR.

In this paper, we provide a comprehensive survey of existing learning-based APR techniques.
We describe the typical learning-based repair framework, involving \text{fault localization}, \text{data pre-processing}, \text{patch generation}, \text{patch ranking}, \text{validation} and \text{correctness} components.
We summarize how existing learning-based techniques design strategies for these crucial components.
We discuss the metrics, datasets and empirical studies in the learning-based APR community. 
Finally, we point out several challenges (such as overfitting issues) and provide possible directions for future study.

\section*{Acknowledgments}
\revise{The authors would like to thank the anonymous reviewers for their insightful comments.}
This work is supported partially by the National Natural Science Foundation of China (61932012, 62141215, 62372228), CCF-Huawei Populus Grove Fund (CCF-HuaweiSE202304, CCF-HuaweiSY202306), and Science, Technology and Innovation Commission of Shenzhen Municipality (CJGJZD20200617103001003).

\bibliographystyle{ACM-Reference-Format}
\bibliography{reference}

\end{document}

%% file: sec/sec_SOTA_seq.tex
\delete{\textit{State-of-the-arts.}
In the following, we discuss these individual sequence-based patch-generation techniques in more detail.}

\delete{
Tufano~\etal~\cite{tufano2019learning} design an NMT model to generate the same patches applied by developers under a narrow context. 
They reduce the vocabulary size by mapping the method of the code to a specific ID and feed the model with pairs of methods before and after the patch. 
The model can replicate up to 36\% of the buggy code. 
Moreover, it can be applied to refactoring and other code relating activities.
}

\delete{
As early as 2019, Tufano~\etal~\cite{tufano2019learning} conduct this first attempt to investigate the ability of NMT models to learn code changes during pull requests.
They first mine pull requests from three large Gerrit repositories and extract the method pairs before and after the pull requests, where each pair serves as an example of a meaningful change.
They then map the identifiers and literals in the source code to specific IDs (\ie code abstraction) to reduce the vocabulary size.
Finally, they train NMT models to translate the method before the pull request into the one after the pull request, to emulate the actual code change.
The experimental results show that NMT models can generate the same patches for 36\% pull requests
Overall, this study demonstrates the potential of NMT models in learning a wide variety of meaningful code changes, especially refactorings and bug-fixing activities.
Further, Tufano~\etal~\cite{tufano2019empirical} perform an empirical study to investigate the potential of NMT models in generating bug-fixing patches in the wild, which is discussed in Section~\ref{sec:empirical}.
}

\delete{
At the same time, Chen~\etal~\cite{chen2019sequencer} propose SequenceR, an end-to-end approach based on sequence-to-sequence learning. 
They combine LSTM encoder-decoder architecture with a copy mechanism to address the problem of a large vocabulary. First, they apply state-of-the-art fault localization techniques to identify the buggy method and the suspicious buggy lines. 
Then, they perform a novel buggy context abstraction process that intelligently organizes the fault localization data into a suitable representation for the deep learning model. Finally, SequenceR generates multiple patches for the buggy code.
Although their approach can only be applied to single-line buggy code, this model outperforms the APR tool of Tufano~\etal~on Defects4J benchmarks. Moreover, they prove that the copy mechanism can improve the accuracy of generated patches.}

\delete{
Current works aim at exploring fixes in a limited search space, which may not contain the correct patches.
\revise{Hata~\etal~\cite{hata2018learning}} follow the recently NMT-based approach and use an encoder-decoder model Ratchet with multi-layer attention to fix bugs.
They perform an empirical study with five large software projects. Moreover, they collect a fine-grained dataset from these projects and try to ignore noisy data. They train and evaluate Ratchet on this dataset. Results show that Ratchet performs at least as well as pattern-based APR tools.
Besides, Ratchet’s output was considered helpful in fixing the bugs on many occasions, even if the fix was not 100\% correct.}

\delete{
Lutellier~\etal~\cite{lutellier2020coconut} propose CoCoNut, a \delete{novel} generate\&validate technique with a new context-aware NMT architecture that separately inputs the buggy line and method context. 
They further combine CNN (FConv architecture) with the NMT model to improve the accuracy of generated patches. 
After collecting a large dataset from four programming languages and training the model on it, CoCoNut is then evaluated on six benchmarks (also from four programming languages), \ie Defects4J of Java, QuixBugs of Java, CodeFlaws of C, ManyBugs of C, QuixBugs of Python and BugAID of JS.
It turns out that CoCoNut outperforms previous APR tools, and is capable of fixing 300 more bugs other APR tools fail to . Moreover, CoCoNut proves that FConv architecture can outperform LSTM.}

\delete{\revise{
However, Tufano~\etal~\cite{tufano2019empirical} and SequenceR~\cite{chen2019sequencer} represent both the buggy line and its context as one input for NMT models, making it difficult to extract long-term relations between code tokens.
In 2020, Lutellier~\etal~\cite{lutellier2020coconut} propose CoCoNut, a context-aware NMT approach that separately inputs the buggy line and method context.
In particular, CoCoNut applies CNN (\ie FConv architecture) in the context-aware NMT architecture, which is able to better extract hierarchical features of source code compared with LSTM used in Tufano~\etal~\cite{tufano2019empirical} and SequenceR~\cite{chen2019sequencer}.
Besides, CoCoNuT trains multiple NMT models to capture the diversity of bug fixes with ensemble learning. 
CoCoNut is evaluated on six well-known benchmarks across four programming languages, \ie Defects4J of Java, QuixBugs of Java, CodeFlaws of C, ManyBugs of C, QuixBugs of Python and BugAID of JS.
The experimental results show that CoCoNut is capable of fixing 509 bugs on the six benchmarks, 309 of which have not been fixed by previous APR tools, such as DLFix, Prophet and TBar.
}}

\delete{
Further, Jiang~\etal~\cite{jiang2021cure} propose CURE, an \delete{novel} NMT-based program repair technique to fix Java bugs. 
They pre-train a programming language model on a large corpus and combine it with NMT architecture to learn code syntax and fix patterns. 
They also apply a code-aware search strategy and a new subword tokenization technique to improve the accuracy of generated patches. This model outperforms SequenceR and CoCoNut \delete{APR tools} on Defects4J and QuixBugs benchmarks under different beam search sizes.}

\delete{In 2021, on top of CoCoNut, Jiang~\etal~\cite{jiang2021cure} propose CURE, an NMT-based APR technique to fix Java bugs.
Compared with CoCoNut, the novelty of CURE mainly coms from three aspects.
First, to better learn developer-like source code, CURE pre-trains a programming language model on a large corpus and combines it with the CoCoNut context-aware architecture.
Second, CURE designs a code-aware beam search strategy to avoid uncompilable patches during patch generation.
Third, to address the OOV problem, CURE introduces a new sub-word tokenization technique to tokenize compound and rare words.
The experimental results demonstrate that CURE is able to fix 57 Defects4J bugs and 26 QuixBugs bugs, outperforming existing learning-based APR approaches under different beam search sizes, such as SequenceR and CoCoNut.
}

\delete{
Yao~\etal~\cite{yao2022bug} propose a new transformer-based APR technique, Bug-Transformer, to fix buggy code snippets. 
It applies a novel token pair encoding (TPE) approach to reduce vocabulary size by compressing code structure while preserving semantic information. 
Besides, they apply a novel rename mechanism to preserve semantic features for code abstraction.
Bug-Transformer leverages the transformer architecture and is fine-tuned for \delete{learning}\revise{program repair} tasks. 
It is then evaluated on Java benchmarks and outperforms other baseline models such as \delete{Bug2Fix}~\revise{Tufano~\etal~\cite{tufano2019empirical}}.
}

\delete{
Unlike Tufano~\etal~\cite{tufano2019empirical} without considering semantic and lexical scope information of code tokens, in 2022, Yao~\etal~\cite{yao2022bug} propose Bug-Transformer, a transformer-based APR technique to fix buggy code snippets.
It is equipped with a token pair encoding (TPE) algorithm and a rename mechanism to preserve crucial information.
First, Bug-Transformer designs a TPE algorithm to reduce vocabulary size by compressing code structure while preserving structural and semantic information.
Second, Bug-Transformer employs a rename mechanism to abstract code tokens (\ie identifiers and literals) with consideration of their semantic and lexical scope knowledge.
Third, Bug-Transformer trains a transformer-based model to learn the structural and semantic information of code snippets and predicts patches automatically.
The experimental results on BPF~\cite{tufano2019empirical} datasets show that Bug-Transformer outperforms baseline models, \eg Tufano~\etal~\cite{tufano2019empirical}.
}

\delete{
Lutellier~\etal~\cite{lutellier2019encore} propose ENCORE, a new end-to-end APR technique that leverages the NMT model to generate bug fixes for Java programs. Evaluating ENCORE on two Java benchmarks proves that it can fix diverse bugs, and further experiments on Python, C++, and JS benchmarks proves that it can handle bugs in different programming languages. They also present attention maps to explain why certain fixes are generated or not by ENCORE.}

\delete{
Ye~\etal~\cite{ye2022neural} introduce RewardRepair as a neural program repair approach for fixing bugs in Java code based on transformer. 
They apply a novel training strategy and feed the model with compiling and testing execution information to improve the quality of generated patches. 
This model is then evaluated on four benchmarks, Defects4J v1 and v2, Bugs.jar, and QuixBugs. 
Results show that this model has lower cross-entropy than previous APR tools. 
Besides, RewardRepair outperforms SequenceR, CoCoNut and CURE in terms of top-k accuracy on Defects4J benchmarks.}

\delete{
Existing learning-based APR techniques are usually limited by the generation of lots of low-quality (\eg non-compilable) patches, due to the employed static loss function based on token similarity.
In 2022, Ye~\etal~\cite{ye2022neural} introduce RewardRepair based on a mixed loss function that considers program compilation and test execution information.
In particular, RewardRepair defines a discriminator to discriminate good patches from low-quality ones based on dynamic execution feedback, rather than static token similarity between the generated patch and the human-written ground truth patch.
The discriminator computes a reward value to gauge the patch quality, and this value is subsequently utilized to update the weights of the patch generation model during the backpropagation process.
A higher reward indicates a higher quality \revise{of} generated patch that is compilable and passes the test cases, while a lower reward suggests potentially unsatisfactory patch quality, such as a non-compilable patch.
Thanks to the compilation and test execution results during training, RewardRepair is able to fix 207 bugs on four benchmarks, i.e., Defects4J-v1.2, Defects4J-v2.0, Bugs.jar and QuixBugs, 121 of which are not repaired by previous approaches, \eg DLFix, CoCoNut and CURE.
More importantly,  RewardRepair achieves a compilable rate of up to 45.3\% among Top-30 candidate patches, an improvement over the 39\% by CURE, demonstrating the potential to generate high-quality patches.
}

\delete{Previous neural program repair approaches focus on supervised training and lack project-specific knowledge.
Ye~\etal~\cite{ye2022selfapr} propose, SelfAPR, a self-supervised training approach with test execution diagnostics based on a transformer neural network. 
SelfAPR consists of two components, training sample generator and neural network optimization. 
The first part generates perturbed programs with a perturbing model and tests it to capture compile errors and execute failures information. 
The second part is fed with the previous information and outputs $n$ best patches with beam search. 
SelfAPR is capable of repairing ten bugs that are never repaired before by the supervised neural repair models. Moreover, evaluation results the effectiveness of self-supervised training and its components.}

\delete{
Besides, previous learning-based APR approaches are dominantly founded on supervised training with massive open-source code repositories, resulting in a lack of project-specific knowledge.
In parallel with RewardRepair, Ye~\etal~\cite{ye2022selfapr} also propose, SelfAPR, a self-supervised training approach with test execution diagnostics based on a transformer neural network. 
SelfAPR consists of two well-designed components, \ie training sample generator and neural network optimization.
The first part generates perturbed programs with a perturbing model and tests it to capture compile errors and execute failures information. 
The second part is fed with the previous information and outputs $n$ best candidate patches with beam search. 
The experimental results show that  SelfAPR is capable of repairing 65bugs from Defects4J-v.12 and 45 bugs from Defects4J-v2.0, 10 of which have never been repaired by the previous supervised neural repair models, such as CoCoNut~\cite{lutellier2020coconut} and CURE~\cite{jiang2021cure}. 
More importantly, SelfAPR highlights the potential and power of self-supervised training and project-specific knowledge in the learning-based APR community.
}

\delete{Rahman~\etal~\cite{rahman2021bidirectional} present a bidirectional LSTM model for code evaluation and repair. 
They first train the model as a Seq2Seq model with abundant source code collected from Aizu Online Judge (AOJ) system. Then they fine-tune it to detect errors and provide suggestions for code repair.
The model is evaluated on \delete{Aizu Online Judge (AOJ)} \revise{AOJ} system.
They focus on two types of solution codes: greatest common divisor (GCD) and insertion sort (IS). Results show that this model outperforms previous RNN and LSTM models such as DeepFix. It also proves to be useful for novice programmers and accelerates the code evaluation process.}

\delete{Huang~\etal~\cite{huang2021application} propose an enhanced transformer-based APR technique by introducing a general pyramid encoder, 
which is added in between layers of regular multi-layer encoders. 
For the purpose of testing the generality of the pyramid encoder, they combined this encoder with different attention mechanisms. 
They conduct experiments on Juliet Test Suite for C/C++ and Java to evaluate \delete{seq2seq}\revise{Seq2Seq} models. 
Results show that \delete{seq2seq}\revise{Seq2Seq} models can be well applied in providing suggestions to potential errors and have a decent correct rate in code auto-correction. Besides, their results on transfer learning point out a way of processing this small dataset using the pre-trained model as an encoder, which boosts the performance by a large amount.}

%% file: sec/sec_SOTA_tree.tex
\delete{
\textit{State-of-the-arts.}
In the following, we discuss these individual tree-based patch-generation techniques in more detail.}

\delete{Chakraborty~\etal~\cite{Chakraborty2020codit} propose a tree-based APR technique CODIT to learn code changes from the wild and generate patches for software bugs.
CODIT transforms the correct (or buggy) code snippet into the parse tree and generates the deleted (or added) subtree.
CODIT then predicts the structural changes using a tree-based translation model among the subtrees and employs token names to concrete the structure using a token generation model.
The former tree-based model takes the previous code tree structure and generates a new tree with the maximum likelihood,
while the latter token generation model takes tokens and types of tokens in the code and generates new tokens with the help of LSTM.
The authors conduct a real-world bug-fixing dataset Code-Change-Data from 48 open-source projects and employ Pull-Request-Data~\cite{chen2019sequencer} and Defects4J~\cite{just2014defects4j}.
The results on these three datasets illustrate CODIT outperforms existing \delete{seq2seq}\revise{Seq2Seq} models, highlighting the potential of the tree-based models in APR.}

\delete{
As early as 2020, Chakraborty~\etal~\cite{Chakraborty2020codit} propose a tree-based neural network, CODIT to learn code changes by encoding code structures from the wild and generate patches for software bugs.
CODIT transforms the correct (or buggy) code snippet into the parse tree and generates the deleted (or added) subtree.
CODIT then predicts the structural changes using a tree-based translation model among the subtrees and employs token names to concrete the structure using a token generation model.
The former tree-based model takes the previous code tree structure and generates a new tree with the maximum likelihood,
while the latter token generation model takes tokens and types of tokens in the code and generates new tokens with the help of LSTM.
To evaluate CODIT, Chakraborty~\etal~\cite{Chakraborty2020codit} construct a real-world bug-fixing benchmark from 48 open-source projects and also employ two well-known benchmarks, Pull-Request-Data~\cite{chen2019sequencer} and Defects4J~\cite{just2014defects4j}.
The experimental results on these three benchmarks illustrate CODIT outperforms existing sequence-based models (\eg SequenceR~\cite{chen2019sequencer}), highlighting the potential of the tree-based models in APR.}

\delete{One variant of LSTM is tree-LSTM architecture which leverages tree structure AST or graph to parse the syntax information of buggy code.
For example, \revise{Yi~\etal~\cite{li2020dlfix}} propose a tree-based model with 2 layers.
After training, filtering and re-ranking (with the help of a CNN layer), the model generates a bunch of patches.
DLFix is designed for single-statement bug fixing, and it shows the potential of a tree-based model in bug fixing.}

\delete{
Despite the tree structure being considered, CODIT mainly employs a sequence-to-sequence NMT model to learn code changes from ASTs, which can still be regarded as an NMT task.
In 2021, Li~\etal~\cite{li2020dlfix} propose DLFix, a two-layer tree-based APR model to learn code transformations on the AST level.
In particular, DLFix first employs a tree-based RNN model to learn the contexts of bug fixes, which is passed to another tree-based RNN model to learn the bug-fixing code transformations.
Besides, a CNN-based classification model is built to re-rank possible patches.
The experimental results on three benchmarks (\ie Defects4J, Bugs.jar and BigFix) demonstrate that DLFix is able to outperform previous learning-based APR approaches (\eg Tufano~\etal~\cite{tufano2019empirical} and achieve comparable performance against pattern-based APR approaches (\eg Tbar~\cite{liu2019tbar}).
Overall, DLFix demonstrates that it is promising and valuable to treat the APR problem as a code transformation learning task over the tree structure rather than an NMT task over code tokens.
}

\delete{
\revise{Li~\etal~\cite{li2022dear} propose DEAR, a learning-based approach for multi-hunk multi-statement fixes.} 
They design a fault localization technique based on traditional SBFL and data flow analysis. 
This technique can acquire multi-hunks that need to be fixed together.
They further design a two-tier tree-based LSTM with an attention layer for fixing multiple statements in the suitable fixing context. Moreover, they apply cycle training 
to learn code transformation and fix need-to-be-fixed-together bugs detected by \delete{the fault localization technique mentioned above} \revise{SBFL}. 
This approach outperforms many DL APR techniques such as DLFix and CoCoNut. Besides, it can fix multi-statement bugs that other APR tools fail to fix.}

\delete{
In 2022, considering that DLFix is able to only fix individual statements at a time, Li~\etal~\cite{li2022dear} propose DEAR, a learning-based approach for multi-hunk multi-statement fixes.
On top of DLFix, DEAR is designed with three key contributions.
First, DEAR introduces an FL technique to acquire multi-hunks that need to be fixed together based on traditional SBFL and data flow analysis.
Second, DEAR develops a compositional approach to generate multi-hunk, multi-statement patches by a divideand-conquer strategy to learn each subtree transformation in ASTs.
Third, DEAR improves the mode architecture of DLFix by designing a two-tier tree-based LSTM with an attention layer learn proper code transformations for fixing multiple statements.
The experimental results on three benchmarks (\ie Defects4J, BigFix, and CPatMiner) demonstrate that DEAR fixes 164 more bugs than DLFix, 61 of which are multi-hunk/multi-statement bugs.
}

\delete{
Devlin~\etal~\cite{devlin2017semantic} present a \delete{novel} model SSC (Share, Specialize, and Compete) to repair semantic bugs, which means fixing non-syntactic bugs in source code.
The input code snippet is encoded with a neural network on the AST level, and each repair type is associated with its own specialized neural module, which emits a score for every repair candidate of that type.
The authors conduct a large-scale corpus by mining code snippets from real-world Python projects on GitHub.
Results indicate that it outperforms existing sequence-to-sequence models with an attention mechanism.
}

\delete{
Yu~\etal~\cite{yu2019learning} present a \delete{novel} approach to predict code transformation at AST level based on structural information for Java programs. 
For structured prediction of source code transforms, they establish a conditional random field (CRF) for the transform prediction, then define the feature functions used in CRF, and finally train the CRF model for prediction. 
They use the learned model to predict transforms for the new, unseen buggy code snippets. 
They conduct a large-scale experimental evaluation on a large dataset of 4,590,679 bug-fixing commits.
The results show the great performance of the proposed technique to generate patches by predicting code structure transforms.
}

%% file: sec/sec_SOTA_graph.tex
\delete{
\textit{State-of-the-arts.}
In the following, we discuss these individual graph-based patch-generation techniques in more detail.}

\delete{Dinella~\etal~\cite{dinella2020hoppity} introduce HOPPITY, an end-to-end learning-based tool for detecting and fixing bugs in JS programs. They apply one step graph edit which is called graph transformation for the model and feed the model with the graph structure of buggy code. This model is then trained to detect and fix more complex and diverse bugs that require adding or deleting code. It outperforms other tools {such as SequenceR and GGNN~\cite{allamanis2017learning}} on the same benchmark with or without the perfect bug locations.}

\delete{
As early as 2020, Dinella~\etal~\cite{dinella2020hoppity} introduce HOPPITY, an end-to-end graph transformation learning-based approach for detecting and fixing bugs in JS programs. 
HOPPITY first represents the buggy program as a graph representation by paring source code into an AST and connecting the leaf nodes.
HOPPITY then performs graph transformation to generate patches by making a sequence of predictions including the position of bug nodes and corresponding graph edits.
The experimental results on a self-constructed benchmark show that HOPPITY outperforms existing repair approaches (\eg SequenceR~\cite{chen2019sequencer}) with or without the perfect FL results.}

\delete{In parallel with HOPPITY, Yasunaga~\etal~\cite{yasunaga2020graph} propose DrRepair to repair C/C++ bugs {based on a program feedback graph}.
They parse the buggy source code into a joint graph representation with diagnostic feedback that captures the semantic structure.
The graph representation takes all identifiers in the source code and any symbols
in the diagnostic feedback as nodes, and connects the same symbols as edges.
They then design a GNN model for learning the graph representation.
Besides, they apply a self-supervised learning paradigm that can generate extra patches by corrupting unlabeled programs. 
They also discover that pre-training on unlabeled programs improves accuracy. 
\delete{The model is evaluated on DeepFix and SPoC datasets and it outperforms existing state-of-art APR tools.}
The experimental results on DeepFix and SPoC datasets demonstrate that DrRepair outperforms three compared APR approaches, \ie DeepFix~\cite{gupta2017deepfix}, RLAssist~\cite{gupta2019deep} and SampleFix~\cite{hajipour2021samplefix}.}

\delete{
Nguyen~\etal~\cite{nguyen2021graphix} propose GRAPHIX, a medium-scale graph edit model which is pre-trained with deleted sub-tree reconstruction.
This model is trained with both abstract and concrete code to learn both structural and semantic code patterns, and it suggests that abstraction may be unnecessary.
\revise{The model} is then evaluated on the Java benchmark from Tufano~\etal~\cite{tufano2019empirical} and it turns out that this model is as competitive as large-scale transformer models and outperforms other state-of-the-art APR tools.}

\delete{
Inspired by HOPPITY, Nguyen~\etal~\cite{nguyen2021graphix} propose GRAPHIX, a graph edit model that is pre-trained with deleted sub-tree reconstruction for program repair.
On top of HOPPITY, GRAPHIX enhances the encoder with multiple graph heads to capture diverse aspects of hierarchical code structures. 
Besides, GRAPHIX introduces a novel pre-training task (\ie deleted sub-tree reconstruction) tolearn implicit program structures from unlabeled data. 
Finally, GRAPHIX is trained with both abstracted and concrete code to learn both structural and semantic code patterns.
The experimental results 
GRAPHIX is evaluated on the Java benchmark from Tufano~\etal~\cite{tufano2019empirical} and it turns out that GRAPHIX is as competitive as large pre-trained models (\eg PLBART~\cite{ahmad2021unified} and CodeT5~\cite{wang2021codet5}) and outperforms the previous learning-based APR approaches (\eg HOPPITY~\cite{dinella2020hoppity} and Tufano~\etal~\cite{tufano2019empirical}).
}

\delete{\revise{In 2021,} Zhu~\etal~\cite{zhu2021syntax} propose Recoder, a syntax-guided edit decoder that uses a novel provider/decider architecture \revise{based on an AST-based graph}.
Recoder takes a buggy statement and its context as input and generates edits as output by 
(1) embedding the buggy statement and its context by a code reader; 
(2) embedding the partial AST of the edits by an AST reader;
(3) embedding a path from the root node to a non-terminal node by a tree path reader;
and (4) producing a probability of each choice for expanding the non-terminal node based on previous embeddings.
In particular, Recoder treats an AST as a directional graph, with its nodes representing AST nodes and its edges connecting a node to its children and its immediate left sibling.
The AST-based graph is then embedded in the form of an adjacency matrix using a Graph Neural Network (GNN) layer.
The authors evaluate Recoder on four widely adopted Java benchmarks: Defects4J v1.2 with 395 bugs, Defects4J v2.0 with 420 bugs, QuixBugs with 40 bugs, and IntroClassJava with 297 bugs.
The results indicate that Recoder is the first learning-based APR technique that outperforms existing traditional techniques on these four Java benchmarks.
The results show that Recoder correctly repairs 53 bugs on Defects4J v1.2, 11 bugs more than TBar~\cite{liu2019tbar} and 19 bugs more than SimFix~\cite{jiang2018shaping}.
Besides, Recoder correctly repairs 19 bugs on Defects4J v2.0, 35 bugs on IntroClassJava and 17 bugs QuixBugs, respectively.
More importantly, Recoder is the first learning-based APR technique that outperforms existing traditional techniques (\eg TBar~\cite{liu2019tbar} and SimFix~\cite{jiang2018shaping}) on these four Java benchmarks.
Recently, Zhu~\etal~\cite{zhu2023tare} further propose Tare built upon Recoder, a type-aware model for program repair to learn the typing rules.
Compared with Recoder, Tare replaces the grammar in Recoder with a T-Grammar that integrates the type information into a standard grammar, and replaces the neural components of Recoder encoding ASTs with neural components encoding T-Graphs, which is a heterogeneous graph with attributes.}

\delete{
Tang~\etal~\cite{tang2021grasp} propose \delete{a novel}\revise{an} end-to-end approach Grasp for repairing buggy Java programs. They represent the buggy method as a graph to retain structural information and apply the Graph-to-Sequence model to capture information from the graph, overcoming the problem of information missing.
Grasp is then evaluated on the Defects4J benchmark as well as real-world bugs from open-source projects and it achieves good results} 

\delete{
Meanwhile, in 2021, Tang~\etal~\cite{tang2021grasp} propose Grasp, an end-to-end graph-to-sequence learning-based approach for repairing buggy Java programs. 
Grasp represents the source code as a graph to retain structural information and applies a graph-to-sequence model to capture information from the graph, overcoming the problem of missing information.
The experimental results on Defects4J show that GrasP is able to generate compilable patches for 75 bugs, 34 of which are correct.
Besides, GrasP achieves better performance than the baseline approach SequenceR with two more correct patches and 11 more plausible patches.}

\delete{Xu~\etal~\cite{xu2022m3v} introduce M3V, a new multi-modal multi-view context embedding approach to predict repair operators for buggy Java code. 
They apply a GNN with multi-view graph-based context structure embedding to capture data and control dependencies. 
They also present a tree-LSTM with tree-based context signature embedding for capturing high-level semantics. 
After M3V is evaluated on repairing two common types of bugs: null pointer exceptions and index out of bounds, results show that M3V is effective in predicting repair operators, achieving over 70\% accuracy on both types of bugs.}

\delete{
In 2022, Xu~\etal~\cite{xu2022m3v} introduce M3V, a new multi-modal multi-view graph-based context embedding approach to predict repair operators for buggy Java code.
Different from previous studies performing patch generation and validation, M3V focuses on repair operator selection.
M3V first applies a GNN with multi-view graph-based context structure embedding to capture data and control dependencies. 
M3V also employs a tree-LSTM model with tree-based context signature embedding for capturing high-level semantics. 
The evaluation experiment is conducted on 20 open-source Java projects with two common types of bugs: null pointer exceptions and index out of bounds.
The results show that M3V is effective in predicting repair operators, achieving 71.45\%$\sim$75.60\% accuracy on both types of bugs, highlighting the future of context embedding in APR.
}

%% file: tab/tab_APCA.tex
\begin{table}[htbp]
\scriptsize
  \centering
  \caption{\revise{A summary and comparison of learning-based APCA studies}}
    \begin{tabular}{cccccm{2.7cm}}
    \toprule
    \multicolumn{1}{l}{Year} & Approach & Language & Feature & Dataset & Repository \\
    \midrule
    2020  & Csuvik~\etal~\cite{csuvik2020utilizing} & Java  & Code Representation & QuixBugs~\cite{lin2017quixbugs} & {\scriptsize \url{https://github.com/AAI-USZ/APR-patch-correctness-IBF2020}} \\ 
    \midrule

    2020  & Tian~\etal~\cite{tian2020evaluating} & Java  & Code Representation & 
    \makecell{Defects4J~\cite{just2014defects4j},Bears~\cite{madeiral2019bears},\\ Bugs.jar~\cite{saha2018bugs}, ManySStubBs4J~\cite{karampatsis2020often},\\ QuixBugs~\cite{lin2017quixbugs},RepairThemAll~\cite{durieux2019empirical}} & {\scriptsize \url{https://github.com/TruX-DTF/DL4PatchCorrectness}} \\ \midrule

    2021  & Csuvik~\etal~\cite{csuvik2021exploring} & JavaScript  & Code Representation & BugsJS~\cite{gyimesi2019bugsjs} & {\scriptsize \url{https://github.com/AAI-USZ/JS-patch-exploration-APR2021}} \\ \midrule

    2021  & ODS~\cite{ye2021automated} & Java  & Engineered Feature & \makecell{Defects4J~\cite{just2014defects4j},\\Bears~\cite{madeiral2019bears},Bugs.jar~\cite{saha2018bugs}} & {\scriptsize \url{https://github.com/SophieHYe/ODSExperiment}} \\ \midrule
    2021  & CACHE~\cite{lin2022context} & Java  & Code Representation & \makecell{Defects4J~\cite{just2014defects4j},ManySStuBs4J~\cite{karampatsis2020often},\\RepairThemAll~\cite{durieux2019empirical}} & {\scriptsize \url{https://github.com/Ringbo/Cache}} \\ \midrule
    
    2022  & Tian~\etal~\cite{tian2022best} & Java  & \makecell{Code Representation\\Engineered Feature} & \makecell{Defects4J~\cite{just2014defects4j},Bears~\cite{madeiral2019bears},\\ Bugs.jar~\cite{saha2018bugs},ManySStubBs4J~\cite{karampatsis2020often},\\ QuixBugs~\cite{lin2017quixbugs},RepairThemAll~\cite{durieux2019empirical}} & {\scriptsize \url{https://github.com/HaoyeTianCoder/Panther}} \\ \midrule
    
    2022  & BATS~\cite{tian2022predicting} & Java  & Test Specification & Defects4J~\cite{just2014defects4j} & {\scriptsize \url{https://github.com/HaoyeTianCoder/BATS}} \\ \midrule    
    2022  & QUATRAIN~\cite{tian2022change} & Java  & Bug Report & \makecell{Defects4J~\cite{just2014defects4j},\\Bears~\cite{madeiral2019bears},Bugs.jar~\cite{saha2018bugs}} & {\scriptsize \url{https://github.com/Trustworthy-Software/Quatrain}} \\ \midrule
    2022  & Shibboleth~\cite{ghanbari2022patch} & Java  & \makecell{Textual similarity \\ Execution Trace \\ Code coverage} & Defects4J~\cite{just2014defects4j} & {\scriptsize \url{https://github.com/ali-ghanbari/shibboleth}} \\ \midrule
    
    2022  & Crex~\cite{yan2022crex} & C     & Execution Trace & CodeFlaws~\cite{tan2017codeflaws} & {\scriptsize \url{https://github.com/1993ryan/crex}} \\
    \bottomrule
    \end{tabular}%
  \label{tab:apca}%
\end{table}%

%% file: APR/APCA/2020-IBF-Utilizing.tex
\delete{Csuvik~\etal~\cite{csuvik2020utilizing} present a Doc2Vec model to explore the nature of similarity-based approach for patch correctness assessment. 
They feed the model with a token sequence under a simple rule and the model measure the similarity between plausible patches and original programs. 
They find that plain source code embeddings fail to capture nuanced code semantics, thus a more sophisticated technique is needed to validate patches correctly.}

\revise{
As early as 2020, inspirited by the similarity-based strategy in state-of-the-art traditional APCA techniques (\eg the behavior similarity of execution traces in PATCH-SIM~\cite{xiong2018identifying}), Csuvik~\etal~\cite{csuvik2020utilizing} present the first study to explore the nature of similarity-based approach based on static code representation for patch correctness assessment.
They leverage embedding models (\ie Doc2vec and BERT) to calculate the similarity between the buggy and patched code snippets and classify patch correctness by a pre-defined similarity threshold.
The experimental results on the QuixBugs dataset show that the proposed approach successfully filters out 45\% (16/35) of the incorrect patches.
In 2021, Csuvik~\etal~\cite{csuvik2021exploring} further adapt the similarity-based method based on static code representation to JavaScript with quantitative and qualitative analysis in depth.
}

%% file: APR/APCA/2020-ASE-Evaluating.tex
\delete{
Different from traditional APCA techniques relying on dynamic information or
manually-crafted heuristics, Tian~\etal~\cite{tian2020evaluating} investigate the feasibility of code representation learning to encode the properties of patch correctness.
They consider different representation learning techniques (\ie Doc2Vec, BERT, code2vec, and CC2Vec) to get embedding vectors for code changes, including pre-trained models and the retraining of models.
They also investigate the discriminative power of learned features in a classification training pipeline (\ie Decision tree, Logistic regression, and Naive Bayes) for patch correctness.
Based on previous work~\cite{tian2020evaluating}, Tian~\etal~\cite{tian2022best} further leverage representation learning models and supervised learning algorithms to investigate the feasibility of statically predicting patch correctness. 
They implement two patch correctness predicting frameworks, Leopard and Panther (upgraded version of Leopard), to investigate the discriminative power of the deep learned features by training machine learning classifiers to predict correct patches.
Besides, they run exploratory experiments assessing the possibility of selecting cutoff similarity scores between learned embeddings of buggy code and patched code snippets for heuristically filtering out incorrect patches. 
After evaluating several models on the same dataset, they find that the performance of these models on learned embedding features is promising when compared against the state-of-the-art techniques which apply dynamic execution traces.}

\revise{
However, the study of Csuvik~\etal~\cite{csuvik2020utilizing} is preliminary and small-scale, only with a single BERT model on 40 one-line bugs.
Tian~\etal~\cite{tian2020evaluating} further conduct a more large-scale empirical study to investigate the feasibility of code representation learning to encode the properties of patch correctness.
They consider different representation learning techniques (\ie Doc2Vec, BERT, code2vec, and CC2Vec) to get embedding vectors for code changes, including pre-trained models and the retraining of models.
They also investigate the discriminative power of learned features in a classification training pipeline (\ie Decision tree, Logistic regression, and Naive Bayes) for patch correctness.
Overall, this work demonstrates the promising future of representation learning in patch correctness assessment, and is valuable fo follow-up works~\cite{tian2022best,lin2022context}.
Later, Tian~\etal~\cite{tian2022best} further extend their previous work~\cite{tian2020evaluating} by examining the effectiveness of code representation, engineered features, and their combination for predicting patch correctness.
They first introduce Leopard, a classification training pipeline to investigate the discriminative power of learned embeddings by training machine learning classifiers to predict correct patches.
The experimental results demonstrate the potential of Leopard to reason about patch correctness based on representation learning models and supervised learning algorithms, \eg BERT associated with XGBoost on 2,147 labeled patches achieves an AUC value of about 0.803, outperforming state-of-the-art APCA techniques PATCH-SIM.
They then introduce Panther, an upgraded version of Leopard to explore the combination of the learned embeddings and the engineered features to improve the performance of identifying patch correctness with more accurate classification.
Panther is proven to outperform Leopard with higher scores in terms of AUC, +Recall and -Recall by combining deep
learned embeddings and engineered features.
}

%% file: APR/APCA/2021-TSE-ODS.tex
\delete{
Ye~\etal~\cite{ye2021automated} propose ODS, a learning-based approach to identify overfitting patches based on static code features and supervised learning.
ODS first defines and extracts a set of 202 static code features from the AST to represent a candidate patch.
ODS then adopts the gradient boosting with the captured code features and patch correctness labels to train a classifier for patch correctness classification.
They conduct on three benchmarks (\ie Defects4J, Bugs.jar and Bears) and the results show that ODS achieves an accuracy of 71.9\% in detecting overfitting patches from 26 projects, and outperforms other state-of-the-art techniques, \revise{\eg PATCH-SIM~\cite{xiong2018identifying}}.}

\revise{
In 2021, different from Tian~\etal~\cite{tian2020evaluating} focusing on code representation, Ye~\etal~\cite{ye2021automated} propose ODS, a learning-based approach to identify overfitting patches based on hand-crafted features and supervised learning.
ODS first defines and extracts a set of 202 static code features from the AST to represent a candidate patch.
ODS then adopts gradient boosting with the captured code features and patch correctness labels to train a classifier for patch correctness classification.
They conduct on three benchmarks (\ie Defects4J, Bugs.jar and Bears) and the results show that ODS achieves an accuracy of 71.9\% in detecting overfitting patches from 26 projects, and outperforms other state-of-the-art techniques, \revise{\eg PATCH-SIM~\cite{xiong2018identifying}}.
}

%% file: APR/APCA/2021-TOSEM-Cache.tex
\delete{Lin~\etal~\cite{lin2022context} propose Cache, a novel context-aware code change embedding technique for the patch correctness task. 
They leverage context information of unchanged code and parse the AST nodes to capture the code structure information. 
They conduct various experiments to evaluate Cache on diverse patch benchmarks. 
The results show that Cache achieves significantly better performance than both previous representation learning techniques and existing APCA techniques.}

\revise{
Despite promising, Tian~\etal~\cite{tian2020evaluating} only focus on the buggy and patched statements while ignoring the surrounding context information.
In 2021, Lin~\etal~\cite{lin2022context} propose CACHE, a context-aware code change embedding technique for the patch correctness task.
CACHE leverages context information of unchanged code and captures the code structure information with the AST path technique.
CACHE then trains a deep learning-based classifier to predict the correctness of the patch based on several pre-defined heuristics.
The experimental results on three benchmarks (\ie Defects4J~\cite{just2014defects4j}, ManySStuBs4J~\cite{karampatsis2020often} and RepairThemAll~\cite{durieux2019empirical}) show that CACHE achieves significantly better performance than both previous representation learning techniques (\ie Tian~\etal~\cite{tian2020evaluating}) and traditional APCA techniques (\eg dynamic-based PATCH-SIM~\cite{xiong2018identifying} and static-based Anti-patterns~\cite{tan2016anti}).
}

%% file: APR/APCA/2022-TOSEM-BATS.tex
\revise{In 2022, unlike previous studies~\cite{tian2020evaluating,ye2021automated} only considering buggy and patched code snippets,} Tian~\etal~\cite{tian2022predicting} introduce BATS, an unsupervised learning-based approach to predict patch correctness based on failing test specifications.
BATS first constructs a search space of historical patches with failing test cases.
Given a plausible patch, BATS identifies similar failing test cases in the search space.
BATS then calculates the similarity of historical patches and the plausible patch based on the failing test cases.
The plausible patch is predicted as correct if the similarity score is larger than a predefined threshold; otherwise, it is predicted as incorrect. 
After collecting plausible patches from 32 APR tools to construct a large dataset, they evaluate the performance of BATS on Defects4J benchmarks with some standard classification metrics (\eg recall). 
BATS outperforms existing techniques in identifying correct patches and filtering out incorrect patches.

%% file: APR/APCA/2022-ASE-QUATRAIN.tex
Besides, Tian~\etal~\cite{tian2022change} attempt to formulate the patch correctness assessment problem as a question-answering problem, which can assess the semantic correlation between a bug report (question) and a patch description (answer).
They introduce QUATRAIN, a supervised learning approach that exploits a deep NLP model to predict patch correctness based on the relatedness of a bug report with a patch description.
QUATRAIN first mines bug reports for bug datasets automatically and generates patch descriptions by existing commit message generation models.
QUATRAIN then leverages an NLP model to capture the semantic correlation between bug reports and patch descriptions.
They evaluate QUATRAIN on a large dataset of 9135 patches from three Java datasets (\ie Defects4j, Bugs.jar, and Bears). 
The results demonstrate that QUATRAIN achieves comparable or better performance against other state-of-the-art dynamic and static techniques, \revise{such as PATCH-SIM~\cite{xiong2018identifying} and BATS~\cite{tian2022predicting}}.
Besides, QUATRAIN is proven practical in learning the relationship between bug reports and code change descriptions for the patch prediction task.

%% file: APR/APCA/2022-IST-Crex.tex
Different from most existing studies focusing on Java programs, Yan~\etal~\cite{yan2022crex} propose Crex to predict patch correctness in C programs based on execution semantics. 
They first leverage transfer learning to extract semantics from micro-traces in buggy C code on the function level.
They then perform semantic similarity computation to denote patch correctness.
They evaluate Crex on a set of 212 patches generated by the CoCoNut APR tool on CodeFlaws programs.
The experimental results indicate that Crex can achieve high precision and recall in predicting patch correctness.

%% file: APR/APCA/2022-ISSTA-Shibboleth.tex
\delete{
Ghanbari~\etal~\cite{ghanbari2022patch} propose Shibboleth, a learning-based technique for patch correctness assessment via ranking and classification. It leverages the impact of the patches on both production code and test suite coverage and relies on a simpler set of assumptions. 
They collect a curated and annotated data set of generated and human-written patches, and they evaluate the model on this dataset. 
The experimental results show that Shibboleth outperforms existing patch classification techniques, \revise{such as ODS~\cite{ye2021automated} and PATCH-SIM~\cite{xiong2018identifying}}.}

\revise{
At the same time, considering that previous studies~\cite{tian2020evaluating,ye2021automated} training a patch prediction classifier with static features (\eg code representation or hand-crafted features), Ghanbari~\etal~\cite{ghanbari2022patch} propose Shibboleth, a hybrid learning-based technique by considering static and dynamic measures from both production and test code to assess the correctness of the patches.
Shibboleth measures the impact of the patches by static syntactic feature (\ie token-level textual similarity), dynamic semantic feature (\ie execution traces similarity) on production code, and code coverage on test code (\ie branch coverage of the passing test cases).
Shibboleth then assesses the correctness of patches via both ranking (\ie prioritizing the patches that are more likely to be correct before the ones that are less likely to be correct) and classification (\ie categorizing patches into two classes of likely correct and likely incorrect) modes. 
The experimental results show that Shibboleth outperforms existing patch ranking (\eg an Ochiai-based sterategy~\cite{wong2016survey}) and classification techniques, such as ODS~\cite{ye2021automated} and PATCH-SIM~\cite{xiong2018identifying}.
}

%% file: tab/tab_APR.tex
\setlength{\rotFPtop}{0pt plus 1fil}
\setlength{\rotFPbot}{0pt plus 1fil}

\begin{sidewaystable}[htbp]
\scriptsize
  \centering
  \caption{A summary and comparison of representative learning-based APR approaches}
    \begin{tabular}{ccccccccccc}
    \toprule
    \revise{Year} & \revise{Technique} & \revise{Type} & \revise{Language} & \revise{Localization} & \revise{Abstraction} & \revise{Context} & \revise{Tokenization} & \revise{Representation} & \revise{Model} & \revise{Ranking} \\
    \midrule

    \revise{2016} & \revise{Bhatia~\etal~\cite{bhatia2016automated}} & \revise{Syntax} & \revise{Python} & \revise{Perfect} & \revise{No} & \revise{Method} & \revise{word} & \revise{token} & \revise{RNN} & \revise{N.A.} \\
    \revise{2017} & \revise{Deepfix~\cite{gupta2017deepfix}} & \revise{Syntax} & \revise{C} & \revise{SD} & \revise{No} & \revise{Method} & \revise{N.A.} & \revise{token} & \revise{GRU} & \revise{N.A.} \\
    \revise{2017} & \revise{Wang~\etal~\cite{wang2018dynamic}} & \revise{Semantic} & \revise{C} & \revise{N.A.} & \revise{No} & \revise{Method} & \revise{N.A.} & \revise{token} & \revise{RNN} & \revise{N.A.} \\
    \revise{2017} & \revise{VuRLE~\cite{ma2017vurle}} & \revise{Vulnerability} & \revise{Java} & \revise{SD} & \revise{Yes} & \revise{Statement} & \revise{N.A.} & \revise{graph} & \revise{N.A.} & \revise{N.A.} \\
    \revise{2018} & \revise{Harer~\etal~\cite{harer2018learning}} & \revise{Vulnerability} & \revise{C,C++} & \revise{N.A.} & \revise{No} & \revise{Method} & \revise{N.A.} & \revise{token} & \revise{GAN} & \revise{N.A.} \\
    \revise{2018} & \revise{TRACER~\cite{ahmed2018compilation}} & \revise{Syntax} & \revise{C} & \revise{SD} & \revise{Yes} & \revise{Method} & \revise{N.A.} & \revise{token} & \revise{RNN} & \revise{beam search} \\
    \revise{2018} & \revise{Santos~\etal~\cite{santos2018syntax}} & \revise{Syntax} & \revise{Java} & \revise{SD} & \revise{Yes} & \revise{Method} & \revise{N.A.} & \revise{token} & \revise{LSTM} & \revise{patch re-ranking} \\
    \revise{2018} & \revise{Bhatia~\etal~\cite{bhatia2018neuro}} & \revise{Syntax} & \revise{Python} & \revise{N.A.} & \revise{No} & \revise{Method} & \revise{N.A.} & \revise{token} & \revise{RNN} & \revise{patch re-ranking} \\
    \revise{2018} & \revise{Sarfgen~\cite{wang2018search}} & \revise{Syntax} & \revise{C} & \revise{N.A.} & \revise{No} & \revise{Method} & \revise{N.A.} & \revise{tree} & \revise{N.A.} & \revise{patch filtering \& re-ranking} \\
    \revise{2019} & \revise{SequenceR~\cite{chen2019sequencer}} & \revise{Semantic} & \revise{Java} & \revise{Perfect} & \revise{Yes} & \revise{Class} & \revise{word} & \revise{token} & \revise{LSTM} & \revise{beam search} \\
    \revise{2019} & \revise{Codit~\cite{Chakraborty2020codit}} & \revise{Syntax} & \revise{Java} & \revise{Perfect} & \revise{Yes} & \revise{Method} & \revise{N.A.} & \revise{tree} & \revise{Tree-LSTM} & \revise{beam search} \\
    \revise{2019} & \revise{Tufano~\etal~\cite{tufano2019learning}} & \revise{Semantic} & \revise{Java} & \revise{N.A.} & \revise{Yes} & \revise{Method} & \revise{word} & \revise{token} & \revise{RNN} & \revise{beam search} \\
    \revise{2019} & \revise{Tufano~\etal~\cite{tufano2019empirical}} & \revise{Semantic} & \revise{Java} & \revise{Perfect} & \revise{perfect} & \revise{Method} & \revise{word} & \revise{token} & \revise{RNN} & \revise{RNN} \\
    \revise{2019} & \revise{Chen~\etal~\cite{chen2019sequencer}} & \revise{Semantic} & \revise{Java} & \revise{N.A.} & \revise{No} & \revise{Class} & \revise{N.A.} & \revise{token} & \revise{RNN} & \revise{N.A.} \\
    \revise{2019} & \revise{DeepDelta~\cite{mesbah2019deepdelta}} & \revise{Syntax} & \revise{Java} & \revise{Perfect} & \revise{Yes} & \revise{Method} & \revise{N.A.} & \revise{tree} & \revise{LSTM} & \revise{beam search} \\
    \revise{2019} & \revise{RLAssitst~\cite{gupta2019deep}} & \revise{Syntax} & \revise{C} & \revise{SD} & \revise{No} & \revise{Method} & \revise{N.A.} & \revise{token} & \revise{LSTM} & \revise{N.A.} \\
    \revise{2020} & \revise{CoCoNut~\cite{lutellier2020coconut}} & \revise{Semantic} & \revise{Java,C,Python,JS} & \revise{Perfect} & \revise{Yes} & \revise{Method} & \revise{word} & \revise{token} & \revise{FConv} & \revise{beam search} \\
    \revise{2020} & \revise{DLFix~\cite{li2020dlfix}} & \revise{Semantic} & \revise{Java} & \revise{SBFL} & \revise{Yes} & \revise{Method} & \revise{word} & \revise{tree} & \revise{Tree-LSTM} & \revise{patch filtering \& re-ranking} \\
    \revise{2020} & \revise{DrRepair~\cite{yasunaga2020graph}} & \revise{Syntax} & \revise{C,C++} & \revise{SD} & \revise{No} & \revise{Method} & \revise{N.A.} & \revise{graph} & \revise{LSTM} & \revise{N.A.} \\
    \revise{2020} & \revise{Hoppity~\cite{dinella2020hoppity}} & \revise{Semantic} & \revise{JS} & \revise{SD} & \revise{No} & \revise{Statement} & \revise{N.A.} & \revise{graph} & \revise{LSTM} & \revise{beam search} \\
    \revise{2020} & \revise{Yang~\etal~\cite{yang2020applying}} & \revise{Syntax} & \revise{C} & \revise{SD} & \revise{N.A.} & \revise{Method} & \revise{subword} & \revise{token} & \revise{SeqGAN} & \revise{patch re-ranking} \\
    \revise{2020} & \revise{GGF~\cite{wu2020ggf}} & \revise{Syntax} & \revise{C} & \revise{SD} & \revise{No} & \revise{Method} & \revise{N.A.} & \revise{token,graph} & \revise{GGNN} & \revise{N.A.} \\
    \revise{2021} & \revise{CURE~\cite{jiang2021cure}} & \revise{Semantic} & \revise{Java} & \revise{Perfect} & \revise{No} & \revise{Method} & \revise{subword} & \revise{token} & \revise{GPT} & \revise{code-aware beam search} \\
    \revise{2021} & \revise{Recoder~\cite{zhu2021syntax}} & \revise{Syntax} & \revise{Java} & \revise{SBFL,Perfect} & \revise{No} & \revise{Method} & \revise{word} & \revise{graph} & \revise{Tree-LSTM} & \revise{beam search} \\
    \revise{2021} & \revise{TFix~\cite{berabi2021tfix}} & \revise{Semantic} & \revise{JS} & \revise{Perfect} & \revise{No} & \revise{Statement} & \revise{subword} & \revise{token} & \revise{T5} & \revise{beam search} \\
    \revise{2021} & \revise{GrasP~\cite{tang2021grasp}} & \revise{Semantic} & \revise{Java} & \revise{Perfect} & \revise{No} & \revise{Method} & \revise{word} & \revise{graph} & \revise{RNN,GNN} & \revise{beam search} \\
    \revise{2021} & \revise{SampleFix~\cite{hajipour2021samplefix}} & \revise{Syntax} & \revise{C} & \revise{SD} & \revise{No} & \revise{Method} & \revise{N.A.} & \revise{token} & \revise{LSTM} & \revise{beam search} \\
    \revise{2022} & \revise{CIRCLE~\cite{yuan2022circle}} & \revise{Semantic} & \revise{Java,C,JS,Python} & \revise{Perfect} & \revise{No} & \revise{Method} & \revise{subword} & \revise{token} & \revise{T5} & \revise{beam search} \\
    \revise{2022} & \revise{DEAR~\cite{li2022dear}} & \revise{Semantic} & \revise{Java} & \revise{SBFL} & \revise{Yes} & \revise{Statement} & \revise{word} & \revise{tree} & \revise{Tree-LSTM} & \revise{patch filtering \& re-ranking} \\
    \revise{2022} & \revise{Graphix~\cite{nguyen2021graphix}} & \revise{Semantic} & \revise{Java} & \revise{Perfect} & \revise{Yes} & \revise{Method} & \revise{N.A.} & \revise{graph,tree} & \revise{Tree-LSTM} & \revise{N.A.} \\
    \revise{2022} & \revise{SelfAPR~\cite{ye2022selfapr}} & \revise{Semantic} & \revise{Java} & \revise{Perfect} & \revise{No} & \revise{Method} & \revise{subword} & \revise{token} & \revise{Transformer} & \revise{beam search} \\
    \revise{2022} & \revise{VRepair~\cite{chen2022neural}} & \revise{Vulnerability} & \revise{C} & \revise{Perfect} & \revise{No} & \revise{Method} & \revise{word} & \revise{token} & \revise{Transformer} & \revise{beam search} \\
    \revise{2022} & \revise{SeqTrans~\cite{chi2022seqtrans}} & \revise{Vulnerability} & \revise{Java} & \revise{Perfect} & \revise{Yes} & \revise{Statement} & \revise{subword} & \revise{token} & \revise{Transformer} & \revise{beam search} \\
    \revise{2022} & \revise{AlphaRepair~\cite{xia2022less}} & \revise{Semantic} & \revise{Java,Python} & \revise{Perfect} & \revise{No} & \revise{Class} & \revise{subword} & \revise{token} & \revise{CodeBERT} & \revise{CodeBERT re-ranking} \\
    \revise{2022} & \revise{VulRepair~\cite{fu2022vulrepair}} & \revise{Vulnerability} & \revise{C} & \revise{Perfect} & \revise{No} & \revise{Method} & \revise{subword} & \revise{token} & \revise{T5} & \revise{beam search} \\
    \revise{2022} & \revise{Bug-Transformer~\cite{yao2022bug}} & \revise{Semantic} & \revise{Java} & \revise{Perfect} & \revise{Yes} & \revise{Method} & \revise{subword} & \revise{token} & \revise{Transformer} & \revise{beam search} \\
    \revise{2022} & \revise{SPVF~\cite{zhou2022spvf}} & \revise{Vulnerability} & \revise{C++,C,Python} & \revise{Perfect} & \revise{No} & \revise{Method} & \revise{N.A.} & \revise{tree} & \revise{Transformer} & \revise{beam search,patch filtering} \\
    \revise{2022} & \revise{SYNSHINE~\cite{ahmed2022synshine}} & \revise{Syntax} & \revise{Java} & \revise{SD} & \revise{Yes} & \revise{Class} & \revise{subword} & \revise{token} & \revise{Transformer} & \revise{N.A.} \\
    \revise{2022} & \revise{MMAPR~\cite{zhang2022repairing}} & \revise{Semantic,Syntax} & \revise{Python} & \revise{Perfect} & \revise{No} & \revise{Class} & \revise{subword} & \revise{token} & \revise{Codex} & \revise{N.A.} \\
    \revise{2022} & \revise{RING~\cite{joshi2022repair}} & \revise{Syntax} & \revise{Python,JS,C} & \revise{SD} & \revise{No} & \revise{Method} & \revise{subword} & \revise{token} & \revise{Codex} & \revise{patch re-ranking} \\
    \revise{2022} & \revise{RewardRepair~\cite{ye2022neural}} & \revise{Semantic} & \revise{Java} & \revise{SBFL,Perfect} & \revise{No} & \revise{Statement} & \revise{subword} & \revise{token} & \revise{Transformer} & \revise{beam search} \\
    \revise{2021} & \revise{BIFI~\cite{yasunaga2021break}} & \revise{Syntax} & \revise{Python,C} & \revise{N.A.} & \revise{No} & \revise{Method} & \revise{N.A.} & \revise{token,graph} & \revise{LSTM} & \revise{beam search} \\

    \bottomrule
    \end{tabular}%
  \label{tab:apr}%
\end{sidewaystable}%

%% file: sec/sec_sematic.tex
\subsubsection{Semantic error repair}

Semantic errors usually refer to any case where the actual program behavior is not expected by developers, and can be detected by functional test cases.
Considering that the vast majority of existing learning-based studies are concentrated in this field of semantic error, we group them based on the form of code representation.
In the following, we discuss and summarize existing individual learning-based APR techniques that focus on semantic bugs in detail.

\ding{182} \textbf{Sequence-based Approaches.}

\input{APR/semantic/2019-ICSE-Learning}

\input{APR/semantic/2021-TSE-SEQUENCER}

\input{APR/semantic/2019-arXiv-Learning}

\input{APR/semantic/2020-ISSTA-CoCoNuT}

\input{APR/semantic/2021-ICSE-CURE}

\input{APR/semantic/2022-JCSC-Bug-Transformer}

\input{APR/semantic/2019-arxiv-ENCORE}

\input{APR/semantic/2022-ICSE-Neural}

\input{APR/semantic/2022-ASE-SelfAPR}

\ding{183} \textbf{Tree-based Approaches.}

\input{APR/semantic/tree/2020-TSE-CODIT}

\input{APR/semantic/tree/2020-ICSE-DLFix}

\input{APR/semantic/tree/2022-ICSE-DEAR}

\ding{184} \textbf{Graph-based Approaches.}

\input{APR/semantic/graph/2020-ICLR-HOPPITY}

\input{APR/semantic/graph/2020-PMLR-DrRepair}

\input{APR/semantic/graph/2021-arXiv-GRAPHIX}

\input{APR/semantic/graph/2021-FSE-Recoder}

\input{APR/semantic/graph/2021-QRS-GrasP}

\input{APR/semantic/graph/2022-CGO-M3V}

%% file: APR/semantic/2019-ICSE-Learning.tex
\delete{
Tufano~\etal~\cite{tufano2019learning} design an NMT model to generate the same patches applied by developers under a narrow context. 
They reduce the vocabulary size by mapping the method of the code to a specific ID and feed the model with pairs of methods before and after the patch. 
The model can replicate up to 36\% of the buggy code. 
Moreover, it can be applied to refactoring and other code relating activities.
}

\revise{As early as 2019, Tufano~\etal~\cite{tufano2019learning} conduct this first attempt to investigate the ability of NMT models to learn code changes during pull requests.
They first mine pull requests from three large Gerrit repositories and extract the method pairs before and after the pull requests, where each pair serves as an example of a meaningful change.
They then map the identifiers and literals in the source code to specific IDs (\ie code abstraction) to reduce the vocabulary size.
Finally, they train NMT models to translate the method before the pull request into the one after the pull request, to emulate the actual code change.
The experimental results show that NMT models can generate the same patches for 36\% pull requests
Overall, this study demonstrates the potential of NMT models in learning a wide variety of meaningful code changes, especially refactorings and bug-fixing activities.
Further, Tufano~\etal~\cite{tufano2019empirical} perform an empirical study to investigate the potential of NMT models in generating bug-fixing patches in the wild, which is discussed in Section~\ref{sec:empirical}.
}

%% file: APR/semantic/2021-TSE-SEQUENCER.tex
\revise{
At the same time, Chen~\etal~\cite{chen2019sequencer} propose SequenceR, an end-to-end approach based on sequence-to-sequence learning. 
They combine LSTM encoder-decoder architecture with a copy mechanism to address the problem of a large vocabulary. First, they apply state-of-the-art fault localization techniques to identify the buggy method and the suspicious buggy lines. 
Then, they perform a novel buggy context abstraction process that intelligently organizes the fault localization data into a suitable representation for the deep learning model. Finally, SequenceR generates multiple patches for the buggy code.
Although their approach can only be applied to single-line buggy code, this model outperforms the APR tool of Tufano~\etal~on Defects4J benchmarks. Moreover, they prove that the copy mechanism can improve the accuracy of generated patches.}

%% file: APR/semantic/2019-arXiv-Learning.tex
\delete{
Current works aim at exploring fixes in a limited search space, which may not contain the correct patches.
\revise{Hata~\etal~\cite{hata2018learning}} follow the recently NMT-based approach and use an encoder-decoder model Ratchet with multi-layer attention to fix bugs.
They perform an empirical study with five large software projects. Moreover, they collect a fine-grained dataset from these projects and try to ignore noisy data. They train and evaluate Ratchet on this dataset. Results show that Ratchet performs at least as well as pattern-based APR tools.
Besides, Ratchet’s output was considered helpful in fixing the bugs on many occasions, even if the fix was not 100\% correct.}

%% file: APR/semantic/2020-ISSTA-CoCoNuT.tex
\delete{
Lutellier~\etal~\cite{lutellier2020coconut} propose CoCoNut, a \delete{novel} generate\&validate technique with a new context-aware NMT architecture that separately inputs the buggy line and method context. 
They further combine CNN (FConv architecture) with the NMT model to improve the accuracy of generated patches. 
After collecting a large dataset from four programming languages and training the model on it, CoCoNut is then evaluated on six benchmarks (also from four programming languages), \ie Defects4J of Java, QuixBugs of Java, CodeFlaws of C, ManyBugs of C, QuixBugs of Python and BugAID of JS.
It turns out that CoCoNut outperforms previous APR tools, and is capable of fixing 300 more bugs other APR tools fail to . Moreover, CoCoNut proves that FConv architecture can outperform LSTM.}

\revise{
However, Tufano~\etal~\cite{tufano2019empirical} and SequenceR~\cite{chen2019sequencer} represent both the buggy line and its context as one input for NMT models, making it difficult to extract long-term relations between code tokens.
In 2020, Lutellier~\etal~\cite{lutellier2020coconut} propose CoCoNut, a context-aware NMT approach that separately inputs the buggy line and method context.
In particular, CoCoNut applies CNN (\ie FConv architecture) in the context-aware NMT architecture, which is able to better extract hierarchical features of source code compared with LSTM used in Tufano~\etal~\cite{tufano2019empirical} and SequenceR~\cite{chen2019sequencer}.
Besides, CoCoNuT trains multiple NMT models to capture the diversity of bug fixes with ensemble learning. 
CoCoNut is evaluated on six well-known benchmarks across four programming languages, \ie Defects4J of Java, QuixBugs of Java, CodeFlaws of C, ManyBugs of C, QuixBugs of Python and BugAID of JS.
The experimental results show that CoCoNut is capable of fixing 509 bugs on the six benchmarks, 309 of which have not been fixed by previous APR tools, such as DLFix, Prophet and TBar.
At the same time, different CoConut only considering patch generation, Yang~\etal~\cite{yang2020applying} propose a sequence-basd technique for both fault localization and patch generation.
They first employ a CNN-based autoencoder to rank suspicious buggy code by extracting various information from bug reports and the program source code.
They then convert the program source code into multiple lines with tokens and apply the SeqGAN model to generate the candidate patches.
}

%% file: APR/semantic/2021-ICSE-CURE.tex
\delete{
Further, Jiang~\etal~\cite{jiang2021cure} propose CURE, an \delete{novel} NMT-based program repair technique to fix Java bugs. 
They pre-train a programming language model on a large corpus and combine it with NMT architecture to learn code syntax and fix patterns. 
They also apply a code-aware search strategy and a new subword tokenization technique to improve the accuracy of generated patches. This model outperforms SequenceR and CoCoNut \delete{APR tools} on Defects4J and QuixBugs benchmarks under different beam search sizes.}

\revise{In 2021, on top of CoCoNut, Jiang~\etal~\cite{jiang2021cure} propose CURE, an NMT-based APR technique to fix Java bugs.
Compared with CoCoNut, the novelty of CURE mainly coms from three aspects.
First, to better learn developer-like source code, CURE pre-trains a programming language model on a large corpus and combines it with the CoCoNut context-aware architecture.
Second, CURE designs a code-aware beam search strategy to avoid uncompilable patches during patch generation.
Third, to address the OOV problem, CURE introduces a new sub-word tokenization technique to tokenize compound and rare words.
The experimental results demonstrate that CURE is able to fix 57 Defects4J bugs and 26 QuixBugs bugs, outperforming existing learning-based APR approaches under different beam search sizes, such as SequenceR and CoCoNut.
}

%% file: APR/semantic/2022-JCSC-Bug-Transformer.tex
\delete{
Yao~\etal~\cite{yao2022bug} propose a new transformer-based APR technique, Bug-Transformer, to fix buggy code snippets. 
It applies a novel token pair encoding (TPE) approach to reduce vocabulary size by compressing code structure while preserving semantic information. 
Besides, they apply a novel rename mechanism to preserve semantic features for code abstraction.
Bug-Transformer leverages the transformer architecture and is fine-tuned for \delete{learning}\revise{program repair} tasks. 
It is then evaluated on Java benchmarks and outperforms other baseline models such as \delete{Bug2Fix}~\revise{Tufano~\etal~\cite{tufano2019empirical}}.
}

\revise{
In 2022, unlike Tufano~\etal~\cite{tufano2019empirical} without considering semantic and lexical scope information of code tokens, Yao~\etal~\cite{yao2022bug} propose Bug-Transformer, a transformer-based APR technique to fix buggy code snippets.
It is equipped with a token pair encoding (TPE) algorithm and a rename mechanism to preserve crucial information.
First, Bug-Transformer designs a TPE algorithm to reduce vocabulary size by compressing code structure while preserving structural and semantic information.
Second, Bug-Transformer employs a rename mechanism to abstract code tokens (\ie identifiers and literals) with consideration of their semantic and lexical scope knowledge.
Third, Bug-Transformer trains a transformer-based model to learn the structural and semantic information of code snippets and predicts patches automatically.
The experimental results on BPF~\cite{tufano2019empirical} datasets show that Bug-Transformer outperforms baseline models, \eg Tufano~\etal~\cite{tufano2019empirical}.
}

%% file: APR/semantic/2019-arxiv-ENCORE.tex
\delete{
Lutellier~\etal~\cite{lutellier2019encore} propose ENCORE, a new end-to-end APR technique that leverages the NMT model to generate bug fixes for Java programs. Evaluating ENCORE on two Java benchmarks proves that it can fix diverse bugs, and further experiments on Python, C++, and JS benchmarks proves that it can handle bugs in different programming languages. They also present attention maps to explain why certain fixes are generated or not by ENCORE.}

%% file: APR/semantic/2022-ICSE-Neural.tex
\delete{
Ye~\etal~\cite{ye2022neural} introduce RewardRepair as a neural program repair approach for fixing bugs in Java code based on transformer. 
They apply a novel training strategy and feed the model with compiling and testing execution information to improve the quality of generated patches. 
This model is then evaluated on four benchmarks, Defects4J v1 and v2, Bugs.jar, and QuixBugs. 
Results show that this model has lower cross-entropy than previous APR tools. 
Besides, RewardRepair outperforms SequenceR, CoCoNut and CURE in terms of top-k accuracy on Defects4J benchmarks.}

\revise{
Existing learning-based APR techniques are usually limited by the generation of lots of low-quality (\eg non-compilable) patches, due to the employed static loss function based on token similarity.
In 2022, Ye~\etal~\cite{ye2022neural} introduce RewardRepair based on a mixed loss function that considers program compilation and test execution information.
In particular, RewardRepair defines a discriminator to discriminate good patches from low-quality ones based on dynamic execution feedback, rather than static token similarity between the generated patch and the human-written ground truth patch.
The discriminator computes a reward value to gauge the patch quality, and this value is subsequently utilized to update the weights of the patch generation model during the backpropagation process.
A higher reward indicates a higher quality \revise{of} generated patch that is compilable and passes the test cases, while a lower reward suggests potentially unsatisfactory patch quality, such as a non-compilable patch.
Thanks to the compilation and test execution results during training, RewardRepair is able to fix 207 bugs on four benchmarks, i.e., Defects4J-v1.2, Defects4J-v2.0, Bugs.jar and QuixBugs, 121 of which are not repaired by previous approaches, \eg DLFix, CoCoNut and CURE.
More importantly,  RewardRepair achieves a compilable rate of up to 45.3\% among Top-30 candidate patches, an improvement over the 39\% by CURE, demonstrating the potential to generate high-quality patches.
}

%% file: APR/semantic/2022-ASE-SelfAPR.tex
\delete{Previous neural program repair approaches focus on supervised training and lack project-specific knowledge.
Ye~\etal~\cite{ye2022selfapr} propose, SelfAPR, a self-supervised training approach with test execution diagnostics based on a transformer neural network. 
SelfAPR consists of two components, training sample generator and neural network optimization. 
The first part generates perturbed programs with a perturbing model and tests it to capture compile errors and execute failures information. 
The second part is fed with the previous information and outputs $n$ best patches with beam search. 
SelfAPR is capable of repairing ten bugs that are never repaired before by the supervised neural repair models. Moreover, evaluation results the effectiveness of self-supervised training and its components.}

\revise{
Besides, previous learning-based APR approaches are dominantly founded on supervised training with massive open-source code repositories, resulting in a lack of project-specific knowledge.
In parallel with RewardRepair, Ye~\etal~\cite{ye2022selfapr} also propose, SelfAPR, a self-supervised training approach with test execution diagnostics based on a transformer neural network. 
SelfAPR consists of two well-designed components, \ie training sample generator and neural network optimization.
The first part generates perturbed programs with a perturbing model and tests it to capture compile errors and execute failures information. 
The second part is fed with the previous information and outputs $n$ best candidate patches with beam search. 
The experimental results show that  SelfAPR is capable of repairing 65bugs from Defects4J-v.12 and 45 bugs from Defects4J-v2.0, 10 of which have never been repaired by the previous supervised neural repair models, such as CoCoNut~\cite{lutellier2020coconut} and CURE~\cite{jiang2021cure}. 
More importantly, SelfAPR highlights the potential and power of self-supervised training and project-specific knowledge in the learning-based APR community.
}

%% file: APR/semantic/tree/2020-TSE-CODIT.tex
\delete{Chakraborty~\etal~\cite{Chakraborty2020codit} propose a tree-based APR technique CODIT to learn code changes from the wild and generate patches for software bugs.
CODIT transforms the correct (or buggy) code snippet into the parse tree and generates the deleted (or added) subtree.
CODIT then predicts the structural changes using a tree-based translation model among the subtrees and employs token names to concrete the structure using a token generation model.
The former tree-based model takes the previous code tree structure and generates a new tree with the maximum likelihood,
while the latter token generation model takes tokens and types of tokens in the code and generates new tokens with the help of LSTM.
The authors conduct a real-world bug-fixing dataset Code-Change-Data from 48 open-source projects and employ Pull-Request-Data~\cite{chen2019sequencer} and Defects4J~\cite{just2014defects4j}.
The results on these three datasets illustrate CODIT outperforms existing \delete{seq2seq}\revise{Seq2Seq} models, highlighting the potential of the tree-based models in APR.}

\revise{
As early as 2020, Chakraborty~\etal~\cite{Chakraborty2020codit} propose a tree-based neural network, CODIT to learn code changes by encoding code structures from the wild and generate patches for software bugs.
CODIT transforms the correct (or buggy) code snippet into the parse tree and generates the deleted (or added) subtree.
CODIT then predicts the structural changes using a tree-based translation model among the subtrees and employs token names to concrete the structure using a token generation model.
The former tree-based model takes the previous code tree structure and generates a new tree with the maximum likelihood,
while the latter token generation model takes tokens and types of tokens in the code and generates new tokens with the help of LSTM.
To evaluate CODIT, Chakraborty~\etal~\cite{Chakraborty2020codit} construct a real-world bug-fixing benchmark from 48 open-source projects and also employ two well-known benchmarks, Pull-Request-Data~\cite{chen2019sequencer} and Defects4J~\cite{just2014defects4j}.
The experimental results on these three benchmarks illustrate CODIT outperforms existing sequence-based models (\eg SequenceR~\cite{chen2019sequencer}), highlighting the potential of the tree-based models in APR.}

%% file: APR/semantic/tree/2020-ICSE-DLFix.tex
\delete{One variant of LSTM is tree-LSTM architecture which leverages tree structure AST or graph to parse the syntax information of buggy code.
For example, \revise{Yi~\etal~\cite{li2020dlfix}} propose a tree-based model with 2 layers.
After training, filtering and re-ranking (with the help of a CNN layer), the model generates a bunch of patches.
DLFix is designed for single-statement bug fixing, and it shows the potential of a tree-based model in bug fixing.}

\revise{
Despite the tree structure being considered, CODIT mainly employs a sequence-to-sequence NMT model to learn code changes from ASTs, which can still be regarded as an NMT task.
In 2021, Li~\etal~\cite{li2020dlfix} propose DLFix, a two-layer tree-based APR model to learn code transformations on the AST level.
In particular, DLFix first employs a tree-based RNN model to learn the contexts of bug fixes, which is passed to another tree-based RNN model to learn the bug-fixing code transformations.
Besides, a CNN-based classification model is built to re-rank possible patches.
The experimental results on three benchmarks (\ie Defects4J, Bugs.jar and BigFix) demonstrate that DLFix is able to outperform previous learning-based APR approaches (\eg Tufano~\etal~\cite{tufano2019empirical} and achieve comparable performance against pattern-based APR approaches (\eg Tbar~\cite{liu2019tbar}).
Overall, DLFix demonstrates that it is promising and valuable to treat the APR problem as a code transformation learning task over the tree structure rather than an NMT task over code tokens.
}

%% file: APR/semantic/tree/2022-ICSE-DEAR.tex
\delete{
\revise{Li~\etal~\cite{li2022dear} propose DEAR, a learning-based approach for multi-hunk multi-statement fixes.} 
They design a fault localization technique based on traditional SBFL and data flow analysis. 
This technique can acquire multi-hunks that need to be fixed together.
They further design a two-tier tree-based LSTM with an attention layer for fixing multiple statements in the suitable fixing context. Moreover, they apply cycle training 
to learn code transformation and fix need-to-be-fixed-together bugs detected by \delete{the fault localization technique mentioned above} \revise{SBFL}. 
This approach outperforms many DL APR techniques such as DLFix and CoCoNut. Besides, it can fix multi-statement bugs that other APR tools fail to fix.}

\revise{
In 2022, considering that DLFix is able to only fix individual statements at a time, Li~\etal~\cite{li2022dear} propose DEAR, a learning-based approach for multi-hunk multi-statement fixes.
On top of DLFix, DEAR is designed with three key contributions.
First, DEAR introduces an FL technique to acquire multi-hunks that need to be fixed together based on traditional SBFL and data flow analysis.
Second, DEAR develops a compositional approach to generate multi-hunk, multi-statement patches by a divideand-conquer strategy to learn each subtree transformation in ASTs.
Third, DEAR improves the mode architecture of DLFix by designing a two-tier tree-based LSTM with an attention layer learn proper code transformations for fixing multiple statements.
The experimental results on three benchmarks (\ie Defects4J, BigFix, and CPatMiner) demonstrate that DEAR fixes 164 more bugs than DLFix, 61 of which are multi-hunk/multi-statement bugs.
}

%% file: APR/semantic/graph/2020-ICLR-HOPPITY.tex
\delete{Dinella~\etal~\cite{dinella2020hoppity} introduce HOPPITY, an end-to-end learning-based tool for detecting and fixing bugs in JS programs. They apply one step graph edit which is called graph transformation for the model and feed the model with the graph structure of buggy code. This model is then trained to detect and fix more complex and diverse bugs that require adding or deleting code. It outperforms other tools \revise{such as SequenceR and GGNN~\cite{allamanis2017learning}} on the same benchmark with or without the perfect bug locations.}

\revise{
As early as 2020, Dinella~\etal~\cite{dinella2020hoppity} introduce HOPPITY, an end-to-end graph transformation learning-based approach for detecting and fixing bugs in JS programs. 
HOPPITY first represents the buggy program as a graph representation by paring source code into an AST and connecting the leaf nodes.
HOPPITY then performs graph transformation to generate patches by making a sequence of predictions including the position of bug nodes and corresponding graph edits.
The experimental results on a self-constructed benchmark show that HOPPITY outperforms existing repair approaches (\eg SequenceR~\cite{chen2019sequencer}) with or without the perfect FL results.}

%% file: APR/semantic/graph/2020-PMLR-DrRepair.tex
\revise{In parallel with HOPPITY,} Yasunaga~\etal~\cite{yasunaga2020graph} propose DrRepair to repair C/C++ bugs \revise{based on a program feedback graph}.
They parse the buggy source code into a joint graph representation with diagnostic feedback that captures the semantic structure.
The graph representation takes all identifiers in the source code and any symbols
in the diagnostic feedback as nodes, and connects the same symbols as edges.
They then design a GNN model for learning the graph representation.
Besides, they apply a self-supervised learning paradigm that can generate extra patches by corrupting unlabeled programs. 
They also discover that pre-training on unlabeled programs improves accuracy. 
\delete{The model is evaluated on DeepFix and SPoC datasets and it outperforms existing state-of-art APR tools.}
\revise{The experimental results on DeepFix and SPoC datasets demonstrate that DrRepair outperforms three compared APR approaches, \ie DeepFix~\cite{gupta2017deepfix}, RLAssist~\cite{gupta2019deep} and SampleFix~\cite{hajipour2021samplefix}}.

%% file: APR/semantic/graph/2021-arXiv-GRAPHIX.tex
\delete{
Nguyen~\etal~\cite{nguyen2021graphix} propose GRAPHIX, a medium-scale graph edit model which is pre-trained with deleted sub-tree reconstruction.
This model is trained with both abstract and concrete code to learn both structural and semantic code patterns, and it suggests that abstraction may be unnecessary.
\revise{The model} is then evaluated on the Java benchmark from Tufano~\etal~\cite{tufano2019empirical} and it turns out that this model is as competitive as \delete{large-scale transformer models and outperforms other state-of-the-art APR tools}\revise{large pre-trained models (\eg PLBART~\cite{ahmad2021unified} and CodeT5~\cite{wang2021codet5}) and outperforms the previous APR approaches (\eg HOPPITY~\cite{dinella2020hoppity} and Tufano~\etal~\cite{tufano2019empirical})}.}

\revise{
Inspired by HOPPITY, Nguyen~\etal~\cite{nguyen2021graphix} propose GRAPHIX, a graph edit model that is pre-trained with deleted sub-tree reconstruction for program repair.
On top of HOPPITY, GRAPHIX enhances the encoder with multiple graph heads to capture diverse aspects of hierarchical code structures. 
Besides, GRAPHIX introduces a novel pre-training task (\ie deleted sub-tree reconstruction) tolearn implicit program structures from unlabeled data. 
Finally, GRAPHIX is trained with both abstracted and concrete code to learn both structural and semantic code patterns.
The experimental results 
GRAPHIX is evaluated on the Java benchmark from Tufano~\etal~\cite{tufano2019empirical} and it turns out that GRAPHIX is as competitive as large pre-trained models (\eg PLBART~\cite{ahmad2021unified} and CodeT5~\cite{wang2021codet5}) and outperforms the previous learning-based APR approaches (\eg HOPPITY~\cite{dinella2020hoppity} and Tufano~\etal~\cite{tufano2019empirical}).
}

%% file: APR/semantic/graph/2021-FSE-Recoder.tex
\revise{In 2021,} Zhu~\etal~\cite{zhu2021syntax} propose Recoder, a syntax-guided edit decoder that uses a novel provider/decider architecture \revise{based on an AST-based graph}.
Recoder takes a buggy statement and its context as input and generates edits as output by 
(1) embedding the buggy statement and its context by a code reader; 
(2) embedding the partial AST of the edits by an AST reader;
(3) embedding a path from the root node to a non-terminal node by a tree path reader;
and (4) producing a probability of each choice for expanding the non-terminal node based on previous embeddings.
\revise{In particular, Recoder treats an AST as a directional graph, with its nodes representing AST nodes and its edges connecting a node to its children and its immediate left sibling.
The AST-based graph is then embedded in the form of an adjacency matrix using a Graph Neural Network (GNN) layer.}
The authors evaluate Recoder on four widely adopted Java benchmarks: Defects4J v1.2 with 395 bugs, Defects4J v2.0 with 420 bugs, QuixBugs with 40 bugs, and IntroClassJava with 297 bugs.
\delete{The results indicate that Recoder is the first learning-based APR technique that outperforms existing traditional techniques on these four Java benchmarks.}
\revise{The results show that Recoder correctly repairs 53 bugs on Defects4J v1.2, 11 bugs more than TBar~\cite{liu2019tbar} and 19 bugs more than SimFix~\cite{jiang2018shaping}.
Besides, Recoder correctly repairs 19 bugs on Defects4J v2.0, 35 bugs on IntroClassJava and 17 bugs QuixBugs, respectively.
More importantly, Recoder is the first learning-based APR technique that outperforms existing traditional techniques (\eg TBar~\cite{liu2019tbar} and SimFix~\cite{jiang2018shaping}) on these four Java benchmarks.}

%% file: APR/semantic/graph/2021-QRS-GrasP.tex
\delete{
Tang~\etal~\cite{tang2021grasp} propose \delete{a novel}\revise{an} end-to-end approach Grasp for repairing buggy Java programs. They represent the buggy method as a graph to retain structural information and apply the Graph-to-Sequence model to capture information from the graph, overcoming the problem of information missing.
Grasp is then evaluated on the Defects4J benchmark as well as real-world bugs from open-source projects and it achieves good results} 

\revise{
Meanwhile, in 2021, Tang~\etal~\cite{tang2021grasp} propose Grasp, an end-to-end graph-to-sequence learning-based approach for repairing buggy Java programs. 
Grasp represents the source code as a graph to retain structural information and applies a graph-to-sequence model to capture information from the graph, overcoming the problem of missing information.
The experimental results on Defects4J show that GrasP is able to generate compilable patches for 75 bugs, 34 of which are correct.
Besides, GrasP achieves better performance than the baseline approach SequenceR with two more correct patches and 11 more plausible patches.}

%% file: APR/semantic/graph/2022-CGO-M3V.tex
\delete{Xu~\etal~\cite{xu2022m3v} introduce M3V, a new multi-modal multi-view context embedding approach to predict repair operators for buggy Java code. 
They apply a GNN with multi-view graph-based context structure embedding to capture data and control dependencies. 
They also present a tree-LSTM with tree-based context signature embedding for capturing high-level semantics. 
After M3V is evaluated on repairing two common types of bugs: null pointer exceptions and index out of bounds, results show that M3V is effective in predicting repair operators, achieving over 70\% accuracy on both types of bugs.}

\revise{
In 2022, Xu~\etal~\cite{xu2022m3v} introduce M3V, a new multi-modal multi-view graph-based context embedding approach to predict repair operators for buggy Java code.
Different from previous studies performing patch generation and validation, M3V focuses on repair operator selection.
M3V first applies a GNN with multi-view graph-based context structure embedding to capture data and control dependencies. 
M3V also employs a tree-LSTM model with tree-based context signature embedding for capturing high-level semantics. 
The evaluation experiment is conducted on 20 open-source Java projects with two common types of bugs: null pointer exceptions and index out of bounds.
The results show that M3V is effective in predicting repair operators, achieving 71.45\%$\sim$75.60\% accuracy on both types of bugs, highlighting the future of context embedding in APR.
}

%% file: sec/sec_syntax.tex
\subsubsection{Syntax error repair}

Most existing learning-based APR techniques usually expect that the programs under repair are syntactically correct and these techniques are not applicable for syntax errors.
Novice programmers are more likely to make syntax errors (\eg replacing a ``$*$'' with an ``$x$'') that make compilers fail.
Previous studies have indicated the long-term challenge from a wide range of syntax mistakes, consuming a lot of time for novices and their instructors.
Recently, the release of high-quality novice error data and the emergence of trustworthy deep learning models have raised the possibility of designing and training DL models to fix syntax errors automatically.

Now, we list and summarize the recent learning-based APR studies that focus on syntax errors as follows.

\input{APR/syntax/2017-AAAI-Deepfix}

\input{APR/syntax/2018-SANER-Syntax}
\input{APR/syntax/2018-ICSE-Neuro}

\input{APR/syntax/2019-FSE-DeepDelta}

\input{APR/syntax/2019-AAAI-Deep}

\input{APR/syntax/2020-ICPC-GGF}

\input{APR/syntax/2021-ICML-Break}

\input{APR/syntax/2021-ECML-SampleFix}

%% file: APR/syntax/2017-AAAI-Deepfix.tex
\revise{As early as 2017, Gupta~\etal~\cite{gupta2017deepfix} propose a sequence-based approach, DeepFix to fix common programming errors.
DeepFix is regarded as the first end-to-end solution using a sequence-to-sequence model for localizing and fixing errors.
In particular, DeepFix applies an RNN-based encoder-decoder with gated recurrent units (GRUs) to serve as the Seq2Seq model.
Beside, DeepFix attempts to fix multiple errors iteratively by repairing one bug each time and using an oracle to decide whether to accept the patch or not.
The evaluation experiment is conducted on 6971 C erroneous programs written by students for 93 different programming tasks in an introductory programming course.
More importantly, as the pioneering end-to-end sequence-based approach in this field, DeepFix demonstrates the potential of Seq2Seq models in fixing syntax errors and serves as a catalyst for follow-up works, detailed in the following paragraphs.
}

%% file: APR/syntax/2018-SANER-Syntax.tex
\revise{In 2018, different from DeepFix focusing on C program, Santos~\etal~\cite{santos2018syntax} propose to leverage language models for repairing syntax errors in Java programs. 
They compare n-gram with LSTM models trained on a large corpus of Java projects from GitHub about localizing bugs and repairing them. Besides, their methodology does not rely on buggy code from the same domain as the training data. Evaluation results show that their improved LSTM configuration outperforms n-gram considerably.
Thus, this tool can localize and suggest corrections for syntax errors, and it is especially useful to novice programmers.}

%% file: APR/syntax/2018-ICSE-Neuro.tex
\delete{Bhatia~\etal~\cite{bhatia2018neuro} propose an approach for repairing programs committed by students. 
They first apply an RNN to repair syntax errors and then formalize the problem of syntax corrections in programs as a token sequence prediction problem. 
Then they leverage the constraint-based technique to find minimal repairs for semantic correctness. 
This approach is then evaluated on a Python dataset and results show that their approach outperforms the n-gram baseline model, demonstrating the effectiveness of their system.}

\revise{Meanwhile, Bhatia~\etal~\cite{bhatia2018neuro} propose a Neuro-symbolic approach to repair programs committed by students based on neural networks with constraint-based reasoning. 
They first apply an RNN to repair syntax errors and then formalize the problem of syntax corrections in programs as a token sequence prediction problem. 
They further leverage the constraint-based technique to find minimal repairs for the semantic correctness of syntactically-fixed programs. 
This approach is then evaluated on a Python dataset and results show that their approach outperforms the n-gram baseline model, demonstrating the potential of RNNs with constraint-based reasoning in repair syntax errors.}

%% file: APR/syntax/2019-FSE-DeepDelta.tex
\delete{In 2019, Mesbah~\etal~\cite{mesbah2019deepdelta} propose DeepDelta to repair the most costly classes of build-time compilation failures in Java programs. They perform a large-scale study of compilation errors and collect a large dataset from logs in Google. They further classify different compilation errors and target repairing these errors following specific patterns learned from the AST diff files in the dataset. For the two most prevalent and costly classes of Java compilation errors: missing symbols and mismatched method signatures, evaluation results show that DeepDelta generates over half of the correct patches.}

\revise{
In 2019, unlike DeepFix targeting syntax errors in C from small student programs, Mesbah~\etal~\cite{mesbah2019deepdelta} propose DeepDelta to repair the most costly classes of build-time compilation failures in Java programs from real-world developer-written programs. 
They perform a large-scale study of compilation errors and collect a large dataset from logs in Google. They further classify different compilation errors and target repairing these errors following specific patterns learned from the AST diff files in the dataset.
DeepDelta is then evaluated on two most prevalent and costly classes of Java compilation errors: missing symbols and mismatched method signatures.
The experimental results demonstrate that DeepDelta generates over half of the correct patches for unseen compilation
errors.}

%% file: APR/syntax/2019-AAAI-Deep.tex
\delete{Gupta~\etal~\cite{gupta2019deep} propose RLAssist to address the problem of syntactic error repair in student programs. They leverage reinforcement learning and train the model using Asynchronous Advantage Actor-Critic (A3C)~\cite{mnih2016asynchronous}. A3C uses multiple asynchronous parallel
actor-learner threads to update a shared model, stabilizing the learning process by reducing the correlation of an agent’s experience. After they evaluate RLAssist on the C benchmark from~\cite{gupta2017deepfix}, results show that this model outperforms the APR tool DeepFix~\cite{gupta2017deepfix} without using any labeled data for training and can help novice programmers.}

\revise{
Meanwhile, different from DeepFix employing fully supervised learning, Gupta~\etal~\cite{gupta2019deep} propose RLAssist, a deep reinforcement learning-based technique to address the problem of syntactic error repair in student programs. 
RLAssist is able to learn syntactic error repair directly from raw source code through self-exploration,\ie without any supervision.
They leverage reinforcement learning and train the model using Asynchronous Advantage Actor-Critic (A3C)~\cite{mnih2016asynchronous}.
A3C uses multiple asynchronous parallel
actor-learner threads to update a shared model, stabilizing the learning process by reducing the correlation of an agent’s experience. After they evaluate RLAssist on the C benchmark from DeepFix~\cite{gupta2017deepfix}, results show that this model outperforms the APR tool DeepFix~\cite{gupta2017deepfix} without using any labeled data for training, showing the potential to help novice programmers.
}

%% file: APR/syntax/2020-ICPC-GGF.tex
\delete{Wu~\etal~\cite{wu2020ggf} propose a deep supervise learning model, Graph-based Grammar Fix (GGF), to localize and fix syntax errors. They first parse the erroneous code into ASTs. Since the parser may crash in the parsing process due to syntax errors, they create a so-called sub-AST and build the graph based on it. To tackle the problem of isolated points and some error edges in the generated graph, they treat the code snippet as a mixture of token sequences and graphs. Thus, GGF utilizes a mixture of the GRU and the GGNN as the encoder module and a token replacement mechanism as the decoder module. The evaluation shows that GGF is able to fix 50.12\% of the erroneous code, outperforming DeepFix~\cite{gupta2017deepfix}.
Besides, the ablation study proves that the architecture used in GGF is quite helpful for the programming language syntax error correction task.}

\revise{
In 2020, unlike most existing techniques that use Seq2Seq models, Wu~\etal~\cite{wu2020ggf} propose GGF, a graph-based eep supervised learning model to localize and fix syntax errors.
They first parse the erroneous code into ASTs. Since the parser may crash in the parsing process due to syntax errors, they create a so-called sub-AST and build the graph based on it. To tackle the problem of isolated points and some error edges in the generated graph, they treat the code snippet as a mixture of token sequences and graphs. Thus, GGF utilizes a mixture of the GRU and the GGNN as the encoder module and a token replacement mechanism as the decoder module. The evaluation shows that GGF is able to fix 50.12\% of the erroneous code, outperforming DeepFix~\cite{gupta2017deepfix}.
Besides, the ablation study proves that the architecture used in GGF is quite helpful for the programming language syntax error correction task.
}

%% file: APR/syntax/2021-ICML-Break.tex
\delete{Existing work often applies heuristics on generating buggy code to construct buggy-fixed pairs. Such synthetically-generated data may not improve the model and generate low-quality patches.
Yasunaga~\etal~\cite{yasunaga2021break} propose Break-It-Fix-It (BIFI), a novel APR tool to address this problem. 
They first try to train a breaker with real-world buggy-fixed pairs to generate more realistic. 
They also leverage correct paired data to train the fixer. 
BIFI does not simply collect data, it is also capable of turning raw unlabeled data into usable paired data with the help of a critic. 
They then evaluate this approach on both Python and C language benchmarks and it outperforms \revise{previous}\delete{state-of-the-art} APR tools, \revise{such as DeepFix~\cite{gupta2017deepfix}}.}

\revise{
However, most of the existing APR techniques employ supervised learning to train repair models with labeled bug-fixing datasets, and their performance may be limited by the quantity and quality of labeled data.
In 2021, Yasunaga~\etal~\cite{yasunaga2021break} propose an unsupervised learning-based approach, Break-It-Fix-It (BIFI) to fix syntax errors.
BIFI first uses a fixer to generate patched code for buggy code and uses a critic to check the patched code.
BIFI then trains a breaker with real-world bug-fixing code pairs to generate more realistic buggy code.
Different from previous approaches, BIFI is capable of turning raw unlabeled data into usable paired data with the help of a critic, which is then used to train the fixer continuously.
The experimental results on both Python and C language benchmarks show that BIFI outperforms the previous repair approach DeepFix~\cite{gupta2017deepfix}.
}

%% file: APR/syntax/2021-ECML-SampleFix.tex
\delete{Hajipour~\etal~\cite{hajipour2021samplefix} propose an efficient method to fix common programming errors by learning the distribution over potential patches. To encourage the model to generate diverse fixes even with a limited number of samples, they propose a novel regularizer that aims to increase the distance between the two closest candidate fixes.
They prove that this approach is capable of generating multiple diverse fixes with different functionalities \revise{for 65\% of repaired programs}. 
After evaluating the approach on real-world datasets, they show that this approach outperforms \delete{DeepFix}\revise{previous APR tools such as DeepFix~\cite{gupta2017deepfix} and RLAssist~\cite{gupta2019deep}}.}

\revise{
At the same time, considering that previous approaches (\eg DeepFix~\cite{gupta2017deepfix}) usually ignore the true intend of the programmer during the patch generation process, Hajipour~\etal~\cite{hajipour2021samplefix} propose SampleFix, an efficient method to fix common programming errors by learning the distribution over potential patches. To encourage the model to generate diverse fixes even with a limited number of samples, they propose a novel regularizer that aims to increase the distance between the two closest candidate fixes.
They prove that this approach is capable of generating multiple diverse fixes with different functionalities for 65\% of repaired programs. 
After evaluating the approach on real-world datasets, they show that this approach outperforms previous APR tools such as DeepFix~\cite{gupta2017deepfix} and RLAssist~\cite{gupta2019deep}.}

%% file: sec/sec_vul.tex
\subsubsection{Security vulnerability repair}

Software vulnerability generally refers to the security flaws in the concrete implementation of hardware, software, or protocols.
Malicious attackers can exploit unresolved security vulnerabilities to get access to the system without authorization or even paralyze the system.
Such vulnerabilities open a range of threats to cyber security, resulting in severe economic damage and fatal consequences.
For example, the \textit{Log4Shell} vulnerability (CVE-2021-44228) from Apache Log4j library\footnote{\url{https://logging.apache.org/log4j/2.x/}} allows attackers to run arbitrary code on any affected system\footnote{\url{https://www.ftc.gov/policy/advocacy-research/tech-at-ftc/2022/01/ftc-warns-companies-remediate-log4j-security-vulnerability}} and is widely recognized as the most severe vulnerability in the last decade (\eg 93\% of the cloud enterprise environment are vulnerable to \textit{Log4Shell}\footnote{\url{https://www.wiz.io/blog/10-days-later-enterprises\\-halfway-through-patching-log4shell}}).
Nowadays, the number of exposed security vulnerabilities recorded by the National Vulnerability Database
(NVD)\footnote{\url{https://www.nist.gov/}} has been increasing at a striking speed, affecting millions of software systems annually.

However, it is incredibly time-consuming and labor-intensive for security experts to repair such security vulnerabilities manually due to the strikingly increasing number of detected vulnerabilities and the complexity of modern software systems~\cite{zhang2022program,gao2021beyond}.
For example, previous studies report that the average time for repairing severe vulnerabilities is 256 days\footnote{\url{https://www.securitymagazine.com/articles/95929-average-time-to-fix-severe-vulnerabilities-is-256-days}} and the life spans of 50\% of vulnerabilities even exceed 438 days~\cite{li2017large}.
It is incredibly time-critical to patch reported security vulnerabilities, as a belated vulnerability repair could expose software systems to attack~\cite{liu2020large,li2021sysevr}, posing enormous risks to millions of users around the globe and costing billions of dollars in financial losses~\cite{kshetri2006simple}.
Given the potentially disastrous effect when software vulnerabilities are exploited, a mass of learning-based studies has recently been conducted on automated software vulnerability repair~\cite{fu2022vulrepair,chen2022neural}.

We list and summarize the recent learning-based vulnerability repair studies in detail as follows.

\input{APR/vul/2017-ESORICS-VuRLE}

\input{APR/vul/2018-NeurIPS-Learning}

\input{APR/vul/2022-TSE-Neural}

\input{APR/vul/2022-TSE-SeqTrans}

\input{APR/vul/2022-ESE-SPVF}

%% file: APR/vul/2017-ESORICS-VuRLE.tex
\delete{Ma~\etal~\cite{ma2017vurle} introduce a novel tool, VuRLE, to automatically detect and repair vulnerabilities in Java programs. 
In the learning phase, it generates templates by analyzing edits from repair examples. First, it extracts edit blocks by performing AST diff. Then, it compares each edit block with the other edit blocks, and produces groups of similar edit blocks. Finally, for each edit group, VuRLE generates a repair template for each pair of edit blocks that are adjacent to each other. 
In the repairing phase, VuRLE detects and repairs vulnerabilities by selecting the most appropriate template. It applies repair templates in order of their matching score until it detects no redundant code.
Evaluation results on real-world vulnerabilities show that VuRLE outperforms another APR tool in fixing vulnerabilities.}

\revise{
As early as 2017, Ma~\etal~\cite{ma2017vurle} introduce a learning-based vulnerability repair tool, VuRLE, to automatically detect and repair vulnerabilities in Java programs. 
In the learning phase, it generates templates by analyzing edits from repair examples. First, it extracts edit blocks by performing AST diff. Then, it compares each edit block with the other edit blocks, and produces groups of similar edit blocks. Finally, for each edit group, VuRLE generates a repair template for each pair of edit blocks that are adjacent to each other. 
In the repairing phase, VuRLE detects and repairs vulnerabilities by selecting the most appropriate template. It applies repair templates in order of their matching score until it detects no redundant code.
Evaluation results on real-world vulnerabilities show that VuRLE successfully repaired 101 vulnerabilities, achieving an accuracy of 55.19\%.
}

%% file: APR/vul/2018-NeurIPS-Learning.tex
\delete{Harer~\etal~\cite{harer2018learning} propose a GAN-based approach to train an NMT model for learning to automatically repair the source code containing security vulnerabilities.
They apply an NMT model as the generator and employ two novel generator loss functions instead of the traditional negative likelihood loss.
They also design a discriminator to distinguish the output generated by the NMT model and oracle output.
This approach can be used in the absence of paired bug-fixing datasets, thus reducing the requirements of datasets.
The authors evaluate the proposed approach on the SATE IV dataset and prove the promising results in fixing vulnerabilities.
They also demonstrate the proposed approach can be applicable to other tasks, such as grammatical error correction.}

\revise{In 2018, to get rid of the previous work's dependence on labeled datasets, Harer~\etal~\cite{harer2018learning} propose a GAN-based approach to automatically repair security vulnerabilities based on adversarial learning without requiring labeled code samples.
They first apply an NMT model as the generator and employ two novel generator loss functions instead of the traditional negative likelihood loss.
They then design a discriminator to distinguish the output generated by the NMT model and oracle output.
This approach can be used in the absence of paired bug-fixing datasets, thus reducing the requirements of datasets.
The authors evaluate the proposed approach on the SATE IV dataset and prove the promising results in fixing vulnerabilities compared with the original Seq2Seq model.
Besides, the proposed approach is proven to be applicable to other repair tasks, such as grammatical error correction.}

%% file: APR/vul/2022-TSE-Neural.tex
\revise{In 2022, based on the transformer and transfer learning,} Chen~\etal~\cite{chen2022neural} propose VRepair, a learning-based approach to repair security vulnerabilities \delete{based on the transformer and transfer learning}. 
VRepair is first trained on a large bug-fixing dataset and is then transferred to a relatively small vulnerability-fixing dataset. 
VRepair uses a transformer neural network model to generate potential patches that are likely to be correct based on the training data.
The results show that VRepair trained on a bug-fixing dataset already fix some vulnerabilities.
Besides, they demonstrate the knowledge learned from the program repair task can be transferred to the vulnerability repair task.
In particular, VRepair with the transfer learning gains a better repair performance than that only trained on a vulnerability-fixing or bug-fixing dataset.

%% file: APR/vul/2022-TSE-SeqTrans.tex
\revise{Different from VRepair focusing on C code}, Chi~\etal~\cite{chi2022seqtrans} propose SeqTrans, a learning-based \delete{appraoach}\revise{approach} to provide suggestions for automatically repairing Java vulnerability.
SeqTrans first uses Gumtree to search for differences between different commits and then traverses the whole AST to label the variables.
SeqTrans then traverses up the leaf nodes, localizes the statement with vulnerability and generates code change pairs, which is fed into the NMT model.
As SeqTrans requires a massive amount of training data, SeqTrans is first trained on a bug-fixing dataset (\ie source domain) and fine-tuned on a vulnerability-fixing dataset (\ie target domain).
SeqTrans is proven to achieve better repair accuracy than existing techniques (\eg SequenceR) and performs very well in certain kinds of vulnerabilities (\eg CWE-287).

%% file: APR/vul/2022-ESE-SPVF.tex
\delete{
Zhou~\etal~\cite{zhou2022spvf} propose a novel approach SFVP for automatically fixing vulnerabilities based on the attention-mechanism model. 
SPVF first extracts the security properties from descriptions of the vulnerabilities (\eg CWE category).
SPVF then designs the pointer generator network to combine the AST representation and the security properties.
The authors evaluate SPVF on two public C/C++ and Python vulnerability-fixing datasets and results show that it outperforms state-of-the-art SeqTrans~\cite{chi2022seqtrans}. 
}

\revise{
However, previous approaches~\cite{chi2022seqtrans,chen2022neural} usually only consider source code while ignoring the valuable vulnerability characteristics.
Zhou~\etal~\cite{zhou2022spvf} propose an attention-based approach SFVP for automatically fixing vulnerabilities by capturing the security property.
SPVF first extracts the security properties from NL descriptions of the vulnerabilities (\eg CWE category).
SPVF then designs the pointer generator network to combine the AST representation and the security properties.
The authors evaluate SPVF on two public C/C++ and Python vulnerability-fixing datasets and results show that it outperforms existing vulnerability repair technique SeqTrans~\cite{chi2022seqtrans}. 
}

%% file: sec/sec_programming.tex
\subsubsection{Programming error repair}

\revise{With the emergence of programming competition websites (\eg LeetCode), developers frequently submit solutions, resulting in a vast amount of source code.
A portion of solutions contain flaws that prevent developers from solving the programming challenges successfully. 
These programming flaws are usually simple types of errors, \eg fail to compile and execute due to syntax errors, or pass the corresponding test cases due to semantic errors.
In the following, we discuss and summarize existing individual learning-based APR techniques that focus on programming errors in detail.}

\delete{
Since previous works fail to parse ASTs for student programs with syntax errors, Bhatia~\etal~\cite{bhatia2016automated} present a technique to apply RNN for repairing syntax errors in student programs. 
They first train the model with syntactically correct programs. Then, they query the trained model with student submissions with syntax errors and feed the model with the prefix token sequence. Finally, the model would predict suffix tokens and repair the syntax error. 
Evaluation on a dataset obtained from a MOOC course shows that this approach can provide automated feedback on syntax errors for students.}

\revise{
As early as 2016, since previous works fail to parse ASTs for student programs with syntax errors, Bhatia~\etal~\cite{bhatia2016automated} present a technique to apply RNN for repairing syntax errors in student programs. 
They first train the model with syntactically correct programs. Then, they query the trained model with student submissions with syntax errors and feed the model with the prefix token sequence. Finally, the model would predict suffix tokens and repair the syntax error. 
Evaluation on a dataset obtained from a MOOC course shows that this approach outperforms the baseline models (\eg RNN and LSTM with different configurations) 
 and can provide automated feedback on syntax errors for students.}

\delete{
Wang~\etal~\cite{wang2018dynamic} present dynamic program embeddings that learn from runtime execution traces to predict error patterns that students would make in their online programming submissions. They define three program embedding models: 1) variable trace model to obtain a sequence of variables; 2) state trace model to embed each program state as a numerical vector and feed all program state embeddings as a sequence to another RNN encoder; 3) dependency enforcement model to combine the advantages of the previous two approaches.
They have proved that dynamic embeddings overcome critical problems with syntax-based program representations and outperform other syntactic program embeddings.}

\revise{
In 2017, considering previous approaches focusing on static program representation, Wang~\etal~\cite{wang2018dynamic} present dynamic program embeddings that learn from runtime execution traces to predict error patterns that students would make in their online programming submissions.
They define three program embedding models: 1) variable trace model to obtain a sequence of variables; 2) state trace model to embed each program state as a numerical vector and feed all program state embeddings as a sequence to another RNN encoder; 3) dependency enforcement model to combine the advantages of the previous two approaches.
They conduct experiments to prove htat dynamic embeddings overcome critical problems with syntax-based program representations and significantly outperform the syntactic program embeddings based on token sequences and AST.}
At the same time, on top of dynamic program embeddings, Wang~\etal~\cite{wang2018search} further propose Sarfgen, a high-level data-driven framework to fix student-submitted programs for introductory programming exercises. They develop novel program embeddings and the associated distance metric to efficiently and precisely identify similar programs and compute program alignment. They also conduct an extensive evaluation of Sarfgen on thousands of student submissions on 17 different programming exercises from the Microsoft DEV204-.1x edX course and the Microsoft CodeHunt platform. Results show that Sarfgen is effective and it improves existing systems automation, capability, and scalability.

\delete{Ahmed et al. \cite{ahmed2018compilation} introduce TRACER to generate targeted repairs for novice programmers in C programs. They leverage buggy student programs in Prutor and conduct experiments on single-line and multi-line bugs. TRACER first localizes the buggy line, then abstracts the program, and finally converts it into fixed code. Evaluation on the dataset collected from IIT-K shows that TRACER achieves high accuracy and student-friendliness of the repair.}

\revise{
In 2018, with the aim of offering more informative error messages that would aid programmers in easier diagnoses, Ahmed~\etal~\cite{ahmed2018compilation} introduce TRACER to generate targeted repairs for novice programmers in C programming courses. 
They leverage buggy student programs in Prutor and conduct experiments on single-line and multi-line bugs. TRACER first localizes the buggy line, then abstracts the program, and finally converts it into fixed code. Evaluation on the dataset collected from IIT-K shows that TRACER achieves high accuracy up to 68\% and outperforms DeepFix~\cite{gupta2017deepfix}, proving to be a student-friendliness repair tool.}

\delete{Zhang~\etal~\cite{zhang2022repairing} propose a novel approach on top of to repair both semantic and syntactic bugs in Python programs. 
They apply a large language model trained on code (LLMC). 
They also leverage multimodal prompts, iterative querying, test-case-based few-shot selection, and program chunking to repair bugs in students' committed programs. 
Zhang~\etal~implement it in MMAPR through Codex as LLMC. 
After evaluating MMAPR on real student programs and another baseline (BIFI and Refactory), it outperforms other state-of-art tools.}

\revise{
In 2022, different from previous works using a basic transformer, Zhang~\etal~\cite{zhang2022repairing} propose MMAPR, a pre-trained model-based repair approach to repair both semantic and syntactic bugs in Python programming assignments. 
MMAPR applies a large language model trained on code (\ie Codex) for introductory Python programming assignments. 
In particular, MMAPR leverages multimodal prompts, iterative querying, test-case-based few-shot selection, and program chunking to repair bugs in students' committed programs. 
The experimental results on 18 assignments demonstrate MMAPR is able to outperform a transformer-based syntax repair tool BIFI~\cite{yasunaga2021break}, and a re-factoring-based semantics repair tool Refactory~\cite{hu2019re}.}

%% file: sec/sec_domain_old.tex
\delete{
\textbf{Domain Repair}

In the learning-based APR field, semantic error (\ie test-triggering error) has attracted considerable attention from researchers, which is the most general application of repair techniques discussed in the previous sections.
Such errors usually refer to any case where the actual program behavior is not expected by developers.
A living review~\cite{monperrus2020living} summarizes and categorizes existing APR techniques into different repair scenarios during software development, including static errors, concurrency errors, \etc
In this section, we discuss some other scenarios where learning-based repair techniques are usually applied, listed as follows.
}

\delete{
\textbf{Vulnerability Repair}}

\delete{Software vulnerability generally refers to the security flaws in the concrete implementation of hardware, software, or protocols.
Malicious attackers can exploit unresolved security vulnerabilities to get access to the system without authorization or even paralyze the system.
Such vulnerabilities open a range of threats to cyber security, resulting in severe economic damage and fatal consequences.
For example, the \textit{Log4Shell} vulnerability (CVE-2021-44228) from Apache Log4j library allows attackers to run arbitrary code on any affected system\} and is widely recognized as the most severe vulnerability in the last decade (\eg 93\% of the cloud enterprise environment are vulnerable to \textit{Log4Shell}).
Nowadays, the number of exposed security vulnerabilities recorded by the National Vulnerability Database
(NVD) has been increasing at a striking speed, affecting millions of software systems annually.}

\delete{
However, it is incredibly time-consuming and labor-intensive for security experts to repair such security vulnerabilities manually due to the strikingly increasing number of detected vulnerabilities and the complexity of modern software systems~\cite{zhang2022program,gao2021beyond}.
For example, previous studies report that the average time for repairing severe vulnerabilities is 256 days and the life spans of 50\% of vulnerabilities even exceed 438 days~\cite{li2017large}.
It is incredibly time-critical to patch reported security vulnerabilities, as a belated vulnerability repair could expose software systems to attack~\cite{liu2020large,li2021sysevr}, posing enormous risks to millions of users around the globe and costing billions of dollars in financial losses~\cite{kshetri2006simple}.
Given the potentially disastrous effect when software vulnerabilities are exploited, a mass of learning-based studies has recently been conducted on automated software vulnerability repair.}

\delete{We list the recent learning-based vulnerability repair studies in detail as follows.}

\delete{Ma~\etal~\cite{ma2017vurle} introduce a novel tool, VuRLE, to automatically detect and repair vulnerabilities in Java programs. 
In the learning phase, it generates templates by analyzing edits from repair examples. First, it extracts edit blocks by performing AST diff. Then, it compares each edit block with the other edit blocks, and produces groups of similar edit blocks. Finally, for each edit group, VuRLE generates a repair template for each pair of edit blocks that are adjacent to each other. 
In the repairing phase, VuRLE detects and repairs vulnerabilities by selecting the most appropriate template. It applies repair templates in order of their matching score until it detects no redundant code.
Evaluation results on real-world vulnerabilities show that VuRLE outperforms another APR tool in fixing vulnerabilities.}

\delete{
As early as 2017, Ma~\etal~\cite{ma2017vurle} introduce a learning-based vulnerability repair tool, VuRLE, to automatically detect and repair vulnerabilities in Java programs. 
In the learning phase, it generates templates by analyzing edits from repair examples. First, it extracts edit blocks by performing AST diff. Then, it compares each edit block with the other edit blocks, and produces groups of similar edit blocks. Finally, for each edit group, VuRLE generates a repair template for each pair of edit blocks that are adjacent to each other. 
In the repairing phase, VuRLE detects and repairs vulnerabilities by selecting the most appropriate template. It applies repair templates in order of their matching score until it detects no redundant code.
Evaluation results on real-world vulnerabilities show that VuRLE successfully repaired 101 vulnerabilities, achieving an accuracy of 55.19\%.
}

\delete{Harer~\etal~\cite{harer2018learning} propose a GAN-based approach to train an NMT model for learning to automatically repair the source code containing security vulnerabilities.
They apply an NMT model as the generator and employ two novel generator loss functions instead of the traditional negative likelihood loss.
They also design a discriminator to distinguish the output generated by the NMT model and oracle output.
This approach can be used in the absence of paired bug-fixing datasets, thus reducing the requirements of datasets.
The authors evaluate the proposed approach on the SATE IV dataset and prove the promising results in fixing vulnerabilities.
They also demonstrate the proposed approach can be applicable to other tasks, such as grammatical error correction.}

\delete{\delete{Based on the transformer and transfer learning,} Chen~\etal~\cite{chen2022neural} propose VRepair, a learning-based approach to repair security vulnerabilities \delete{based on the transformer and transfer learning}. 
VRepair is first trained on a large bug-fixing dataset and is then transferred to a relatively small vulnerability-fixing dataset. 
VRepair uses a transformer neural network model to generate potential patches that are likely to be correct based on the training data.
The results show that VRepair trained on a bug-fixing dataset already fix some vulnerabilities.
Besides, they demonstrate the knowledge learned from the program repair task can be transferred to the vulnerability repair task.
In particular, VRepair with the transfer learning gains a better repair performance than that only trained on a vulnerability-fixing or bug-fixing dataset.}

\delete{\delete{Unlike VRepair employing a basic transformer}, Fu~\etal~\cite{fu2022vulrepair} propose VulRepair, a T5-based automated vulnerability repair technique based on subword tokenization and pre-training components.
They compare VulRepair with two competitive baseline approaches, VRepair and CodeBERT on a C benchmark -- CVEFixes.
Besides, they analyze the impact of adopted components (\ie tokenization and pre-training) and conduct an ablation study to investigate the contribution of each component. 
The results show that VulRepair outperforms \delete{other state-of-the-art vulnerability repair techniques} \delete{the vulnerability repair technique VRepair~\cite{chen2022neural}} and it is capable of repairing the Top-10 most dangerous CWEs.
}

\delete{
\delete{Different from VRepair focusing on C code}, Chi~\etal~\cite{chi2022seqtrans} propose SeqTrans, a learning-based \delete{appraoach}\delete{approach} to provide suggestions for automatically repairing Java vulnerability.
SeqTrans first uses Gumtree to search for differences between different commits and then traverses the whole AST to label the variables.
SeqTrans then traverses up the leaf nodes, localizes the statement with vulnerability and generates code change pairs, which is fed into the NMT model.
As SeqTrans requires a massive amount of training data, SeqTrans is first trained on a bug-fixing dataset (\ie source domain) and fine-tuned on a vulnerability-fixing dataset (\ie target domain).
SeqTrans is proven to achieve better repair accuracy than existing techniques (\eg SequenceR) and performs very well in certain kinds of vulnerabilities (\eg CWE-287).}

\delete{
Zhou~\etal~\cite{zhou2022spvf} propose a novel approach SFVP for automatically fixing vulnerabilities based on the attention-mechanism model. 
SPVF first extracts the security properties from descriptions of the vulnerabilities (\eg CWE category).
SPVF then designs the pointer generator network to combine the AST representation and the security properties.
The authors evaluate SPVF on two public C/C++ and Python vulnerability-fixing datasets and results show that it outperforms state-of-the-art SeqTrans~\cite{chi2022seqtrans}. 
}

\delete{
Zhou~\etal~\cite{zhou2022spvf} propose a novel approach SFVP for automatically fixing vulnerabilities based on the attention-mechanism model. 
SPVF first extracts the security properties from descriptions of the vulnerabilities (\eg CWE category).
SPVF then designs the pointer generator network to combine the AST representation and the security properties.
The authors evaluate SPVF on two public C/C++ and Python vulnerability-fixing datasets and results show that it outperforms existing vulnerability repair technique SeqTrans~\cite{chi2022seqtrans}. 
}

\delete{To overcome the shortcomings of learning-based APR techniques, Huang~\etal~\cite{huang2022repairing} propose to apply large pre-trained models for vulnerability repair.
They compare the performance of CodeBERT and GraphCodeBERT on a C/C++ vulnerability dataset with five CWE types. 
They discover that GraphCodeBERT with a data flow graph is significantly better than CodeBERT without documenting code dependencies. 
They also demonstrate that such pre-trained models outperform learning-based APR techniques (\eg CoCoNut ~\cite{lutellier2020coconut} and DLFix~\cite{li2020dlfix}) and more data-dependent features (\eg data flow and control flow) will help to repair more complex vulnerabilities.}

\delete{
\textbf{Syntax Errors}}

\delete{
In the learning-based APR field, semantic errors (\ie test-triggering errors) have attracted considerable attention.
Such errors usually refer to any case where the actual program behavior is not expected by developers.
Existing learning-based APR techniques usually expect that the programs under repair are syntactically correct and these techniques are not applicable for syntax errors.
Novice programmers are more likely to make syntax errors (\eg replacing a ``$*$'' with an ``$x$'') that make compilers fail.
Previous studies have indicated the long-term challenge from a wide range of syntax mistakes, consuming a lot of time for novices and their instructors.
Recently, the release of high-quality novice error data and the emergence of trustworthy deep learning models have raised the possibility of designing and training DL models to fix syntax errors automatically.}

\delete{Now, we list the recent learning-based APR studies that focus on syntax errors as follows.}

\delete{
Previous works mainly focus on logical errors and assume that the program should be compiled successfully.
Gupta~\etal~\cite{gupta2017deepfix}. propose DeepFix to fix multiple errors in a program interactively.
They apply the RNN encoder-decoder to serve as the seq2seq network.
To implement the iterative repair for multiple errors, they decide to repair one bug each time. An oracle is applied after the decoder to decide whether the program needs further repair after one patch is generated.
Although this approach is only evaluated on C program language written by students in an introductory programming course, the result shows that DeepFix can fix a variety of errors and has potential in other languages.
}

\delete{
As early as 2017, Gupta~\etal~\cite{gupta2017deepfix} propose a sequence-based approach, DeepFix to fix common programming errors.
DeepFix is regarded as the first end-to-end solution using a sequence-to-sequence model for localizing and fixing errors.
In particular, DeepFix applies an RNN-based encoder-decoder with gated recurrent units (GRUs) to serve as the Seq2Seq model.
Beside, DeepFix attempts to fix multiple errors iteratively by repairing one bug each time and using an oracle to decide whether to accept the patch or not.
The evaluation experiment is conducted on 6971 C erroneous programs written by students for 93 different programming tasks in an introductory programming course.
The result shows that DeepFix is able to fix 1881 (27\%) programs completely and 1338 (19\%) programs partially.
More importantly, as the first end-to-end sequence-based approach, DeepFix demonstrates the potential of Seq2Seq models in fixing software bugs.}

\delete{
Santos~\etal~\cite{santos2018syntax} propose to leverage language models for repairing syntax errors in Java programs. They compare n-gram with LSTM models trained on a large corpus of Java projects from Github about localizing bugs and repairing them. Besides, their methodology does not rely on buggy code from the same domain as the training data. Evaluation results show that this tool can localize and suggest corrections for syntax errors, and it is especially useful to novice programmers.}

\delete{
In 2018, Santos~\etal~\cite{santos2018syntax} propose to leverage language models for repairing syntax errors in Java programs. They compare n-gram with LSTM models trained on a large corpus of Java projects from GitHub about localizing bugs and repairing them. Besides, their methodology does not rely on buggy code from the same domain as the training data. Evaluation results show that their improved LSTM configuration outperforms n-gram considerably.
Thus, this tool can localize and suggest corrections for syntax errors, and it is especially useful to novice programmers.}

\delete{
Bhatia~\etal~\cite{bhatia2018neuro} propose a novel approach for repairing programs committed by students. 
They first apply an RNN to repair syntax errors and then formalize the problem of syntax corrections in programs as a token sequence prediction problem. 
Then they leverage the constrain-based technique to find minimal repairs for semantic correctness. 
This approach is then evaluated on a Python dataset and results demonstrate the effectiveness of their system.
}

\delete{ 
Bhatia~\etal~\cite{bhatia2018neuro} propose an approach for repairing programs committed by students. 
They first apply an RNN to repair syntax errors and then formalize the problem of syntax corrections in programs as a token sequence prediction problem. 
Then they leverage the constraint-based technique to find minimal repairs for semantic correctness. 
This approach is then evaluated on a Python dataset and results show that their approach outperforms the n-gram baseline model, demonstrating the effectiveness of their system.
}

\delete{
Mesbah~\etal~\cite{mesbah2019deepdelta} propose DeepDelta to repair the most costly classes of build-time compilation failures in Java programs. They perform a large-scale study of compilation errors and collect a large dataset from logs in Google. They further classify different compilation errors and target repairing these errors following specific patterns learned from the AST diff files in the dataset. For the two most prevalent and costly classes of Java compilation errors: missing symbols and mismatched method signatures, evaluation results show that DeepDelta generates over half of the correct patches.}

\delete{
In 2019, Mesbah~\etal~\cite{mesbah2019deepdelta} propose DeepDelta to repair the most costly classes of build-time compilation failures in Java programs. They perform a large-scale study of compilation errors and collect a large dataset from logs in Google. They further classify different compilation errors and target repairing these errors following specific patterns learned from the AST diff files in the dataset. For the two most prevalent and costly classes of Java compilation errors: missing symbols and mismatched method signatures, evaluation results show that DeepDelta generates over half of the correct patches.
}

\delete{
Gupta~\etal~\cite{gupta2019deep} propose RLAssist to address the problem of syntactic error repair in student programs. They leverage reinforcement learning and train the model using Asynchronous Advantage Actor-Critic (A3C)~\cite{mnih2016asynchronous}. A3C uses multiple asynchronous parallel
actor-learner threads to update a shared model, stabilizing the learning process by reducing the correlation of an agent’s experience. After they evaluate RLAssist on the C benchmark from~\cite{gupta2017deepfix}, results show that this model outperforms the APR tool DeepFix without using any labeled data for training and can help novice programmers.}

\delete{
Gupta~\etal~\cite{gupta2019deep} propose RLAssist to address the problem of syntactic error repair in student programs. They leverage reinforcement learning and train the model using Asynchronous Advantage Actor-Critic (A3C)~\cite{mnih2016asynchronous}. A3C uses multiple asynchronous parallel
actor-learner threads to update a shared model, stabilizing the learning process by reducing the correlation of an agent’s experience. After they evaluate RLAssist on the C benchmark from~\cite{gupta2017deepfix}, results show that this model outperforms the APR tool DeepFix~\cite{gupta2017deepfix} without using any labeled data for training and can help novice programmers.
}

\delete{
Wu~\etal~\cite{wu2020ggf} propose a novel deep supervise learning model, Graph-based Grammar Fix (GGF), to localize and fix syntax errors. They first parse the erroneous code into ASTs. Since the parser may crash in the parsing process due to syntax errors, they create so-called sub-AST and build the graph based on it. To tackle the problem of isolated points and some error edges in the generated graph, they treat the code snippet as a mixture of token sequences and graphs. Thus, GGF utilizes a mixture of the GRU and the GGNN as the encoder module and a token replacement mechanism as the decoder module. The evaluation shows that the architecture used in GGF is quite helpful for the programming language syntax error correction task.}

\delete{
Wu~\etal~\cite{wu2020ggf} propose a deep supervise learning model, Graph-based Grammar Fix (GGF), to localize and fix syntax errors. They first parse the erroneous code into ASTs. Since the parser may crash in the parsing process due to syntax errors, they create a so-called sub-AST and build the graph based on it. To tackle the problem of isolated points and some error edges in the generated graph, they treat the code snippet as a mixture of token sequences and graphs. Thus, GGF utilizes a mixture of the GRU and the GGNN as the encoder module and a token replacement mechanism as the decoder module. The evaluation shows that GGF is able to fix 50.12\% of the erroneous code, outperforming DeepFix~\cite{gupta2017deepfix}.
Besides, the ablation study proves that the architecture used in GGF is quite helpful for the programming language syntax error correction task.}

\delete{
Ahmed~\etal~\cite{ahmed2022synshine} propose SynShine, a machine learning-based approach to fix syntax errors in Java programs. 
They apply a three-stage syntax repair tool: BlockFix for recovering block structure, LineFix for fixing line errors, and UnkFix for recovering unknown tokens. 
SynShine leverages RoBERTa pre-training, uses compiler errors, and generates fixes using multi-label classification.
After being evaluated on a dataset collected from the Blackbox repository, SynShine outperforms other state-of-the-art tools on different token ranges. 
They have also integrated SynShine with the VSCode IDE for public usage.}

\delete{
Ahmed~\etal~\cite{ahmed2021learning} introduce an indirect-supervision approach to leverage GitHub code to create massive amounts of "incorrect-fixed" training pairs for model training. They apply a two-stage approach, with two different neural networks for learning to model block nesting structure and code fragments. This approach performs better on the large and diverse BlackBox dataset than previous work (\eg DeepFix~\cite{gupta2017deepfix}). It also performs well for StackOverflow fragment parsing and helps fix errors for novice programmers.
}

\delete{Existing work often applies heuristics on generating buggy code to construct buggy-fixed pairs. Such synthetically-generated data may not improve the model and generate low-quality patches.
Yasunaga~\etal~\cite{yasunaga2021break} propose Break-It-Fix-It (BIFI), a novel APR tool to address this problem. 
They first try to train a breaker with real-world buggy-fixed pairs to generate more realistic. 
They also leverage correct paired data to train the fixer. 
BIFI does not simply collect data, it is also capable of turning raw unlabeled data into usable paired data with the help of a critic. 
They then evaluate this approach on both Python and C language benchmarks and it outperforms \delete{previous}\delete{state-of-the-art} APR tools, \delete{such as DeepFix~\cite{gupta2017deepfix}}. }

\delete{Berabi~\etal~\cite{berabi2021tfix} present TFix to deal with text-to-text prediction problems. They fine-tune a pre-trained T5 model to generate JavaScript fixes on datasets extracted from GitHub by themselves. By feeding the model with line context and fine-tuning it according to various error types, they obtain multiple fine-tuned T5 models. The evaluation shows that TFix is able to generate 67\% of correct patches, significantly outperforming SequenceR~\cite{chen2019sequencer} and CoCoNut~\cite{lutellier2020coconut}.}

\delete{
Ahmed~\etal~\cite{ahmed2021learning} introduce an indirect-supervision approach to leverage GitHub code to create massive amounts of "incorrect-fixed" training pairs for model training. They apply a two-stage approach, with two different neural networks for learning to model block nesting structure and code fragments. This approach performs better on the large and diverse BlackBox dataset than previous work. It also performs well for StackOverflow fragment parsing and helps fix errors for novice programmers.}

\delete{Hajipour~\etal~\cite{hajipour2021samplefix} propose an efficient method to fix common programming errors by learning the distribution over potential patches. To encourage the model to generate diverse fixes even with a limited number of samples, they propose a novel regularizer that aims to increase the distance between the two closest candidate fixes.
They prove that this approach is capable of generating multiple diverse fixes with different functionalities \delete{for 65\% of repaired programs}. 
After evaluating the approach on real-world datasets, they show that this approach outperforms DeepFix.}

\delete{
In 2022, Ahmed~\etal~\cite{ahmed2022synshine} propose SynShine, a machine learning-based approach to fix syntax errors in Java programs. 
They apply a three-stage syntax repair tool: BlockFix for recovering block structure, LineFix for fixing line errors, and UnkFix for recovering unknown tokens. 
SynShine leverages RoBERTa-based pre-training and information from compiler errors to generate fixes using multi-label classification.
After being evaluated on a dataset collected from the Blackbox repository, SynShine outperforms other state-of-the-art tools on different token ranges.
Importantly, they have also integrated SynShine with the VSCode IDE for public usage, showing the practical value in a real-world development environment.}

\delete{\textbf{Programming Assignments}}

\delete{
Now, we list the recent learning-based APR studies that focus on programming assignments as follows.}

\delete{
Since previous works fail to parse ASTs for student programs with syntax errors, Bhatia~\etal~\cite{bhatia2016automated} present a technique to apply RNN for repairing syntax errors in student programs. 
They first train the model with syntactically correct programs. Then, they query the trained model with student submissions with syntax errors and feed the model with the prefix token sequence. Finally, the model would predict suffix tokens and repair the syntax error. 
Evaluation on a dataset obtained from a MOOC course shows that this approach can provide automated feedback on syntax errors for students.}

\delete{
Since previous works fail to parse ASTs for student programs with syntax errors, Bhatia~\etal~\cite{bhatia2016automated} present a technique to apply RNN for repairing syntax errors in student programs. 
They first train the model with syntactically correct programs. Then, they query the trained model with student submissions with syntax errors and feed the model with the prefix token sequence. Finally, the model would predict suffix tokens and repair the syntax error. 
Evaluation on a dataset obtained from a MOOC course shows that this approach outperforms the baseline models (\eg RNN and LSTM with different configurations) and can provide automated feedback on syntax errors for students.}

\delete{
Wang~\etal~\cite{wang2018dynamic} present dynamic program embeddings that learn from runtime execution traces to predict error patterns that students would make in their online programming submissions. They define three program embedding models: 1) variable trace model to obtain a sequence of variables; 2) state trace model to embed each program state as a numerical vector and feed all program state embeddings as a sequence to another RNN encoder; 3) dependency enforcement model to combine the advantages of the previous two approaches.
They conduct experiments to prove htat dynamic embeddings overcome critical problems with syntax-based program representations and significantly outperform the syntactic program embeddings based on token sequences and AST.}

\delete{Ahmed~\etal~\cite{ahmed2018compilation} introduce TRACER to generate targeted repairs for novice programmers in C programs. They leverage buggy student programs in Prutor and conduct experiments on single-line and multi-line bugs. TRACER first localizes the buggy line, then abstracts the program, and finally converts it into fixed code. Evaluation on the dataset collected from IIT-K shows that TRACER achieves high accuracy \delete{up to 68\% and outperforms DeepFix~\cite{gupta2017deepfix}, proving to be a} student-friendliness \delete{of the repair}\delete{repair tool}.}

\delete{Wang~\etal~\cite{wang2018search} propose Sarfgen, a high-level data-driven framework to fix student-submitted programs for introductory programming exercises. They develop novel program embeddings and the associated distance metric to efficiently and precisely identify similar programs and compute program alignment. They also conduct an extensive evaluation of Sarfgen on thousands of student submissions on 17 different programming exercises from the Microsoft DEV204-.1x edX course and the Microsoft CodeHunt platform. Results show that Sarfgen is effective and it improves existing systems automation, capability, and scalability.}

\delete{
Wang~\etal~\cite{wang2018dynamic} present dynamic program embeddings that learn from runtime execution traces to predict error patterns that students would make in their online programming submissions. They define three program embedding models: 1) variable trace model to obtain a sequence of variables; 2) state trace model to embed each program state as a numerical vector and feed all program state embeddings as a sequence to another RNN encoder; 3) dependency enforcement model to combine the advantages of the previous two approaches.
They have proved that dynamic embeddings overcome critical problems with syntax-based program representations and outperform other syntactic program embeddings.}

\delete{
Zhang~\etal~\cite{zhang2022repairing} propose a novel approach  on top of to repair both semantic and syntactic bugs in Python programs. 
They apply a large language model trained on code (LLMC). 
They also leverage multimodal prompts, iterative querying, test-case-based few-shot selection, and program chunking to repair bugs in students' committed programs. 
Zhang~\etal~implement it in MMAPR through Codex as LLMC. 
After evaluating MMAPR on real student programs and another baseline (BIFI and Refactory), it outperforms other state-of-art tools.}

\delete{
Zhang~\etal~\cite{zhang2022repairing} propose an APR approach MMAPR to repair both semantic and syntactic bugs in Python programs. 
MMAPR applies a large language model trained on code (\ie Codex) for introductory Python programming assignments. 
In particular, MMAPR leverages multimodal prompts, iterative querying, test-case-based few-shot selection, and program chunking to repair bugs in students' committed programs. 
The experimental results on 18 assignments demonstrate MMAPR is able to outperform a \delete{state-of-the-art} transformer-based syntax repair tool BIFI~\cite{yasunaga2021break}, and a \delete{state-of-the-art} re-factoring-based semantics repair tool Refactory~\cite{hu2019re}.}

\delete{
Ahmed~\etal~\cite{ahmed2022verifix} propose Verifix as a tool to provide feedback for students in programming tasks. They first align a student-submitted program with a reference solution in terms of control flow. Then the variables of the two programs are automatically aligned. After that, they turn a verification problem into a MaxSMT problem if the above verification attempt fails. 
The solution to the MaxSMT problem leads to a minimal repair. Ahmed~\etal~are the first to espouse verified repair for general-purpose programming education and their approach produces small-sized verified patches as feedback which can be used by struggling students with high confidence.}

\delete{\textbf{Other Domains}}

\delete{\textit{Programming Contests.}
Fan~\etal~\cite{fan2022improving} propose to leverage a large pre-trained model to repair buggy programs generated by Codex model. They collect a Java dataset (LMDefects) from the LeetCode contest containing different levels of tasks. They then compare the performance of Codex-e \delete{(\eg Codex edit mode)} and traditional APR tools (TBar and Recoder) on this dataset. Results show that existing APR techniques (TBar and Recoder) do not perform well at fixing bugs in auto-generated programs. 
Fan~\etal also define three strategies as instructions: 1) fix bugs in the program;2) \delete{fx}\delete{fix} line N; 3) and fix statement S to evaluate their approach. They find that Codex-e performs well under proper instructions.}

\delete{\textit{Program Synthesis.}
Gupta~\etal~\cite{gupta2020synthesize} present SED as a framework incorporating synthesis, execution, and debugging stages. 
SED applies a synthesizer that employs greedy decoding to generate buggy programs for training and the debugger is fed with synthesized bugs as well as execution results. 
SED is then evaluated on the Karel benchmark and it outperforms other beam search techniques.
}

\delete{\textit{Nonidiomatic Snippets.}}
\delete{Szalontai~\etal~\cite{szalontai2021detecting} present a novel algorithm to localize and substitute non-idiomatic code snippets in Python programs. 
They apply a feed-forward and two RNNs to accomplish the task. 
Once the code snippet is localized, the model classifies the type of the nonidiomatic pattern and extracts the key variables. 
Finally, the model substitutes the code snippet with a cleaner and more performant alternative. This model is evaluated on a Python dataset and it achieves good results.}
\delete{Szalontai~\etal~\cite{szalontai2021detecting} propose to localize and substitute non-idiomatic code snippets in Python programs with DL models.
In particular, they first localize the nonidiomatic snippets by considering the localization problem as a sequence tagging problem with LSTM networks.
They then train a feedforward network-based classifier to determine the type of the nonidiomatic pattern and extract the key variables.
Finally, the non-idiomatic code snippet is substituted with a cleaner and more performant alternative.
The experimental results on a Python dataset demonstrate the proposed method is able to achieve 58.92\% precision and 47.27\% recall, highlighting promising performance in fixing nonidiomatic snippets in Python source code.}

%% file: tab/tab_LLM.tex
\begin{table}[htbp]
\small
  \centering
  \caption{A summary and comparison of existing pre-trained model-based APR techniques}
    \begin{tabular}{rllll}
    \toprule
    Year  & Study  & Type  & Model  & Language \\
    \midrule
    2021  & TFix~\cite{berabi2021tfix} & Syntax & T5    & JS \\
    2022  & CIRCLE~\cite{yuan2022circle} & Semantic & T5 & Java,C,JS,Python \\
    2022  & VulRepair~\cite{fu2022vulrepair} & Vulnerability & CodeT5 & C \\
    2022  & AlphaRepair~\cite{xia2022less} & Semantic & CodeBERT & Java,Python \\
    2022  & SYNSHINE~\cite{ahmed2022synshine} & Syntax & RoBERTa & Java \\
    \bottomrule
    \end{tabular}%
  \label{tab:llm}%
\end{table}%

%% file: APR/syntax/2021-PMLR-TFix.tex
\delete{
Berabi~\etal~\cite{berabi2021tfix} present TFix to deal with text-to-text prediction problems. They fine-tune a pre-trained T5 model to generate JavaScript fixes on datasets extracted from GitHub by themselves. By feeding the model with line context and fine-tuning it according to various error types, they obtain multiple fine-tuned T5 models. The evaluation shows that TFix is able to generate 67\% of correct patches, significantly outperforming SequenceR~\cite{chen2019sequencer} and CoCoNut~\cite{lutellier2020coconut}.}

\revise{
In 2021, inspired by the pre-trained T5 model~\cite{raffel2019t5} that converts all text-based language problems into a text-to-text format in the NLP field, Berabi~\etal~\cite{berabi2021tfix} formulates the problem of fixing coding errors as a text-to-text prediction task and propose TFix, a T5-based approach to fix syntax errors. 
They fine-tune a pre-trained T5 model to generate JavaScript fixes on datasets extracted from GitHub by themselves. 
By feeding the model with line context and fine-tuning it according to various error types, they obtain multiple fine-tuned T5 models. 
The evaluation shows that TFix is able to generate 67\% of correct patches, significantly outperforming SequenceR~\cite{chen2019sequencer} and CoCoNut~\cite{lutellier2020coconut}.
}

%% file: APR/syntax/2022-TSE-SYNSHINE.tex
\delete{In 2022, Ahmed~\etal~\cite{ahmed2022synshine} propose SynShine, a machine learning-based approach to fix syntax errors in Java programs. 
They apply a three-stage syntax repair tool: BlockFix for recovering block structure, LineFix for fixing line errors, and UnkFix for recovering unknown tokens. 
SynShine leverages RoBERTa-based pre-training and information from compiler errors to generate fixes using multi-label classification.
After being evaluated on a dataset collected from the Blackbox repository, SynShine outperforms other state-of-the-art tools on different token ranges., DeepFix~\cite{gupta2017deepfix} and SequenceR~\cite{chen2019sequencer} on different token ranges.
Importantly, they have also integrated SynShine with the VSCode IDE for public usage, showing the practical value in a real-world development environment.}

\revise{
In 2022, to address the issue of previous works not performing well on large programs, Ahmed~\etal~\cite{ahmed2022synshine} propose SynShine, a learning-based approach to fix syntax errors in Java programs by innovatively using the diagnostics from a compiler and exploiting the ability to pre-train model. 
SynShine first applies a three-stage syntax repair workflow, \ie BlockFix for recovering block structure, LineFix for fixing line errors, and UnkFix for recovering unknown tokens. 
SynShine then leverages RoBERTa-based pre-training and information from compiler errors to generate fixes using multi-label classification.
The experimental results on the Blackbox dataset show that SynShine outperforms previous repair approaches, \eg DeepFix~\cite{gupta2017deepfix} and SequenceR~\cite{chen2019sequencer} on different token ranges.
Importantly, they have also integrated SynShine with the VSCode IDE for public usage, showing the practical value in a real-world development environment.
}

%% file: APR/semantic/sequence/2021-ASE-MODIT.tex
\delete{
Chakraborty~\etal~\cite{chakraborty2021multi} present MODIT, a novel multi-modal NMT-based tool, to automatically generate fixes for buggy code. They leverage three modalities of information: edit location, edit code context, and commit messages (natural language guidance from the developer).
They conduct many experiments and the evaluation shows that, through pre-training, MODIT improves the ability to generate patches. 
Also, leveraging additional modalities of information could benefit the source code repairing.}

\revise{
Considering that previous NMT-based repair approaches fail to consider DL descriptions about the code context, Chakraborty~\etal~\cite{chakraborty2021multi} present MODIT, a multi-modal pre-trained model-based approach, to automatically generate fixes for buggy code. 
They leverage three modalities of information during training: edit location, edit code context, and commit messages (\ie natural language guidance from the developer).
They then employ the pre-trained PLBART model as the as the starting point to train MODIT.
The experimental results show that MODIT generates 29.99\% correct patches for the BFP-small dataset~\cite{tufano2019empirical}, outperforming CodeBERT by 15.12\%, GraphCodeBERT by 16.82\% and CodeGPT 5.49\%.
Similarly, 23.02\% of patches generated by MODIT on the BFP-medium dataset are correct and the improvement against the three pre-trained models reaches 34.38\%, 25.72\%, and 30.50\%, respectively.}

%% file: APR/semantic/sequence/2022-ISSTA-CIRCLE.tex
\revise{
Existing learning-based APR techniques can only generate patches for a single programming language and most of them are developed offline.
In 2022, Yuan~\etal~\cite{yuan2022circle} propose CIRCLE, a T5-based APR technique targeting multiple programming languages with continual learning.
CIRCLE first employs a pre-trained model as a repair skeleton, then designs a prompt template to bridge the gap between pre-trained tasks and program repair.
To further strengthen the continual learning ability, CIRCLE applies a difficulty-based rehearsal method to achieve lifelong learning without access to the entire historical data and an elastic regularization to resolve catastrophic forgetting.
Finally, to perform the multi-lingual repair, CIRCLE designs a simple but effective re-repairing mechanism to eliminate incorrectly generated patches caused by multiple programming languages.
The experimental results on five benchmarks across four programming languages (\ie C, JAVA, JavaScript, and
Python) show that CIRCLE is able to achieve outperform various previous learning-based APR approaches, such as CoCoNut~\cite{lutellier2020coconut}, DLFix~\cite{li2020dlfix} and CURE~\cite{jiang2021cure}.
More importantly, the results demonstrate the potential of CIRCLE in repairing multiple programming language bugs with a single repair model in the continual learning setting.
}

%% file: APR/semantic/sequence/2022-FSE-AlphaRepair.tex
\revise{
Different from previous learning-based APR approaches (\eg CIRCLE~\cite{yuan2022circle}) that heavily rely on large numbers of high-quality bug-fixing code pairs, in 2022, Xia~\etal~\cite{xia2022less} introduce AlphaRepair as a cloze-style APR tool to directly query a pre-trained model for generating patches. 
They apply the newly pre-trained CodeBERT as an example under zero-shot learning settings. They try to mask the buggy line in the source code with different templates or strategies and feed the whole source code into the model with the buggy line as a ``comment". Then with a large number of patches this model generated, they propose probabilistic patch ranking to determine top-$k$ plausible patches. After evaluating this technique on both Java and Python benchmarks, it outperforms other APR tools \revise{(\eg Recoder~\cite{zhu2021syntax}, DLFix~\cite{li2020dlfix} and TBar~\cite{liu2019tbar})} and proves that a pre-trained model with no fine-tuning is feasible.}

%% file: APR/vul/2022-FSE-VulRepair.tex
\delete{Fu et al. \cite{fu2022vulrepair} propose VulRepair, a T5-based automated vulnerability repair technique based on subword tokenization and pre-training components.
They compare VulRepair with two competitive baseline approaches, VRepair and CodeBERT on a C benchmark -- CVEFixes.
Besides, they analyze the impact of adopted components (i.e., tokenization and pre-training) and conduct an ablation study to investigate the contribution of each component. 
The results show that VulRepair outperforms other state-of-the-art vulnerability repair techniques and it is capable of repairing the Top-10 most dangerous CWEs.}

\revise{Unlike VRepair employing a basic transformer, Fu~\etal~\cite{fu2022vulrepair} propose VulRepair, a T5-based automated vulnerability repair technique based on subword tokenization and pre-training components.
They compare VulRepair with two competitive baseline approaches, VRepair and CodeBERT on a C benchmark CVEFixes.
Besides, they analyze the impact of adopted components (\ie tokenization and pre-training) and conduct an ablation study to investigate the contribution of each component. 
The results show that VulRepair outperforms the previous repair technique VRepair~\cite{chen2022neural} and it is capable of repairing the Top-10 most dangerous CWEs.}

%% file: tab/tab_dataset.tex
\begin{table}[htbp]
\scriptsize
  \centering
  \caption{\revise{Detailed information on collected datasets in existing learning-based APR studies.}}
    \begin{tabular}{c|c|ccccc|m{1.7cm}<{\centering}| c}
    \toprule
    \textbf{ID}    & \textbf{Name}  & \textbf{Language} & \textbf{\#Bugs}  & \textbf{Test Suite} & \textbf{Training} & \textbf{Testing} & \textbf{Techniques} & \textbf{URL}\\
    \midrule
    1     & \textbf{Bears~\cite{madeiral2019bears}} & Java  & 251   & \ding{52}   & \ding{52}   & \ding{52}   &~\cite{li2022improving,ye2022neural}& \href{https://github.com/bears-bugs/bears-benchmark}{\revise{Link}}\\
    2     & \textbf{BFP medium~\cite{tufano2019empirical}} & Java  & 65454 & \ding{56}    & \ding{52}   & \ding{52}   &~\cite{chakraborty2021multi,chi2022seqtrans,drain2021generating,nguyen2021graphix,hu2022fix,tang2021grasp,tang2021grammar,tufano2019empirical,yao2022bug}& \href{https://sites.google.com/view/learning-fixes/data}{\revise{Link}}\\
    3     & \textbf{BFP small~\cite{tufano2019empirical}} & Java  & 58350 & \ding{56}    & \ding{52}   & \ding{52}   &~\cite{chakraborty2021multi,chi2022seqtrans,drain2021generating,nguyen2021graphix,hu2022fix,tang2021grasp,tang2021grammar,tufano2019empirical,yao2022bug}& \href{https://sites.google.com/view/learning-fixes/data}{\revise{Link}}\\
    4     & \textbf{BigFix~\cite{li2020dlfix}} & Java  & 1.824 M & \ding{56}    & \ding{52}   & \ding{52}   &~\cite{li2020dlfix,li2022dear}& \href{https://github.com/OOPSLA-2019-BugDetection/OOPSLA-2019-BugDetection}{\revise{Link}}\\
    5     & \textbf{Bugs2Fix~\cite{lu2021codexglue}} & Java  & 92849 & \ding{56}    & \ding{52}   & \ding{52}   &~\cite{chen2019sequencer,connor2022can}& \href{https://github.com/microsoft/CodeXGLUE/tree/main/Code-Code/code-refinement}{\revise{Link}}\\
    6     & \textbf{Bugs.jar~\cite{saha2018bugs}} & Java  & 1158  & \ding{52}   & \ding{52}   & \ding{52}   &~\cite{li2020dlfix,ye2022neural,tian2020evaluating}& \href{https://github.com/bugs-dot-jar/bugs-dot-jar}{\revise{Link}}\\
    7     & \textbf{Code-Change-Data~\cite{Chakraborty2020codit}} & Java  & 44372 & \ding{56}    & \ding{52}   & \ding{52}   &~\cite{Chakraborty2020codit}& \href{https://drive.google.com/file/d/1wSl_SN17tbATqlhNMO0O7sEkH9gqJ9Vr/edit}{\revise{Link}}\\
    8     & \textbf{CodeXGlue~\cite{lu2021codexglue}} & Java  & 122 K & \ding{56}    & \ding{56}    & \ding{52}   &~\cite{connor2022can}& \href{https://github.com/microsoft/CodeXGLUE}{\revise{Link}}\\
    9     & \textbf{CodRep~\cite{chen2018codrep}} & Java  & 58069 & \ding{56}    & \ding{52}   & \ding{52}   &~\cite{chen2019sequencer,ye2022neural}& \href{https://github.com/ASSERT-KTH/CodRep}{\revise{Link}}\\
    10    & \textbf{CPatMiner~\cite{nguyen2019graph}} & Java  & 44 K  & \ding{56}    & \ding{52}   & \ding{52}   &~\cite{li2022dear}& \href{https://nguyenhoan.github.io/CPatMiner/patterns.html}{\revise{Link}}\\
    11    & \textbf{DeepRepair~\cite{white2019sorting}} & Java  & 374   & \ding{56}    & \ding{52}   & \ding{56}    &~\cite{white2019sorting}& \href{https://sites.google.com/view/deeprepair}{\revise{Link}}\\
    12    & \textbf{Defects4J~\cite{just2014defects4j}} & Java  & 835   & \ding{52}   & \ding{52}   & \ding{52}   &~\cite{Chakraborty2020codit,chen2022program,li2022improving,li2022dear,lutellier2019encore,ye2022selfapr,mamatha2022oapr,tang2021grasp,tian2022predicting,tian2020evaluating,wang2022leveraging}& \href{https://github.com/rjust/defects4j}{\revise{Link}}\\
    13    & \textbf{Function-SStuBs4J~\cite{ni2022defect}} & Java  & 21047 & \ding{56}    & \ding{52}   & \ding{52}   &~\cite{ni2022defect}& \href{https://zenodo.org/record/5353354#.YS8iYtMzZhE}{\revise{Link}}\\
    14    & \textbf{IntroClassJava~\cite{durieux2016introclassjava}} & Java  & 998   & \ding{52}   & \ding{52}   & \ding{52}   &~\cite{chen2022program,zhu2021syntax}& \href{https://github.com/Spirals-Team/IntroClassJava}{\revise{Link}}\\
    15    & \textbf{Java-med~\cite{alon2019code2vec}} & Java  & 7454  & \ding{56}    & \ding{52}   & \ding{56}    &~\cite{kang2022language}& \href{https://s3.amazonaws.com/code2vec/data/java14m_data.tar.gz}{\revise{Link}}\\
    16    & \textbf{ManySStuBs4J large~\cite{karampatsis2020often}} & Java  & 63923 & \ding{56}    & \ding{52}   & \ding{52}   &~\cite{mashhadi2021applying}& \href{https://zenodo.org/record/3653444}{\revise{Link}}\\
    17    & \textbf{ManySStuBs4J small~\cite{karampatsis2020often}} & Java  & 10231 & \ding{56}    & \ding{52}   & \ding{52}   &~\cite{mashhadi2021applying,tian2020evaluating}& \href{https://zenodo.org/record/3653444}{\revise{Link}}\\
    18    & \textbf{MegaDiff~\cite{monperrus2021megadiff}} & Java  & 663029 & \ding{56}    & \ding{52}   & \ding{56}    &~\cite{chen2022neural}& \href{https://zenodo.org/record/5013515}{\revise{Link}}\\
    19    & \textbf{Ponta\newrevise{~\etal}~\cite{ponta2019manually}} & Java  & 624   & \ding{56}    & \ding{52}   & \ding{52}   &~\cite{chi2022seqtrans}& \href{https://github.com/SAP/project-kb/tree/main/MSR2019}{\revise{Link}}\\
    20    & \textbf{Pull-Request-Data~\cite{tufano2019learning}} & Java  & 10666 & \ding{56}    & \ding{52}   & \ding{52}   &~\cite{Chakraborty2020codit,tufano2019learning}& \href{https://sites.google.com/view/learning-codechanges/data}{\revise{Link}}\\
    21    & \textbf{Ratchet~\cite{hata2018learning}} & Java  & 35 K  & \ding{56}    & \ding{52}   & \ding{52}   &~\cite{hata2018learning}& \href{https://github.com/hideakihata/NMTbasedCorrectivePatchGenerationDataset}{\revise{Link}}\\
    22    & \textbf{Recoder~\cite{zhu2021syntax}} & Java  & 103585 & \ding{56}    & \ding{52}   & \ding{56}    &~\cite{zhu2021syntax}& \href{https://drive.google.com/drive/folders/1ECNX98qj9FMdRT2MXOUY6aQ6-sNT0b_a}{\revise{Link}}\\
    23    & \textbf{TRANSFER~\cite{meng2022improving}} & Java  & 408091 & \ding{56}    & \ding{52}   & \ding{56}    &~\cite{meng2022improving}& \href{https://mega.nz/file/u0wQzRga#Q2BHCuRD2aW_61vshVbcxj-ObYh2cyGhqOAmAXNn-T0}{\revise{Link}}\\
    24    & \textbf{\newdelete{Mesbah}\newrevise{Deepdelta}~\cite{mesbah2019deepdelta}} & Java  & 4.8 M & \ding{56}    & \ding{52}   & \ding{52}   &~\cite{mesbah2019deepdelta}& {\revise{N.A.}}\\
    \midrule
    25    & \textbf{\newdelete{AOJ}\newrevise{Rahman~\etal}~\cite{rahman2021bidirectional}}   & C     & 2482  & \ding{56}    & \ding{52}   & \ding{52}   &~\cite{rahman2021bidirectional}& \href{https://onlinejudge.u-aizu.ac.jp/home}{\revise{N.A.}}\\
    26    & \textbf{Big-Vul~\cite{fan2020ac}} & C     & 3745  & \ding{56}    & \ding{52}   & \ding{52}   &~\cite{chen2022neural}~\cite{zhou2022spvf}& \href{https://github.com/ZeoVan/MSR_20_Code_Vulnerability_CSV_Dataset}{\revise{Link}}\\
    27    & \textbf{Code4Bench~\cite{majd2019code4bench}} & C     & 25 K  & \ding{52}   & \ding{52}   & \ding{52}   &~\cite{valueian2022siturepair}& \href{https://zenodo.org/record/2582968}{\revise{Link}}\\
    28    & \textbf{\newdelete{CodeHunt}\newrevise{Wang~\etal}~\cite{tillmann2014code}} & C     & 195 K & \ding{52}   & \ding{52}   & \ding{52}   &~\cite{wang2018dynamic}& \href{https://www.microsoft.com/en-us/research/project/code-hunt/}{\revise{N.A.}}\\
    29    & \textbf{CVEFixes~\cite{bhandari2021cvefixes}} & C     & 8482  & \ding{56}    & \ding{52}   & \ding{52}   &~\cite{fu2022vulrepair,wang2018search}& \href{https://github.com/secureIT-project/CVEfixes}{\revise{Link}}\\
    30    & \textbf{DeepFix~\cite{gupta2017deepfix}} & C     & 6971  & \ding{52}   & \ding{52}   & \ding{52}   &~\cite{gupta2017deepfix,gupta2019deep,yasunaga2020graph,yasunaga2021break,hajipour2021samplefix,joshi2022repair}& \href{https://bitbucket.org/iiscseal/deepfix}{\newrevise{Link\tablefootnote{\newrevise{The link provided in the original paper has expired. We find and provide a new link on Bitbucket, which is also maintained by the authors.}}}}\\
    31    & \textbf{ManyBugs~\cite{le2015manybugs}} & C     & 185   & \ding{52}   & \ding{52}   & \ding{52}   &~\cite{lutellier2020coconut,yuan2022circle,wang2022leveraging}& \href{https://repairbenchmarks.cs.umass.edu/ManyBugs/}{\revise{Link}}\\
    32    & \textbf{Prophet~\cite{long2016automatic}} & C     & 69    & \ding{52}   & \ding{52}   & \ding{52}   &~\cite{long2016automatic,lutellier2019encore}& \href{https://github.com/epicosy/prophet}{\revise{Link}}\\
    33    & \textbf{Prutor~\cite{das2016prutor}} & C     & 6971  & \ding{52}   & \ding{52}   & \ding{52}   &~\cite{mohan2019automatic,yang2020applying}& \href{https://www.cse.iitk.ac.in/users/karkare/prutor/}{\revise{Link}}\\
    \midrule
    34    & \textbf{BugAID~\cite{hanam2016discovering}} & JS    & 105133 & \ding{56}    & \ding{52}   & \ding{52}   &~\cite{lutellier2019encore,lutellier2020coconut,wang2022leveraging,yuan2022circle}& \href{http://salt.ece.ubc.ca/software/bugaid/}{\revise{Link}}\\
    35    & \textbf{BugsJS~\cite{gyimesi2019bugsjs}} & JS    & 453   & \ding{52}   & \ding{52}   & \ding{52}   &~\cite{lajko2022towards}& \href{https://bugsjs.github.io/#nav-download}{\revise{Link}}\\
    36    & \textbf{HOPPITY~\cite{dinella2020hoppity}} & JS    & 363 K & \ding{56}    & \ding{52}   & \ding{52}   &~\cite{dinella2020hoppity}& \href{https://github.com/AI-nstein/hoppity}{\revise{Link}}\\
    37    & \textbf{KATANA~\cite{sintaha2023katana}} & JS    & 114 K & \ding{56}    & \ding{52}   & \ding{52}   &~\cite{sintaha2023katana}& \href{https://github.com/saltlab/Katana}{\revise{Link}}\\
    38    & \textbf{REPTORY~\cite{namavar2022controlled}} & JS    & 407 K & \ding{56}    & \ding{52}   & \ding{52}   &~\cite{namavar2022controlled}& \href{https://github.com/annon-reptory/reptory}{\revise{Link}}\\
    39    & \textbf{TFix~\cite{berabi2021tfix}}  & JS    & 100 K & \ding{56}    & \ding{52}   & \ding{52}   &~\cite{berabi2021tfix}& \href{https://github.com/eth-sri/TFix}{\revise{Link}}\\
    \midrule
    40    & \textbf{ETH Py150~\cite{raychev2016probabilistic}} & Python & 150 K & \ding{56}    & \ding{52}   & \ding{52}   &~\cite{hellendoorn2019global,richter2022can,vasic2019neural}& \href{https://www.sri.inf.ethz.ch/py150}{\revise{Link}}\\
    41    & \textbf{GitHub-Python~\cite{yasunaga2021break}} & Python & 3 M   & \ding{56}    & \ding{52}   & \ding{52}   &~\cite{yasunaga2021break}& \href{https://github.com/michiyasunaga/bifi}{\revise{Link}}\\
    42    & \textbf{\newdelete{Mester}\newrevise{Szalontai~\etal}~\cite{szalontai2021detecting}} & Python & 13 K  & \ding{56}    & \ding{52}   & \ding{52}   &~\cite{szalontai2021detecting}& {\revise{N.A.}}\\
    43    & \textbf{PyPIBugs~\cite{allamanis2021self}} & Python & 2374  & \ding{56}    & \ding{52}   & \ding{52}   &~\cite{allamanis2021self,richter2022can}& \href{https://github.com/microsoft/neurips21-self-supervised-bug-detection-and-repair}{\revise{Link}}\\
    44    & \textbf{SSB-9M~\cite{richter2022tssb}} & Python & 9 M   & \ding{56}    & \ding{52}   & \ding{56}    &~\cite{richter2022can}& \href{https://cedricrupb.github.io/TSSB3M/}{\revise{Link}}\\
    45    & \textbf{VUDENC~\cite{wartschinski2022vudenc}} & Python & 10 K  & \ding{56}    & \ding{52}   & \ding{52}   &~\cite{zhou2022spvf}& \href{https://zenodo.org/record/3559203#.XeRoytVG2Hs}{\revise{Link}}\\
    46    & \textbf{\newdelete{Chhatbar}\newrevise{Macer}~\cite{chhatbar2020macer}} & Python & 286   & \ding{52}   & \ding{56}    & \ding{52}   &~\cite{zhang2022repairing}& \href{https://github.com/purushottamkar/macer/}{\revise{Link}}\\
    \midrule
    47    & \textbf{SPoC~\cite{kulal2019spoc}}  & C++   & 18356 & \ding{52}   & \ding{52}   & \ding{52}   &~\cite{yasunaga2020graph}& \href{https://sumith1896.github.io/spoc/}{\revise{Link}}\\
    \midrule
    48    & \textbf{QuixBugs~\cite{lin2017quixbugs}} & Java,Python & 40    & \ding{52}   & \ding{52}   & \ding{52}   &~\cite{chen2022program,chen2022neural,drain2021deepdebug,wang2022leveraging,jiang2021cure,lutellier2019encore,lutellier2020coconut,tian2020evaluating,zhu2021syntax,yuan2022circle,xia2022less}& \href{https://github.com/jkoppel/QuixBugs}{\revise{Link}}\\
    49    & \textbf{DeepDebug~\cite{drain2021deepdebug}} & Java,Python & 523   & \ding{56}    & \ding{52}   & \ding{52}   &~\cite{drain2021deepdebug,drain2021generating}& {\revise{N.A.}}\\
    50    & \textbf{CoCoNut~\cite{lutellier2020coconut}} & \makecell{C,Java,JS,Python} & 24 M  & \ding{52}   & \ding{52}   & \ding{56}    &~\cite{jiang2021cure,lutellier2020coconut,yuan2022circle,ye2022neural}& \href{https://github.com/lin-tan/CoCoNut-Artifact}{\revise{Link}}\\
    51    & \textbf{CodeFlaws~\cite{tan2017codeflaws}} & C,Python & 3902  & \ding{52}   & \ding{52}   & \ding{52}   &~\cite{bohme2020human,lutellier2020coconut,yan2022crex}& \href{https://codeflaws.github.io/}{\revise{Link}}\\
    52    & \textbf{ENCORE~\cite{lutellier2019encore}} & \makecell{Java,JS,Python,C++} & 9.2 M & \ding{56}    & \ding{52}   & \ding{56}    &~\cite{lutellier2019encore}& {\revise{N.A.}}\\
    \bottomrule
    \end{tabular}%
  \label{tab:dataset}%

\end{table}%

%% file: tab/tab_empirical.tex
\begin{table}[htbp]
  \scriptsize
  \centering
  \caption{\revise{A summary and comparison of empirical studies in learning-based APR}}
    \begin{tabular}{ccccm{6cm}}
    \toprule
    Year  &  Study  & Scope  & Language & Description \\
    \midrule
    2019  & Tufano~\etal~\cite{tufano2019empirical} & Program Repair & Java  & the first empirical study to assess the feasibility of using NMT techniques for learning bug-fixing patches. \\

    2020  & Ding~\etal~\cite{ding2020patching} & Program Repair & Java  & investigate how language translation models perform in APR, specifically on the concept of “patching as translation”. \\

    2021  & Mashhadi~\etal~\cite{mashhadi2021applying} & Program Repair & Java & investigate the performance of CodeBERT in fixing bugs from the ManySStuBs4J banchmark. \\\

    2022  & Kolak~\etal~\cite{kolak2022patch} & Program Repair & Java & investigate the performance of the pre-trained model in fixing bugs from the QuixBugs benchmark. \\
    
    2022  & Namavar~\etal~\cite{namavar2022controlled} & Code Representation &  JavaScript & investigate the impact of code representations in APR with 21 models and 14 code representation methods. \\

    2022  & Xia~\etal~\cite{xia2022practical} & Program Repair & Java,Python,C & the first extensive study on directly applying nine pre-trained models for APR across three programming languages. \\

    2022  & Wang~\etal~\cite{wang2022attention} & Patch Correctness & Java & an extensive study of learning-based patch correctness assessment techniques on the Defects4J dataset. \\
        
    2022  & Kim~\etal~\cite{kim2022empirical} & Kotlin Repair & Kotlin & investigate the performance of the pre-trained model in fixing defects in the Samsung Kotlin projects. \\

    2022  & Huang~\etal~\cite{huang2022repairing} & Vulnerability Repair & C/C++ & a preliminary study to investigate the performance of pre-trained models in repairing security vulnerabilities. \\

    \bottomrule
    \end{tabular}%
  \label{tab:empirical}%
\end{table}%

%% file: APR/APCA/2022-arxiv-Attention.tex
Considering most existing APCA techniques evaluated on limited datasets, Wang~\etal~\cite{wang2022attention} conduct an extensive empirical study of patch correctness on Java programs. 
First, they collect a large-scale real-world dataset for patch correctness, containing 1,988 patches generated by the recent PraPR APR tool~\cite{ghanbari2019practical}. 
\revise{Then they revisit state-of-the-art APCA techniques on the new dataset, including static-based (\eg Anti-patterns:~\cite{tan2016anti}), dynamic-based (\eg PATCH-SIM~\cite{xiong2018identifying}), and learning-based (\eg ODS~\cite{ye2021automated}).}
\delete{Results show that learning-based techniques tend to suffer from the overfitting.}
\revise{Results show that learning-based APCA techniques tend to suffer from the dataset overfitting issue~\cite{wang2022attention}.
For example, the embedding-based techniques~\cite{tian2020evaluating} underperform on patches sourced from subjects outside the training set, thereby highlighting the need for cross-dataset evaluation in future learning-based APCA research.}
Besides, the performance of dynamic techniques significantly drops when encountering patches with more complicated changes.

%% file: APR/vul/2022-DSN-W-Repairing.tex
\revise{Meanwhile, to explore the real-world performance of pre-trained models for vulnerability repair, Huang~\etal~\cite{huang2022repairing} conduct a preliminary stucy to apply large pre-trained models for vulnerability repair.
They compare the performance of CodeBERT and GraphCodeBERT on a C/C++ vulnerability dataset with five CWE types. 
They discover that GraphCodeBERT with a data flow graph is significantly better than CodeBERT without documenting code dependencies. 
They also demonstrate that such pre-trained models outperform learning-based APR techniques (\eg CoCoNut ~\cite{lutellier2020coconut} and DLFix~\cite{li2020dlfix}) and more data-dependent features (\eg data flow and control flow) will help to repair more complex vulnerabilities.}

%% file: tab/tab_dl4tapr.tex
\begin{table}[htbp]
\footnotesize
  \centering
  \caption{\revise{A summary and comparison of APR studies combining traditional repair techniques and machine learning techniques}}
  
    \begin{tabular}{ccccm{6cm}}
    \toprule
    Year  & Approach & Base & Language  & Description \\
    \midrule
    2016  & Prophet~\cite{long2016automatic} & SPR  & C  & 
    Training a ranking model to assign a high probability to correct patches based on designed features.
    \\

    2017  & ACS~\cite{xiong2017precise} & N.A.  & Java & Inferring which predicates should be used with a given variable.
    \\
    
    2019  & DeepRepair~\cite{white2019sorting} & Astor & Java  &
    Learning to rank the repair ingredients based on code similarity with representation learning.
    \\
    
    2022  & LIANA~\cite{chen2022program} & RESTORE & Java   &   
    Employing a machine learning model to rank candidate patches based on their static (\eg the number of variables) and dynamic (the number of passing tests) features.
    \\
    
    2022  & TRANSFER~\cite{meng2022improving} & TBar & Java  & 
    Training a BiLSTM-based multi-classifier model to predict which fix template should be tried to repair one suspicious statement.
    \\
    
    2022  & ARJANMT~\cite{li2022improving} & ARJA  & Java  &  
    Employing a Seq2Seq model to generate patches as potential fix ingredients that are manipulated by a multi-objective evolutionary search algorithm.
    \\

    2022  & SituRepair~\cite{valueian2022siturepair} & N.A. & C  &   
    Training a machine-learning model to predict the types of bugs based on static features and apply modifications to the faulty program according to the types.
    \\
    
    \bottomrule
    \end{tabular}%
  \label{tab:dl4apr}%
\end{table}%

%% file: APR/2019-SANER-DeepRepair.tex
\revise{
In 2019, White~\etal~\cite{white2019sorting} propose DeepRepair to  intelligently select repair ingredients via deep learning code similarities.
In particular, DeepRepair is implemented on top of Astor~\cite{martinez2016astor}, a traditional heuristic-based APR approach, and consists of three phases, \ie  language recognition, machine learning, and program repair.
First, the language recognition phase processes the source code to create ASTs and maps the literal tokens to their respective type.
Second, the machine learning phase trains a neural network language model from the file-level corpus to  representations for each term, and then trains an encoder to encode arbitrary streams of embeddings.
Third, the program repair phase leverages the trained encoder to query and transform code snippets for patch generation.
In this step, DeepRepair sorts the repair ingredients based on code similarity and applies repair operators (“addition of statement” and “replacement of statement”) to repair the code snippet.
The experimental results on Defects4J demonstrate that  DeepRepair achieves comparable performance against jGenProg~\cite{martinez2017automatic} in terms of the number of plausible patches with a faster discovery speed of compilable ingredients.
More importantly, as the first approach to expand the fix space by transforming ingredients, DeepRepair generates some patches that cannot be generated by jGenProg, highlighting the differences between the nature of DeepRepair and jGenProg.
}

%% file: tab/tab_tool.tex
\setlength{\rotFPtop}{0pt plus 1fil}
\setlength{\rotFPbot}{0pt plus 1fil}

\begin{sidewaystable}[htbp]
  \centering
  \scriptsize
  \caption{Results on tool availability}
  \resizebox{\linewidth}{!}{
    \begin{tabular}{lcccccccm{6cm}}
    \toprule
    Tool  & Language & Dataset & Hosting Site & Link Accessibility & SA & DA & TA & URL \\
    \midrule
    Wang \etal~\cite{wang2018dynamic} & C & \revise{CodeHunt} & \delete{Github}\revise{GitHub} & valid & \ding{52}   & \ding{56}    & \ding{56}    & \url{https://github.com/keowang/dynamic-program-embedding} \\
    Tufano~\etal~\cite{tufano2019learning} & Java  &  \revise{Pull-Request-Data} & Google & valid & \ding{52}   & \ding{52}   & \ding{52}   & \url{https://sites.google.com/view/learning-codechanges} \\
    Tufano~\etal~\cite{tufano2019empirical} & Java  &  \revise{BFP-small, BFP-medium} & Google & valid & \ding{52}   & \ding{52}   & \ding{56}    & \url{https://sites.google.com/view/learning-fixes} \\
    RLAssitst~\cite{gupta2019deep} & C     & \revise{DeepFix} & bitbucket & valid & \ding{52}   & \ding{52}   & \ding{52}   & \url{https://bitbucket.org/iiscseal/rlassist} \\
    CoCoNut~\cite{lutellier2020coconut} & Java,C,Python,JS & \revise{\makecell{Defects4J, QuixBugs, ManyBugs,\\ CodeFlaw, BugAID}} & \delete{Github}\revise{GitHub} & valid & \ding{52}   & \ding{52}   & \ding{56}    & \url{https://github.com/lin-tan/CoCoNut-Artifact} \\
    DLFix~\cite{li2020dlfix} & Java  & \revise{Defects4J, Bugs.jar, BigFix} & \delete{Github}\revise{GitHub} & valid & \ding{52}   & \ding{52}   & \ding{56}    & \url{https://github.com/ICSE-2019-AUTOFIX/ICSE-2019-AUTOFIX} \\
    Hellendoorn \etal~\cite{hellendoorn2019global} & Python & \revise{ETH Py150} & \delete{Github}\revise{GitHub} & valid & \ding{52}   & \ding{52}   & \ding{52}   & \url{https://github.com/VHellendoorn/ICLR20-Great} \\
    DrRepair~\cite{yasunaga2020graph} & C,C++ & \revise{DeepFix, SPoC} & \delete{Github}\revise{GitHub} & valid & \ding{52}   & \ding{52}   & \ding{52}   & \url{https://github.com/michiyasunaga/DrRepair} \\
    Learn2Fix~\cite{bohme2020human} & Python & \revise{CodeFlaw} & \delete{Github}\revise{GitHub} & valid & \ding{52}   & \ding{52}   & \ding{52}   & \url{https://github.com/mboehme/learn2fix} \\
    Tian \etal~\cite{tian2020evaluating} & Java  & \revise{Defects4J, QuixBugs} & \delete{Github}\revise{GitHub} & valid & \ding{52}   & \ding{52}   & \ding{52}   & \url{https://github.com/TruX-DTF/DL4PatchCorrectness} \\
    BIFI~\cite{yasunaga2021break} & Python,C & \revise{GitHub-Python, DeepFix} & \delete{Github}\revise{GitHub} & valid & \ding{52}   & \ding{52}   & \ding{52}   & \url{https://github.com/michiyasunaga/bifi} \\
    Recoder~\cite{zhu2021syntax} & Java  & \revise{Defects4J, QuixBugs, IntroClassJava} & \delete{Github}\revise{GitHub} & valid & \ding{52}   & \ding{52}   & \ding{56}    & \url{https://github.com/pkuzqh/Recoder} \\
    SequenceR~\cite{chen2019sequencer} & Java  & \revise{Defects4J, Bugs2Fix, CodRep} & \delete{Github}\revise{GitHub} & valid & \ding{52}   & \ding{52}   & \ding{52}   & \url{https://github.com/kth/SequenceR} \\
    TFix~\cite{berabi2021tfix} & JS    & \revise{TFix} & \delete{Github}\revise{GitHub} & valid & \ding{52}   & \ding{52}   & \ding{52}   & \url{https://github.com/eth-sri/TFix} \\
    \textcolor[rgb]{0.141, 0.161, 0.184}{BugLab~\cite{allamanis2021self}} & Python & \revise{PyPIBug} & \delete{Github}\revise{GitHub} & valid & \ding{52}   & \ding{52}   & \ding{56}    & \url{https://github.com/microsoft/neurips21-self-supervised-bug-detection-and-repair} \\
    Reptory~\cite{namavar2022controlled} & JS    & \revise{REPTORY} & \delete{Github}\revise{GitHub} & valid & \ding{52}   & \ding{52}   & \ding{56}    & \url{https://github.com/annon-reptory/reptory} \\
    RewardRepair~\cite{ye2022neural} & Java  & \revise{Defects4J, QuixBugs, Bugs.jar} & \delete{Github}\revise{GitHub} & valid & \ding{52}   & \ding{52}   & \ding{52}   & \url{https://github.com/SophieHYe/RewardRepair} \\
 CodeBERT~\cite{mashhadi2021applying} & Java & \revise{ManySStuBs4J-small, ManySStuBs4J-large} & \delete{Github}\revise{GitHub} & valid & \ding{52} & \ding{52} & \ding{56} & \url{https://github.com/EhsanMashhadi/MSR2021-ProgramRepair} \\
    R-HERO~\cite{baudry2021software} &  &  \revise{Not Known} & \delete{Github}\revise{GitHub} & valid & \ding{56} & \ding{52} & \ding{56} &  \url{https://github.com/repairnator/open-science-repairnator/tree/master/data/2020-r-hero} \\
    Ahmed \etal~\cite{ahmed2021learning} & Java  &  \revise{Ahmed} & zenodo & valid & \ding{52}   & \ding{52}   & \ding{56}    & \url{https://doi.org/10.5281/zenodo.3374019} \\
    ODS~\cite{ye2021automated} & Java  & \revise{Defects4J, Bugs.jar, Bears} & \delete{Github}\revise{GitHub} & valid & \ding{52}   & \ding{52}   & \ding{52}   & \url{https://github.com/SophieHYe/ODSExperiment} \\
    CIRCLE~\cite{yuan2022circle} & Java,C,JS,Python & \revise{Defects4J, QuixBugs, ManyBugs, BugAID} & \delete{Github}\revise{GitHub} & valid & \ding{56}    & \ding{56}    & \ding{56}    & \url{https://github.com/2022CIRCLE/CIRCLE} \\
    TRANSFER~\cite{meng2022improving} & Java  & \revise{Defects4J} & \delete{Github}\revise{GitHub} & valid & \ding{52}   & \ding{52}   & \ding{52}   & \url{https://github.com/mxx1219/TRANSFER} \\
    DEAR~\cite{li2022dear} & Java  & \revise{Defects4J, CPatMiner, BigFix} & \delete{Github}\revise{GitHub} & valid & \ding{52}   & \ding{52}   & \ding{52}   & \url{https://github.com/AutomatedProgramRepair-2021/dear-auto-fix} \\
    Cornor \etal~\cite{connor2022can} & Java  & \revise{CodeXGlue} & \delete{Github}\revise{GitHub} & valid & \ding{52}   & \ding{52}   & \ding{56}    & \url{https://github.com/WM-SEMERU/hephaestus} \\
    BATS~\cite{tian2022predicting} & Java  & \revise{Defects4J} & \delete{Github}\revise{GitHub} & valid & \ding{52}   & \ding{52}   & \ding{52}   & \url{https://github.com/HaoyeTianCoder/BATS} \\
    T5~\cite{mastropaolo2022using} & Java  & \revise{BFP-small, BFP-medium} & \delete{Github}\revise{GitHub} & valid & \ding{52}   & \ding{52}   & \ding{52}   & \url{https://github.com/antonio-mastropaolo/TransferLearning4Code} \\
    CompDefect~\cite{ni2022defect} & Java  &  \revise{Function-SStuBs4J} & zenodo & valid & \ding{52}   & \ding{52}   & \ding{56}    & \url{https://zenodo.org/record/5353354\#.Y4CVdhRByUl} \\
    VRepair~\cite{chen2022neural} & C     & \revise{Big-Vul, CVEfixes} & \delete{Github}\revise{GitHub} & valid & \ding{52}   & \ding{52}   & \ding{52}   & \url{https://github.com/SteveKommrusch/VRepair} \\
    SeqTrans~\cite{chi2022seqtrans} & Java  & \revise{BFP-small, BFP-medium, Ponta} & \delete{Github}\revise{GitHub} & valid & \ding{52}   & \ding{52}   & \ding{52}   & \url{https://github.com/chijianlei/SeqTrans} \\
    VulRepair~\cite{fu2022vulrepair} & C     & \revise{CVEfixes} & \delete{Github}\revise{GitHub} & valid & \ding{52}   & \ding{52}   & \ding{52}   & \url{https://github.com/awsm-research/VulRepair} \\
    Crex~\cite{yan2022crex} & C     & \revise{CodeFlaw} & \delete{Github}\revise{GitHub} & valid & \ding{52}   & \ding{52}   & \ding{52}   & \url{https://github.com/1993ryan/crex} \\
    RealiT~\cite{richter2022can} & Python & \revise{PyPIBug} & \delete{Github}\revise{GitHub} & valid & \ding{52}   & \ding{56}    & \ding{52}   & \url{https://github.com/cedricrupb/nbfbaselines} \\
    GPT-2~\cite{lajko2022towards} & JS    & \revise{BugsJS} & \delete{Github}\revise{GitHub} & valid & \ding{52}   & \ding{52}   & \ding{52}   & \url{https://github.com/RGAI-USZ/APR22-JS-GPT} \\
    CoditT5~\cite{zhang2022coditt5} & Java  & \revise{BFP-small, BFP-medium} & \delete{Github}\revise{GitHub} & valid & \ding{52}   & \ding{52}   & \ding{52}   & \url{https://github.com/EngineeringSoftware/CoditT5} \\
    SYNSHINE~\cite{ahmed2022synshine} & Java  &  \revise{BlackBox} & zenodo & valid & \ding{52}   & \ding{52}   & \ding{52}   & \url{https://zenodo.org/record/4572390\#.Y4CY8xRByUk} \\
    Verifix~\cite{ahmed2022verifix} & C     & \revise{ITSP} & \delete{Github}\revise{GitHub} & valid & \ding{52}   & \ding{52}   & \ding{52}   & \url{https://github.com/zhiyufan/Verifix} \\
    Cache~\cite{lin2022context} & Java  & \revise{wang~\etal~\cite{wang2020automated}, Tian~\etal~\cite{tian2020evaluating}, ManySStuBs4J} & \delete{Github}\revise{GitHub} & valid & \ding{52}   & \ding{52}   & \ding{56}    & \url{https://github.com/Ringbo/Cache} \\
    Wang \etal~\cite{wang2022attention} & Java  & \revise{Wang~\etal~\cite{wang2022attention}} & \delete{Github}\revise{GitHub} & valid & \ding{52}   & \ding{52}   & \ding{52}   & \url{https://github.com/anonymous0903/patch\_correctness} \\
    Quatrain~\cite{tian2022change} & Java  & \revise{Defects4J, Bugs.jar, Bears} & \delete{Github}\revise{GitHub} & valid & \ding{52}   & \ding{52}   & \ding{52}   & \url{https://github.com/Trustworthy-Software/Quatrain} \\
    Shibboleth~\cite{ghanbari2022patch} & Java  & \revise{Defects4J} & \delete{Github}\revise{GitHub} & valid & \ding{52}   & \ding{52}   & \ding{52}   & \url{https://github.com/ali-ghanbari/shibboleth} \\
    Tian \etal~\cite{tian2022best} & Java  & \revise{Tian~\etal~\cite{tian2022best}} & \delete{Github}\revise{GitHub} & valid & \ding{52}   & \ding{52}   & \ding{52}   & \url{https://github.com/HaoyeTianCoder/Panther} \\
    SSC~\cite{devlin2017semantic} & Python & \revise{Devlin~\etal~\cite{devlin2017semantic}} & \delete{Github}\revise{GitHub} & valid & \ding{56}    & \ding{56}    & \ding{56}    & \url{https://iclr2018anon.github.io/semantic_code_repair/index.html} \\
    Huang \etal~\cite{huang2021application}  & Java,C,C++ & \revise{Juliet Test Suite} & \delete{Github}\revise{GitHub} & valid & \ding{52}   & \ding{52}   & \ding{52}   & \url{https://github.com/shan-huang-1993/PLC-Pyramid} \\
    \bottomrule
    \end{tabular}%
    }
  \label{tab:tool}%
\end{sidewaystable}%

%% file: sec/sec_guidelines.tex
Our study reveals the following important practical guidelines for future learning-based APR.

\revise{
\textit{\bf I\&G\ding{182}: Multifarious Code Representation.}
As discussed in Section~\ref{sec:patch_generation}, inspired by the advance of neural machine translation in NLP, early learning-based APR work usually treats source code as a sequence of code tokens.
The follow-up work has begun to consider complex code features, such as code edit~\cite{zhu2021syntax}, AST~\cite{li2020dlfix}, and data flow graph~\cite{nguyen2021graphix}.
For example, CIRCLE~\cite{yuan2022circle} which treats the APR as a simple machine translation task on code sequences, and Recoder~\cite{zhu2021syntax} which is equipped with a syntax-guided edit decoder, are able to fix 64 and 65 real-world software bugs from the Defects4J benchmark, respectively.
Such observation indicates that there does not always exist a specific code representation to demonstrate good performance, \eg  a simple sequence representation can also yield excellent results.
It is difficult to directly investigate the advantages and disadvantages of different code representations because each code representation comes with differentiated configurations, such as model architectures.}

\revise{
We recommend that future work can be conducted in the following three directions.
First, it is crucial to conduct a systematic study to explore the impact of different code representations under various configurations, \eg model architectures.
Second, we find there exists a mass of code representation ways in existing learning-based APR techniques, future work needs to design optimal code representation based on specific scenarios.
For example, researchers can design optimal code representations based on specific programming languages, types of bugs and benchmarks.
Third, considering that most existing repair work statically extracts the buggy and contextual features, it is promising to incorporate the static code representation features (\eg AST) and dynamic execution feedback (\eg test results).
In this way, the NMT model and the repair process can be more deeply integrated to fit the APR scenario.
Fourth, with the rise of pre-trained models, the community has seen the usage of prompt-based representation of feeding inputs to pre-trained models to facilitate the repair task, \eg CIRCLE~\cite{yuan2022circle}.
However, research about prompt-based representation in the repair domain is still in its early stages, mainly focusing on fine-tuning~\cite{yuan2022circle}. 
In the future, researchers can draw from other code-related fields~\cite{nashid2023retrieval,zhang2023pre} to further deepen the understanding of how the knowledge of pre-trained models can be stimulated to support repair tasks with appropriate prompt representation.

}

\revise{
\textit{\bf I\&G\ding{183}: Patch Validation Acceleration.}
As discussed in Section~\ref{sec:patch_validation}, dynamic execution is the common practice to validate candidate patches in the APR community.
Although some techniques have been proposed to speed up patch validation~\cite{benton2022towards,chen2021fast}, it is time-consuming to dynamically execute all candidate patches against each test case.
Besides, existing patch validation studies in learning-based APR are general to both traditional and learning-based APR communities.

We recommend that future research can be conducted from three aspects.
First, it is promising to extensively investigate the differences between patches generated by traditional and learning-based APR techniques, based on which more advanced patch validation techniques can be designed that are targeted at learning-based APR techniques.
Second, predictive patch validation can be conducted on top of the code semantic understanding capability of DL techniques, \ie \textit{predictive patch validation}.
For example, automatically learning patched code features and predicting whether a patch is passed by previous failing test cases without dynamic execution is promising.
Third, we notice that other fields also suffer from the problem of dynamic program execution overhead, such as mutation testing (both mutants and patches are considered variants of a program).
Therefore, some advanced techniques from these similar fields can also be migrated into patch validation.
For example, Wang~\etal~\cite{wang2017faster,wang2021faster} detect equivalencies in mutant execution and execute one for each equivalence class, which is general and applicable to patch validation.
}

\revise{
\textit{\bf I\&G\ding{184}: Training Dataset Construction.}
As discussed in Section~\ref{sec:dataset}, in contrast to traditional APR techniques, learning-based techniques heavily rely on the quality of the training dataset.
A majority of existing techniques mine bug-fixing pairs from open-source code repositories (\eg GitHub) and build their own datasets.
However, the training dataset is usually collected by automated tools (\eg extracting commit by fix-related keywords) and then inspected by some filtering rules (\eg more than five Java files)~\cite{zhu2021syntax}, which means the quality of the training dataset can be variant.
Many training datasets contain noise (\eg CoCoNut contains a number of duplicated samples) that may reduce the performance of the model.
Besides, the number of training samples in different techniques varies greatly (\eg 3,241,966 in CoCoNut~\cite{lutellier2020coconut} and 2,000 in DLFix~\cite{li2020dlfix}).
These concerns may introduce bias when comparing and analyzing learning-based techniques.

We recommend approaching future work in two parts. 
First, a unified standard for training datasets should be built to reduce the burden on researchers when they propose a novel learning-based APR technique.
Second, with a standardized training dataset, researchers can uniformly evaluate the performance of different repair models across various settings, such as code representations, model architectures, and training hyperparameters.
}
  
\revise{
\textbf{I\&G\ding{185}: Practical Evaluation Metrics.}
As discussed in Section~\ref{sec:metric}, when evaluating repair performance, dynamic execution-based metrics (\eg plausible patches) are the common practice in the APR community.
However, such metrics may suffer from some drawbacks.
First, they need to execute all available functional test suites against each patched software program, consuming a significant amount of execution time.
Section~\ref{sec:patch_validation} lists some candidate patch validation acceleration techniques to mitigate this issue.
Second, due to the overfitting problem, developers are required to further perform a manual inspection to assess the correctness of plausible patches, which demands a substantial amount of human resources and is prone to errors.
The overfitting problem leads to the development of some patch correctness assessment techniques in Section~\ref{sec:patch_validation}.
Third, the dynamic execution heavily relies on well-constructed datasets including the corresponding fault-triggering test cases. 
However, such test cases are often unavailable in practical scenarios, making it challenging to rely solely on such metrics.
We encourage further work to explore more practical metrics to evaluate the repair performance of NMT models.
For example, it is interesting to design a hybrid metric by combining dynamic execution and static match.}

\revise{
We find an increasing number of recent learning-based APR techniques~\cite{fu2022vulrepair,tufano2019empirical} rely on static match-based metrics (\eg Accuracy and BLUE) to perform evaluation (mentioned in Section~\ref{sec:metric}).
However, such match-based metrics are usually derived from the NLP domain (\eg neural machine translation) and fail to consider that a program's functionality can be implemented in various ways, such as different algorithms, data structures, or data flows.
In the future, the community needs large-scale empirical work to validate whether the match-based metrics can accurately reflect the repair capability of NMT APR models.
Besides, the two types of evaluation metrics (\ie dynamic execution \textit{vs.} static match) are orthogonal and have their own advantages and disadvantages.
We suggest that the relationships between the recent static match-based and the classical dynamic test execution-based metrics need to be studied in the future.}

\revise{
\textit{\bf I\&G\ding{186}: \newrevise{Exploring} Patch Overfitting Issue}.
Similar to traditional APR techniques, learning-based techniques usually adopt available test suites to filter incorrect candidate patches.
However, the test suite is an incomplete specification under the program behavioral space.
The plausible patches passing the existing test suite may not satisfy the expected outputs of potential test suites, leading to a long challenge in APR (\ie the overfitting issue).
Considering the learning-based APR is an end-to-end repair paradigm (in a black-box manner), which is different from traditional techniques adopting test suites to guide the repair process, the overfitting issue in learning-based APR is more significant and severe.
Recently, researchers have adopted DL techniques (\eg code embedding~\cite{tian2020evaluating, lin2022context}) to predict the correctness of plausible patches, which is a promising direction to address overfitting problems.}

\revise{
We recommend that future work can be conducted from three aspects.
The first recommendation lies in the process of patch generation.
It is possible to design advanced code-aware NMT models that incorporate more code information (\eg code structure information or dynamic execution information) to generate high-quality code snippets.
The second recommendation is the process of patch correctness.
Investigating how to better utilize DL techniques to differentiate between correct patches and overfitting patches is worth exploring.
For example, we can incorporate contrastive learning into existing learning-based patch correctness assessment approaches, as contrastive learning is shown to be effective in distinguishing positive samples (i.e., correct patches) and negative samples (i.e., overfitting patches).
The third recommendation is the repair paradigm.
Previous work~\cite{liu2019tbar,xia2022practical} has shown that fix templates can generate higher-quality code snippets with high precision.
We believe that combining DL techniques with fix patterns as a novel repair paradigm can address this issue in previous learning-based APR techniques.
}

\revise{
\textit{\bf I\&G\ding{187}: Unified Localization and Repair workflow}.
As discussed in Section \ref{sec:fl}, similar to traditional APR techniques, existing learning-based techniques usually consider fault localization as an additional step in the repair process and adopt off-the-shelf fault localization tools (\eg SBFL) to identify suspicious code element, which is the input of NMT repair models.
In the literature, these two tasks (\ie fault localization and patch generation) are developing in their own respective fields so far and little work has explored their potential relationship. 
Recently, Ni~\etal~\cite{ni2022defect} propose CompDefect to handle defect prediction and repair simultaneously.
The powerful capacity of DL to learn the semantic information of source code for fault localization~\cite{lou2021boosting,li2021fault} and program repair~\cite{yuan2022circle,zhu2021syntax} makes it possible to combine the two tasks.
}

\revise{We suggest that future works focus on a unified repair process interactively incorporating fault localization and patch generation.
The fault localization results can be improved with the feedback from the patch generation, while the updated localization results can assist in generating patches more effectively.
Different from previous studies that treat the two tasks as separate, the unified repair facilitates interaction between the two tasks, enabling feedback-driven improvements in both localization and repair performance iteratively.
}

\revise{
\textbf{\bf I\&G\ding{188}: Combination with Traditional APR Techniques}.
As discussed in Section~\ref{sec:patch_generation}, existing DL techniques are usually adopted as a patch generator in the learning-based APR workflow, which takes the buggy code snippets as inputs and returns a ranked list of candidate patches.
Despite remarkable progress, such learning-based APR techniques need to generate correct code snippets from scratch and are developed separately from traditional APR techniques.
Previous work~\cite{zhu2021syntax} has demonstrated that learning-based APR is complementary to traditional repair techniques in terms of fixed bugs.}

\revise{
Future work can be conducted in two aspects.
First, it is interesting to design a predictive APR technique to predict the optimal traditional or learning-based APR technique for a given buggy project based on the program analysis. 
Second, it is flexible to integrate DL techniques into traditional APR techniques as a component instead of developing a brand-new end-to-end patch generator.
For example, a state-of-the-art template-based APR tool TBar retrieves relevant donor code from the local buggy file and may fail to generate correct patches with inappropriate donor code with the correct fix pattern.
Researchers can boost existing template-based APR techniques (\eg TBar) via pre-trained models, which contain generic knowledge pre-trained with millions of code snippets from open-source projects, and provide a variety of donor code to fix different bugs.
}

\revise{
\textbf{I\&G\ding{189}: Exploring Domain Repair Techniques}.
As discussed in Section~\ref{sec:patch_generation}, a majority of learning-based APR techniques focus on semantic bugs, which have been investigated intensively in the literature.
Section~\ref{sec:patch_generation} also summarizes a number of existing repair techniques considering other types of bugs, such as security vulnerabilities and programming assignments.
However, these studies only account for a small proportion of existing techniques, and the types of investigated bugs are also very limited.
}

\revise{
We recommend that future work can be carried out from two perspectives.
First, it is promising to design more domain-specific learning-based APR techniques in repairing other diverse scenarios, \eg test repair, concurrency program repair, and API misuse repair.
Second, we find the community usually treats fixing these types of bugs as separate tasks.
SequenceR~\cite{chen2019sequencer} has demonstrated that NMT-based models only trained on a limited bug-fixing corpus can already fix notable vulnerabilities.
These results indicate that bug fixing and vulnerability repair both aiming to fix errors in the source code have a high degree of similarity, and the knowledge learned from bug fixing can be well transferred to vulnerability repair.
Such observation motivates that some bugs with different types are very similar in both code patterns and repair workflow.
Thus, future researchers are recommended  to explore their potential relationship and investigate whether these bugs can benefit each other.
Besides, it is promising to conduct some empirical studies to migrate existing mature learning-based APR techniques to other scenarios, such as automated vulnerability repair.
}

\revise{
\textbf{I\&G\ding{190}: Explainable Patch Generation}.
As discussed in Section~\ref{sec:patch_generation}, existing learning-based APR techniques usually perform an end-to-end patch generation in a black-box manner, \ie automatically transforming the buggy code snippets into correct ones on top of an NMT model.
The developers are unaware of why NMT models predict such results, thus unsure about the reliability of these generated patches, hindering the adoption of repair NMT models in practice.
In the literature, a majority of studies focus on improving repair accuracy, while minor focus on improving the explainability of such NMT models.
In the future, advanced explainable techniques can be considered to make the predictions of NMT repair models more practical, explainable, and actionable.

We suggest that future work should concentrate on two aspects to support the understanding of NMT models for program repair: the attention mechanism and input perturbation mechanisms.
As a white-box method, the attention mechanism generates explanations by assigning weights to different parts of the input, thus indicating an attribution of importance for the prediction. 
On the other hand, the input perturbation mechanism is a black-box method to modify the input data and observe variations in the model's output, helping to understand which parts of the input the model deems most crucial.
}

\revise{
\textit{\bf I\&G\ding{191}: Pre-trained Model-based APR Research.}
As discussed in Section~\ref{sec:pre-trained}, an increasing number of APR studies are focusing on employing pre-trained language models to generate patches.
We have already seen pre-trained models being successfully applied to the APR domain with promising results~\cite{xia2022practical,zhang2023critical}.
In the future, pre-trained models will still be the main trend for follow-up research, and there is still a lot of room for further improvement.
We stress the importance of conducting more research into pre-trained models to deepen our understanding of the existing challenges in developing APR techniques. 
We describe the relevant topics in the following.}

\begin{enumerate}[(1)]

\revise{
\item \textit{Patch Correctness via Pre-trained Models.}
Recently, the research for generating patches on top of pre-trained models is developing rapidly.
However, patch correctness, as an important research direction in the APR community, has not benefited much from these pre-trained models. 
For example, Tian~\etal~\cite{tian2020evaluating} simply regard BERT as an embedding representation approach without investigating the benefits of the pre-training component itself. 
We believe that future work can be conducted to employ the rich programming knowledge contained in pre-trained models to identify the relationship between correct patches and overfitting patches.
For example, it is promising to employ the pre-trained model as a component in existing patch validation techniques.
Besides, researchers can directly treat the patch correctness assessment as a code classification task, and fine-tune off-the-shelf pre-trained models on patch-specific datasets.

\item \textit{Repair-oriented Pre-trained Model}
We have seen an increasing number of pre-trained models in the APR field. 
In the literature, the majority of these pre-trained models are designed with a general-purpose pre-training approach to facilitate a variety of downstream tasks.
However, considering the distinct difference between these downstream tasks, the universal pre-trained model may hinder the effectiveness of program repair. 
For example, these models usually focus on code-related tasks to encode a given code snippet, such as code search and code summarization. 
Specifically, the designed pre-training tasks (e.g., masked language modeling) typically deal with a code snippet as the input, and the key challenge is to capture the syntactic and semantic information of the code snippet. 
However, APR deals with two code snippets and the key challenge is to understand the code change patterns in bug-fixing pairs. 
The learned knowledge in existing pre-trained models is generally related to the syntactic and semantic information of code snippets, which can hardly be exploited to encode bug-fixing pairs.
Thus, employing existing pre-trained models for APR will inevitably lead to inconsistent inputs and objectives between pre-training and fine-tuning.
It is sub-optimal to fine-tune existing pre-trained code models for APR due to the natural differences between pre-training objectives and APR.
We recommend future work to explore domain-specific models for APR.
For example, the researcher can propose a repair-oriented pre-trained model, which takes two code snippets as inputs to learn the domain knowledge about code change patterns with bug-fixing specific pre-training objectives.

\item \textit{Trade-off between Effectiveness and Model Size}.
In the literature, recent learning-based APR techniques tend to employ the growing size of models, achieving better performance.
Xia~\etal~\cite{xia2022practical} have demonstrated that larger models usually repair a greater number of software bugs, highlighting the promising future of pre-trained models for APR.
However, such large models are difficult to deploy in the development workflow.
Besides, with the release of ever-larger models, there may exist a barrier in the trade-off between effectiveness and model size.
In fact, most existing pre-trained models in the APR literature (\eg CIRCLE~\cite{yuan2022circle}, AlphaRepair~\cite{xia2022less} and VulRepair~\cite{fu2022vulrepair}) usually treat source code as natural language (\ie code sequence), which cannot capture the code structure features.
In the future, investigating how to bring in code features and program analysis (\eg data flow or control flow) in pre-training may be a flexible strategy instead of employing a larger mode size.

\item
\textit{Practical Pre-trained Repair model}.
As discussed in Section~\ref{sec:pre-trained}, an increasing number of learning-based APR techniques attempt to generate candidate patches by large pre-trained language models.
Although remarkable progress is obtained, such repair models contain millions or even billions of parameters.
For example, CodeBERT has 125 million parameters and 476 MB model size in total.
It is significant to deploy these models in modern IDEs to assist developers during software development and maintenance.
However, these repair models consume huge device resources and run slowly in the development workflow (\eg IDEs), limiting their application in practice.
In the future, it is necessary to reduce the size of these repair models to deploy in real-world scenarios while maintaining comparable prediction accuracy, such as model pruning and knowledge distillation.

\item \textit{Pre-trained Model-based Repair Chatbot.}
At the current stage, the goal of most learning-based APR techniques is to automatically generate patches that pass available test cases without human intervention, similar to traditional APR techniques. 
However, there are some long-term challenges in deploying these APR techniques directly into the development process, such as the low recall of repaired bugs and the low precision of correct patches~\cite{liang2021interactive}.
Recently, the natural language understanding capabilities of large pre-trained models (\eg ChatGPT) have provided a new direction, \ie conversation-driven repair.
Specifically, we can employ the large pre-trained model as a repair chatbot, which can converse with developers just like a human to provide potential fix suggestions.
In such a human-machine conversation process, developers can tell the repair chatbot useful debugging information, such as suspicious code statements and bug reports.
More importantly, the patches generated by the repair chatbot can be validated by developers and external devices (\eg static analysis tools and compilers), and then the feedback (\eg dynamic execution information) can be provided to the chatbot for further optimization.
}

\end{enumerate}

%% file: sec/sec_guidelines_old.tex
\delete{
\textbf{The quality of the training dataset is important.}
In contrast to traditional APR techniques, learning-based techniques heavily rely on the quality of the training dataset.
A majority of existing techniques mine bug-fixing pairs from open-source code repositories (e.g., GitHub) and build their own datasets.
However, the training dataset is usually collected by automated tools (e.g., extracting commit by fix-related keywords) and then inspected by some filtering rules (e.g., more than five Java files) \cite{zhu2021syntax}, which means the quality of the training dataset can be variant.
Many training datasets contain noise (e.g., CoCoNut contains a number of duplicated samples) that may reduce the performance of the model.
Besides, the number of training samples in different techniques varies greatly (e.g., 3,241,966 in CoCoNut \cite{lutellier2020coconut} and 2,000 in DLFix \cite{li2020dlfix}).
These concerns may introduce bias when comparing and analyzing learning-based techniques.
Thus, a unified standard for training datasets should be built to reduce the burden on researchers when they evaluate the performance of different repair models.
}
  
\delete{
\textbf{More practical evaluation metrics are needed.}
Recently, when evaluating repair performance, an increasing number of learning-based techniques rely on static match-based metrics, which are derived from NLP.
However, such metrics fail to consider that a program's functionality can be implemented in various ways, such as different algorithms, data structures, or data flows. 
It is unclear whether the match-based metrics can reflect the repair capability of NMT models.
Besides, the relationships between the static match-based and dynamic test execution-based metrics need to be studied in the future.}

\delete{
\textbf{Code features need to be studied.}
Inspired by the advance of machine translation in NLP, early learning-based APR work treats source code as a sequence of tokens.
The follow-up work has begun to consider complex code features, such as code edit, \cite{zhu2021syntax}, AST \cite{li2020dlfix} and data flow graph \cite{nguyen2021graphix}.
However, the most recent technique CIRCLE, treating the APR as a simple machine translation task on code sequences, still achieves state-of-the-art results. 
Such observation indicates that simple features require more attention in future work. 
Besides, considering the mass of code representation ways in learning-based APR, it is crucial to conduct a systematic study to explore the impact of different code representations under various model architectures and benchmarks.}

\delete{
\textbf{Overfitting issue still exists}.
Similar to traditional APR techniques, learning-based techniques usually adopt available test suites to filter incorrect candidate patches.
However, the test suite is an incomplete specification under the program behavioral space.
The plausible patches passing the existing test suite may not satisfy the expected outputs of potential test suites, leading to a long challenge in APR (i.e., the overfitting issue).
Considering the learning-based APR is an end-to-end repair paradigm (in a black-box manner), which is different from traditional techniques adopting test suites to guide the repair process, the overfitting issue in learning-based APR is more significant and severe.
Recently, researchers have adopted DL techniques (e.g., code embedding \cite{tian2020evaluating, lin2022context}) to predict the correctness of plausible patches, which is a promising direction to address overfitting problems.
We also recommend designing some advanced NMT repair frameworks to generate high-quality patches.}

\delete{
\textbf{Unified repair in urgent}.
As discussed in Section \ref{sec:fl}, similar to traditional APR techniques, existing learning-based techniques usually consider fault localization as an additional step in the repair process and adopt off-the-shelf fault localization tools (e.g., SBFL) to identify suspicious code element, which is the input of NMT repair models.
In the literature, these two tasks (i.e., fault localization and patch generation) are developing in their own respective fields so far and little work has explored their potential relationship. 
Recently, Ni et al. \cite{ni2022defect} propose CompDefect to handle defect prediction and repair simultaneously.
The powerful capacity of DL to learn the semantic information of source code for fault localization \cite{lou2021boosting,li2021fault} and program repair \cite{yuan2022circle,zhu2021syntax} makes it possible to combine the two tasks.} 

\delete{
\textbf{Practical NMT repair model is needed}.
An increasing number of learning-based techniques attempt to generate patches by large language models.
Although remarkable progress is obtained, such NMT models contain millions or even billions of parameters.
For example, CodeBERT has 125 million parameters and 476 MB model size in total.
It is significant to deploy these models in modern IDEs to assist developers during software development and maintenance.
However, these repair models consume huge device resources and run slowly in the development workflow (e.g., IDEs), limiting their application in practice.
In the future, It is promising to reduce the size of these repair models to deploy in real-world scenarios while maintaining comparable accuracy, such as model pruning and knowledge distillation.}

\delete{
\textbf{Model size is not the only option}.
As discussed before, learning-based APR techniques tend to employ the growing size of models, achieving better performance.
Xia et al. \cite{xia2022practical} have demonstrated that larger models usually repair a greater number of bugs, highlighting the promising future of pre-trained models for APR.
However, such large models are difficult to deploy in the development workflow.
Besides, with the release of ever-larger models, there may exist a barrier in the trade-off between effectiveness and model size.
In fact, existing pre-trained models in APR usually treat source code as natural language, which cannot capture the code features.
In the future, investigating how to bring in code structures (e.g., data flow or control flow) in model training may be a flexible strategy instead of employing a larger mode size.}

\delete{
\textbf{Combined with traditional APR techniques}.
Existing DL techniques are usually adopted as a patch generator in APR workflow, which takes the buggy code snippets as inputs and returns a ranked list of candidate patches.
Despite remarkable progress, such learning-based APR techniques are developed separately from traditional techniques.
Previous work \cite{zhu2021syntax} has demonstrated that learning-based is complementary to traditional techniques in terms of fixed bugs.
Thus, it is flexible to integrate DL techniques into traditional APR techniques instead of developing a brand-new end-to-end patch generator.
For example, Meng et al. \cite{meng2022improving} design a multi-classifier to rank the fix templates for TBar.
In the future, researchers can boost existing template-based APR techniques (e.g., TBar) via mask prediction.
}

\delete{
\textbf{Domain repair techniques are needed}.
A majority of learning-based techniques focus on semantic bugs, which are investigated intensively.
However, only a small amount of existing techniques consider other types of bugs, such as security vulnerabilities or programming assignments.
The community usually treats fixing these types of bugs as separate tasks.
SequenceR \cite{chen2019sequencer} has demonstrated that NMT-based models only trained on a limited bug-fixing corpus can already fix notable vulnerabilities.
These results indicate that bug fixing and vulnerability repair both aiming to fix errors in the source code have a high degree of similarity and the knowledge learned from bug fixing can be well transferred to vulnerability repair.
Such observation motivates future researchers to explore their potential relationship and investigate whether these tasks can benefit each other.
Besides, it is promising to migrate existing mature learning-based bug-fixing techniques to automated vulnerability repair.
}

\delete{
\textbf{Explainable Patch Generation}.
Traditional APR techniques generate patches along with a log output, which contains detailed information in the generation process, while learning-based APR techniques perform an end-to-end patch generation due to the interpretability of DL.
Thus, the developers are unaware of why repair models predict such results, hindering the adoption of repair models in practice.
In the literature, a majority of studies focus on improving repair accuracy, while minor focus on improving the explainability of such repair models.
In the future, advanced explainable techniques can be considered to make the predictions of NMT repair models more practical, explainable, and actionable.
}